\documentclass[11pt,aps,showpacs,notitlepage,nofootinbib]{revtex4-1}
\usepackage{amssymb,amsmath}
\usepackage{bm}
\usepackage[pdftex]{graphicx}
\usepackage{subfig}
\usepackage{color}
\usepackage{amsthm}
\usepackage[update,prepend]{epstopdf}

\usepackage{footnote}

\newcommand{\beq}{\begin{equation}}
\newcommand{\beqn}{\begin{eqnarray}}
\newcommand{\eeq}{\end{equation}}
\newcommand{\eeqn}{\end{eqnarray}}

\begin{document}

\title{Preheating after Higgs Inflation: \\ Self-Resonance and Gauge boson production}
\author{Evangelos I. Sfakianakis$^{1,2}$ and Jorinde van de Vis$^1$}
\email{\baselineskip 11pt Email addresses: e.sfakianakis@nikhef.nl ; jorindev@nikhef.nl}
\affiliation{
$^1$Nikhef, Science Park 105, 1098XG Amsterdam, The Netherlands
\\
$^2$Lorentz Institute for Theoretical Physics, Leiden University, 2333CA Leiden, The Netherlands
}
\date{\today}
\begin{abstract}
We perform an extensive analysis of linear fluctuations during preheating in Higgs inflation in the Einstein frame, where the fields are minimally coupled to gravity, but the field-space metric is nontrivial. The self-resonance of the Higgs and the Higgsed gauge bosons are governed by effective masses that scale differently with the nonminimal couplings and evolve differently in time. Coupled metric perturbations enhance Higgs self-resonance and make it possible for Higgs inflation to preheat solely through this channel. For $\xi\gtrsim 100$ the total energy of the Higgs-inflaton condensate can be transferred to Higgs particles within $3$ $e$-folds after the end of inflation. For smaller values of the nonminimal coupling preheating takes longer, completely shutting off at around $\xi\simeq 30$. 
The production of gauge bosons is dominated by the gauge boson mass and the field space curvature. For large values of the nonminimal coupling $\xi \gtrsim 1000$, it is possible for the Higgs condensate to transfer the entirety of its energy into gauge fields within one oscillation. For smaller values of the nonminimal coupling gauge bosons decay very quickly into fermions, thereby shutting off Bose enhancement. Estimates of non-Abelian interactions indicate that they will not suppress preheating into gauge bosons for $\xi \gtrsim 1000$.

\end{abstract}
\pacs{Preprint Nikhef  2018-044 }
\maketitle

   \newpage

\section{Introduction} 

While the discovery of the Higgs boson at CERN \cite{Higgsdiscovery} solidified our understanding of the Standard Model (SM), its behavior in the early universe, above the electroweak symmetry-breaking scale, remains unsure. An intriguing possibility is the identification of the Higgs boson with the scalar field(s) necessary for driving inflation, the rapid acceleration phase of the universe required to both solve the horizon and flatness problems, as well as seed primordial fluctuations necessary for structure formation \cite{LythRiotto,GuthKaiser,Mazumdar}. 

The original attempt to use the Higgs or a Higgs-like sector to drive inflation resulted in an inconsistently large amplitude of fluctuations \cite{Guth:1982ec}, because of the value of the Higgs self-coupling $\lambda$ in the Standard Model.
However, the introduction of a nonminimal coupling between the Higgs field and the Ricci scalar can remedy this \cite{Higgsinfl}.
Such nonminimal couplings are not only generic, since they arise as necessary renormalization counter-terms for scalar fields in curved spacetime \cite{Callan, Bunch, BirrellDavies, Buchbinder, ParkerToms, Odintsov1991, Bounakis:2017fkv, Markkanen2013, Fujii,Faraoni2004,Buchbinder}, but they also grow without a UV fixed point under renormalization-group flow - at least below the Planck scale \cite{Buchbinder}. The inherent ambiguity in the running of the Higgs self-coupling $\lambda$ at high energies, due to our incomplete knowledge of possible new physics between the ${\rm TeV}$ and inflationary scales, leads to an ambiguity for the exact value of the required nonminimal coupling \cite{SimoneHertzbergWilczek, BezrukovMass, Barvinsky}. While simple estimates like $\lambda_{\rm infl} = {\cal O}(0.01)$ lead to the requirement of $\xi = {\cal O}\left(10^{4}\right)$, smaller values of $\lambda$ can allow for much smaller nonminimal couplings. We will remain agnostic about the exact running of the Standard Model couplings at high energies and instead explore a broad parameter range\footnote{For inflation on the flat plateau one should consider $\xi \gtrsim 440$ (e.g. \cite{Allison:2013uaa}). In models of hilltop or inflection point inflation, smaller values of $\xi$ are possible, although UV corrections are expected to be larger. In order to provide a  treatment of Higgs inflation as complete as possible without referring to specific unknown physics, we choose to consider a broad range of non-minimal couplings that go below $\xi\approx400$.}
 covering $10\lesssim \xi \lesssim 10^4$.

A basic feature of inflationary models with nonminimal couplings is that they provide universal predictions for the spectral observables $n_s$ and $r$, largely independent of the exact model parameters and initial conditions \cite{KS, LindeRoest}. These observables fall in line with the Starobinsky model \cite{Starobinsky:1980te} as well as with the large family of $\alpha$-attractors \cite{Lindealpha}\footnote{See \cite{Christodoulidis:2018qdw} for a way to alter the predictions of $\alpha$-attractor models through multi-field effects.}
. Even after the latest Planck release \cite{Higgs2018}, these models, which predict $n_s = 1-2/N_*$ and $r={\cal O}(1/N_*^2)$, continue to be compatible with the data for modes that exit the horizon at $N_* \simeq 55$ $e$-folds before the end of inflation.

While inflation provides a robust framework for computing the evolution of the universe and the generation of fluctuations \cite{GuthKaiser,BTW,LythLiddle,Baumann,MartinRingeval,GKN,LindePlanck,MartinRev},
 the transition from an inflating universe to a radiation bath (as required for big-bang nucleosynthesis \cite{Steigman,FieldsBBN,Cyburt}), 
 known as reheating,  remains a weakly constrained era in the cosmic evolution. Despite the difficulty of directly observing reheating due to the very short length-scales involved, knowledge of how the equation of state of the universe transitioned from $w\simeq-1$ to $w=1/3$ is crucial, since it affects how one relates the observed CMB modes to the time during inflation when they exited the horizon \cite{AdsheadEasther,Dai,Creminelli,MartinReheat,GongLeungPi,CaiGuoWang,Cook,Heisig}. This becomes increasingly relevant, as new data shrink the experimental bounds on primordial observables.

The transfer of energy from the inflaton, which carries (almost) the entirety of the energy-density of the universe during inflation, to radiation degrees of freedom, can occur either through perturbative decays, or through nonperturbative processes. The latter case, denoted as preheating, includes parametric and tachyonic resonances (see Ref.~\cite{AHKK} for a review). The end state of any (p)reheating scenario must be a universe filled with SM and Dark Matter (DM) particles, or at least intermediary particles that decay into the SM and DM sectors. Preheating therefore has the potential to address other long-standing challenges in cosmological theory, such as generating the observed baryon - antibaryon asymmetry \cite{Adshead:2017znw, MPHBaryogenesis,Mustafa,Adshead:2015jza,Adshead:2015kza}, or leaving behind cosmological relics, such as cosmological magnetic fields \cite{Adshead:2015pva, Adshead:2016iae} or primordial black holes \cite{Georg:2017mqk, Carr:2018nkm, Cai:2018rqf}.

Higgs inflation provides a unique opportunity to study the transition from inflation to radiation domination, since the couplings of the Higgs-inflaton to the rest of the SM are known. 
Detailed analyses of reheating in Higgs inflation were first performed in Refs.~\cite{GarciaBellido:2008ab, Bezrukov:2008ut}. However, as discussed later in Ref.~\cite{MultiPreheat1, MultiPreheat2, MultiPreheat3} and independently in Ref.~\cite{Ema:2016dny}, multi-field models of inflation with nonminimal couplings to gravity can exhibit more efficient preheating behavior than previously thought, due to the contribution of the field-space structure to the effective mass of the fluctuations.
Furthermore, it was shown in Refs.~\cite{MultiPreheat1, MultiPreheat2, MultiPreheat3} that, in nonminimally coupled models, preheating efficiency can be vastly different for different values of the nonminimal coupling, even if these values lead to 
 otherwise identical predictions for  CMB observables. We will thus perform a detailed study of preheating in Higgs inflation, extending the results of Ref.~\cite{GarciaBellido:2008ab, Bezrukov:2008ut, MultiPreheat1, MultiPreheat2, MultiPreheat3, Ema:2016dny}, in order to distinguish between Higgs inflation models with different values of the nonminimal coupling.

Because of the appeal of Higgs inflation as an economical model of realizing inflation within the particle content of the Standard Model, the  unitarity cutoff scale has been extensively studied \cite{Burgess,Barvinsky,Hertzberg,unitarity,BezrukovInflaton} (see also Ref.~\cite{Rubio:2018ogq} for a recent review). For large values of the Higgs VEV, like the ones appearing during inflation, the appropriate unitarity cutoff scale is $M_{\rm pl} / \sqrt{\xi}$, while for small values of the Higgs VEV it must be substituted by $M_{\rm pl} / {\xi}$, where $M_{\rm pl} \equiv (8 \pi G)^{-1/2}$ is the reduced Planck mass.

In Section \ref{sec:Formalism} of this work, we introduce a simplified model of a complex Higgs field coupled to an Abelian gauge field. Section \ref{sec:electroweak} describes the generalization of this model to the full electroweak sector of the Standard Model.
In Section \ref{sec:Adiabatic} we study self-resonance of Higgs modes. Section \ref{sec:Isocurvature} deals with the evolution of the gauge fields during and after Higgs inflation. At the end of this section we also address the unitarity scale. 
The decays and scattering processes that involve the produced Higgs and gauge bosons are described in Section \ref{sec:decays} and observational consequences in \ref{sec:observables}.
 Concluding remarks follow in Section \ref{sec:Conclusions}. 


\section{Abelian Model and Formalism}
\label{sec:Formalism}

We build on the formalism of Ref.~\cite{KMS} for the evolution of nonminimally coupled multi-field models, as it was applied in Refs.~\cite{KS, GKS, SSK}  during inflation and in Refs.~\cite{MultiPreheat1, MultiPreheat2, MultiPreheat3} during preheating. The electroweak sector consists of a complex Higgs doublet, expressed using $4$ real-valued scalar fields in $3+1$ spacetime dimensions:
\begin{equation}
	\Phi = \frac {1}{\sqrt 2} \begin{pmatrix} \varphi + h + \theta \\ \phi^3 + \phi^4 \end{pmatrix}\, , 
\end{equation}
where $\varphi$ is the background value of the Higgs field, 
 $h$ denotes the Higgs fluctuations and $\theta, \phi^3$ and $\phi^4$ are the Goldstone modes. We also add the $SU(2)_L$ and $U(1)_Y$ gauge sectors. We will start by closely examining an Abelian simplified model of the full electroweak sector, consisting of the complex scalar field 
\begin{equation}
	\Phi = \frac{1}{\sqrt 2} (\varphi + h + i\theta),
\label{eq:Higgsdecompose}
\end{equation}
and a $U(1)$ gauge field only. The full equations of the Higgsed electroweak sector are given in Section \ref{sec:electroweak}, where we also discuss their relation to the Abelian simplified model.

In order to connect our notation to that of Ref.~\cite{KMS} we identify $\phi^1 = \varphi +h$ and $\phi^2 = \theta$. We will start by deriving the equations of motion for general $\phi^I$-fields for notational simplicity.
  We use upper-case Latin letters to label field-space indices, $I, J = 1, 2, 3, 4$ (or just $I,J = 1, 2$ in the Abelian case); Greek letters to label spacetime indices, $\mu, \nu = 0, 1, 2, 3$; and lower-case Latin letters to label spatial indices, $i, j = 1, 2, 3$. The spacetime metric has signature $(-,+,+,+)$.

We first consider $U(1)$ symmetry with the corresponding gauge field $B_\mu$. The Lagrangian in the Jordan frame is given by:
\begin{equation}
	\begin{aligned}
	S_J = \int d^4x \sqrt{-\tilde g} \Big[&f(\Phi,\Phi^\dagger) \tilde R - \tilde g^{\mu\nu}( \tilde\nabla_\mu \Phi)^\dagger \tilde\nabla_\nu \Phi -\frac{1}{4} \tilde g^{\mu\rho}\tilde g^{\nu\sigma}  F_{\mu\nu} F_{\rho\sigma}  -\tilde V(\Phi,\Phi^\dagger)  \Big]\, .
	\end{aligned}
\end{equation}
The covariant derivative $\tilde\nabla_\mu$ is given by:
\begin{equation}
	\tilde\nabla_\mu = \tilde D_\mu + i e B_\mu,
\end{equation}
where $\tilde D_\mu$ is a covariant derivative with respect to the space-time metric $\tilde g_{\mu\nu}$ and $e$ is the coupling constant. The corresponding field strength tensor\footnote{The tensor $F_{\mu\nu}$ is defined with lower indices. In that case it does not matter whether partial or covariant derivatives are used. However, when working with $\tilde F^{\mu\nu}$ it \emph{does} matter, since the metric does not commute with partial derivatives. So $\tilde F^{\mu\nu}$ is given by
$	\tilde F^{\mu\nu}=\tilde g^{\mu\rho} \tilde g^{\nu\sigma} F_{\rho\sigma} = \tilde g^{\mu\rho} \tilde g^{\nu\sigma} (\tilde D_\rho B_\sigma - \tilde D_\sigma B_\rho) = \tilde D^\mu B^\nu - \tilde D^\nu B^\mu.
$} is:
\begin{equation}
	F_{\mu\nu} = \tilde D_\mu B_\nu - \tilde D_\nu B_\mu.
\end{equation}
By performing a conformal transformation
\beq
\tilde{g}_{\mu\nu} (x) \rightarrow g_{\mu\nu} (x) = \frac{2}{ M_{\rm pl}^2} f(\Phi,\Phi^\dagger) \> \tilde{g}_{\mu\nu} (x) ,
\label{conformal}
\eeq
the action in the Einstein frame becomes
\begin{equation}
	\begin{aligned}
		S=&\int d^4 x  \sqrt{ -g} \Bigg[ \frac{M_\text{pl}^2}{2}R - g^{\mu\nu} \Big( \frac{1}{2} \mathcal G_{I J}(\Phi,\Phi^\dagger)  D_\mu \phi^I D_\nu \phi^J   +\frac{M_\text{pl}^2}{2f(\Phi,\Phi^\dagger)}
		\\&\left(\left(ie  B_\mu \Phi\right)^\dagger \left(ie B_\nu \Phi\right)  + i e (-B_\mu \Phi^\dagger D_\nu \Phi + B_\nu (D_\mu \Phi^\dagger)\Phi)\right) \Big)  -V(\Phi,\Phi^\dagger) 
		 -\frac{1}{4} g^{\mu\rho} g^{\nu\sigma}  F_{\mu\nu}  F_{\rho\sigma}  \Bigg],
	\end{aligned}
\end{equation}
with 
\begin{equation}
	V(\Phi,\Phi^\dagger) = \frac{M_\text{pl}^4}{4 f^2 (\Phi,\Phi^\dagger)} \tilde V(\Phi,\Phi^\dagger),
\end{equation}
and 
\begin{equation}
	\mathcal G_{IJ}(\Phi,\Phi^\dagger) = \frac{M_\text{pl}^2}{2 f (\Phi,\Phi^\dagger)}\left[\delta_{IJ} + \frac{3}{f (\Phi,\Phi^\dagger)} f(\Phi,\Phi^\dagger),_I f(\Phi,\Phi^\dagger),_J \right] \, ,
\end{equation}
as in Refs.~\cite{KMS, GKS}. 
The potential in the Jordan frame is the usual Standard Model Higgs potential
\begin{equation}
\tilde V(\Phi,\Phi^\dagger) = {\lambda \over 4} \left ( \left | \Phi \right |^2  -v^2\right )^2 \simeq  {\lambda \over 4}  \left | \Phi \right |  ^4 \, ,
\end{equation}
where the Higgs vacuum expectation value $v=246 \, {\rm GeV}$ can be safely neglected at field values that arise during inflation and preheating. Hence the Higgs potential can be adequately modeled by a pure quartic term. 

For the sake of readability, we will drop the arguments of $\mathcal G, V \, \text{and} \, f$ from now on.
Varying the action with respect to the scalar fields $\phi^I$, the corresponding equation of motion for $\phi^I$ is:
\begin{equation}
	\begin{aligned}
&\Box \phi^I + g^{\mu\nu}\Gamma^ I_{JK} \partial_\mu \phi^J \partial_\nu \phi^K +\mathcal G^{IJ}\left( \left(\frac{M_\text{pl}^4}{4 \xi f} e^2 B^2  \right),_J - V,_J\right) +ie \frac{M_\text{pl}^2}{2f^2} f,_J  \mathcal G^{IJ}\Big(-B^\mu \Phi^\dagger D_\mu \Phi 
\\&+ B^\mu (D_\mu \Phi^\dagger)\Phi \Big) -ie M_\text{pl}^2 \mathcal G^{IJ}\Bigg (-\frac{1}{2f}B^\mu \Phi^\dagger,_J  D_\mu \Phi 
		 + D_\mu \left(\frac{1}{2f}B^\mu \Phi^\dagger \right) \Phi,_J  
		 \\& -  \Phi^\dagger,_J D_\mu \left(\frac{1}{2f}B^\mu   \Phi\right) +\frac{1}{2f}B^\mu \left(D_\mu \Phi^\dagger\right) \Phi,_J   \Bigg) =0.
	\end{aligned}
\end{equation}
We work to first order in fluctuations, in both the scalar fields and spacetime metric.
The gauge fields have no background component, thus we only treat them as first-order perturbations.
We consider scalar metric perturbations around a spatially flat Friedmann-Lema\^{i}tre-Robertson-Walker (FLRW) metric,
\beqn
\begin{split}
ds^2 &= g_{\mu\nu} (x) \> dx^\mu dx^\nu \\
&= - (1 + 2 A) dt^2 + 2 a \left( \partial_i \mathcal B \right)  dx^i dt + a^2 \left[ ( 1 - 2 \psi ) \delta_{ij} + 2 \partial_i \partial_j E \right] dx^i dx^j ,
\label{eq:ds}
\end{split}
\eeqn
where $a(t)$ is the scale factor. 
We may always choose a coordinate transformation and eliminate two of the four scalar metric functions that appear in Eq.~\eqref{eq:ds}. We work in the longitudinal gauge, where $\mathcal B(x) = E(x)=0$. Furthermore, in the absence of anisotropic pressure perturbations, the remaining two functions are equal $A(x)=\psi(x)$.

We also expand the fields,
\beq
\phi^I (x^\mu) = \varphi^I (t) + \delta \phi^I (x^\mu) . 
\label{phivarphi}
\eeq
Note that for Higgs inflation only $\phi^1$ has a background value, $\varphi(t)$, whereas the background value of $\phi^2$ is zero.

We may then construct generalizations of the Mukhanov-Sasaki variable that are invariant with respect to spacetime
 gauge transformations up to first order in the perturbations
 (see Ref.~\cite{MultiPreheat1} and references therein):
\beq
Q^I = \delta \phi^I + \frac{\dot{\varphi}^I}{H} \psi .
\label{Qdef}
\eeq
The background equation of motion for $\varphi^I$ is unchanged with respect to models with multiple  scalar fields and no gauge bosons 
\begin{equation}
	\mathcal D_t \dot\varphi ^I  + 3 H \dot\varphi^I + \mathcal G^{IJ}V,_J =0 \, ,
\end{equation}
and
\beqn
\begin{split}
H^2 &= \frac{1}{3 M_{\rm pl}^2} \left[ \frac{1}{2} {\cal G}_{IJ} \dot{\varphi}^I \dot{\varphi}^J + V (\varphi^I ) \right] , \\
\dot{H} &= - \frac{1}{ 2 M_{\rm pl}^2 } {\cal G}_{IJ} \dot{\varphi}^I \dot{\varphi}^J ,
\end{split}
\label{Friedmann}
\eeqn
where overdots denote derivatives with respect to $t$, and the Hubble parameter is given by $H (t) = \dot{a} / a$. Covariant derivatives with respect to the field-space metric are given by ${\cal D}_J A^I = \partial_J A^I + \Gamma^I_{\> JK} A^K$ for a field-space vector\footnote{Examples of field-space vectors include $A^I=\delta\phi^I$ and $A^I=\dot\varphi^I$.} $A^I$, from which we may construct the (covariant) directional derivative with respect to cosmic time,
 \beq {\cal D}_t A^I = \dot{\varphi}^J {\cal D}_J A^I = \dot{A}^I + \Gamma^I_{\> JK} \dot{\varphi}^J A^K \, ,
 \eeq
 where the Christoffel symbols $\Gamma^I_{\> JK} (\varphi^L)$ are constructed from ${\cal G}_{IJ} (\varphi^K)$. 

We now specify our analysis to the case of a complex Higgs field with background $\varphi(t)$ and fluctuations $h(t,\vec x)$ and $\theta(t,\vec x)$ as in equation (\ref{eq:Higgsdecompose}).
The equation of motion for the gauge-invariant fluctuation $Q^I$ is identical to the case without the presence of a gauge-field \cite{KMS, MultiPreheat1,MultiPreheat2,MultiPreheat3}, up to terms that mix $\theta$ and $B_\mu$:
\begin{equation}
			\begin{aligned}
		&\mathcal D^2_t Q^I + 3H \mathcal D_t  Q^I + \left[\frac{k^2}{a^2}\delta^I_J + \mathcal M^I\,_J \right] Q^J 
		\\&-e \frac{M_\text{pl}^2}{2f} \mathcal G^{IJ} \frac{d\theta}{d\phi^J}\left(2 B^\mu \partial_\mu \varphi + (D_\mu B^\mu)\varphi + 2fB^\mu \varphi D_\mu\left(\frac{1}{2f}\right)\right) =0,
		\end{aligned}
	\end{equation}
where we 
define the mass-squared matrix by
\beq
{\cal M}^I_{\> J} \equiv {\cal G}^{IK} \left( {\cal D}_J {\cal D}_K V \right) - {\cal R}^I_{\> LMJ} \dot{\varphi}^L \dot{\varphi}^M - \frac{1}{M_{\rm pl}^2 a^3} {\cal D}_t \left( \frac{ a^3}{H} \dot{\varphi}^I \dot{\varphi}_J \right) \, ,
\label{MIJdef}
\eeq
and ${\cal R}^I_{\> LMJ}$ is the Riemann tensor constructed from the field-space metric ${\cal G}_{IJ} (\varphi^K)$. The term in Eq.~(\ref{MIJdef}) proportional to $1 / M_{\rm pl}^2$ arises 
from the coupled metric perturbations through expanding Einstein's field equations to linear order and using Eq.~\eqref{Qdef}.
It hence vanishes in the limit of an infinitely rigid spacetime $M_{\rm Pl}\to \infty$. 
In the single field attractor \cite{KS, GKS, MultiPreheat1}, the background field motion proceeds along a straight single-field trajectory  $\varphi(t)$. $\mathcal G^{IJ}$ and $\mathcal M^{IJ}$ are then diagonal at background order, so the equations of motion for the first order fluctuations $h$ and $\theta$ do not mix:
	\begin{equation}
		\begin{aligned}
					&\mathcal D^2_t Q^h + 3H \mathcal D_t  Q^h + \left[\frac{k^2}{a^2} + \mathcal M^h\,_h \right] Q^h =0,
					\\ & \mathcal D^2_t Q^\theta + 3H \mathcal D_t  Q^\theta + \left[\frac{k^2}{a^2} + \mathcal M^\theta\,_\theta \right] Q^\theta 
					\\&-e \frac{M_\text{pl}^2}{2f} \mathcal G^{\theta\theta} \left(2 B^\mu \partial_\mu \varphi + (D_\mu B^\mu)\varphi+ 2fB^\mu \varphi D_\mu\left(\frac{1}{2f}\right)\right) =0,
		\end{aligned}
	\end{equation}
where 
	\begin{equation}
		Q^h = h + \frac{\dot \varphi}{H}\Psi, \qquad Q^\theta = \theta.
	\end{equation}
We see that only the Higgs fluctuations, generated along the direction of background motion, are coupled to the metric perturbations $\Psi$. In the language of Refs.~\cite{MultiPreheat1, MultiPreheat2, MultiPreheat3}, the Higgs  fluctuations correspond to adiabatic modes.

The equations are simplified if we replace $Q^I \rightarrow {X^I}/{a(t)}$ and use covariant derivatives with respect to conformal time $\tau$ instead of cosmic time. We multiply the equations by $a^3$ and obtain:
	\begin{eqnarray}
			&&\mathcal D_\tau ^2 X^h + (k^2 + a^2 (\mathcal M^ h\,_h - \frac 1 6 R \mathcal G^ h\,_h))X^h =0,
			\label{eq:Xh}
			\\ && \mathcal D_\tau^2 X^\theta +(k^2 + a^2 (\mathcal M^ \theta\,_\theta - \frac 1 6 R \mathcal G^ \theta\,_\theta))X^\theta -e a^3\frac{M_\text{pl}^2}{2f} \mathcal G^{\theta\theta} (2 B^0 \dot \varphi + (D_\mu B^\mu)\varphi -\frac{1}{f} B^0 \varphi \dot f ) =0,
					\label{eq:Xtheta}
	\end{eqnarray}
	where $R$ is the spacetime Ricci curvature.

Variation of the action with respect to the gauge field $S \rightarrow S + \frac{\delta S}{\delta B_\mu} \delta B_\mu$ gives
\begin{equation}
	\begin{aligned}
& D_\nu F^{\nu\mu} -\frac{M_\text{pl}^2 e^2}{f} \Phi^\dagger \Phi B^\mu +ie \frac{M_\text{pl}^2}{2f}g^{\mu\nu}( \Phi^\dagger \partial_\nu \Phi -  (\partial_\nu \Phi^\dagger)\Phi)=0.
		\end{aligned}
		\end{equation}
Since there is no background value for the gauge field\footnote{There has been a growing interest in inflation models where gauge fields acquire a nontrivial background value during inflation. While this is not possible for Abelian fields, $SU(2)$ gauge fields can have nontrivial vacuum configurations during inflation, leading to interesting phenomenology, like violation of the Lyth bound \cite{HGF} and tensor non-Gaussianity \cite{tensorNG}, while providing $n_s$ and $r$ in agreement with CMB observations. 
}, the
first order perturbation equation is:
\begin{equation}
	\begin{aligned}
& D_\nu F^{\nu\mu} -\frac{M_\text{pl}^2 e^2}{2f} \varphi^2  B^\mu + e \frac{M_\text{pl}^2}{2f} g^{\mu\nu} (\theta \partial_\nu \varphi - \varphi \partial_\nu \theta) =0,
		\end{aligned}
\end{equation}
where we used Eq.~\eqref{eq:Higgsdecompose} and we stress again that $F^{\nu\mu}$ is defined using covariant derivatives.

Until now we have worked in full generality, not choosing a gauge. Hence we are in principle working with more degrees of freedom than needed. We will distinguish two frequently used gauges: unitary and Coulomb gauge. The equation of motion of $X^h$ is unaffected by the gauge choice.


\subsection{Unitary gauge}

	In unitary gauge $\theta = 0=X^\theta$. Eq.~\eqref{eq:Xtheta} thus becomes a constraint equation
	\beq
	D_\mu B^\mu = \left(- \frac{2 \dot\varphi}{\varphi} + \frac {\dot f}{f} \right) B^0.
	\label{eq:unitaryconstraint}
	\eeq
	The equations of motion for the gauge fields are rewritten as 
	\beq
	 \frac{1}{\sqrt{-g}} \partial_\nu (\sqrt{-g} g^{\nu\rho}g^{\mu\sigma} F_{\rho\sigma})-\frac{M_\text{pl}^2 e^2}{2f} \varphi^2  B^\mu =0.
	\eeq
Separating the time and space components, the equation for $B_0$ becomes
	\begin{equation}
		 -\frac{1}{a^2} \partial_i (\partial_i B_0 - \partial_0 B_i) + \frac{M_\text{pl}^2 e^2}{2f}\varphi^2 B_0 =0.
	\end{equation}
Performing the analysis in Fourier space, with convention $f(x) = \int \frac{d^3\mathbf k}{(2\pi)^{3/2}} f_\mathbf{k} e^{-i \mathbf{k}\cdot x}$, we  derive an algebraic equation for $B_{0,\mathbf k}$
	\begin{equation}
		B_{0,\mathbf k} = \frac{i k_i \dot B_{i,\mathbf k}}{k^2 + \frac{a^2 M_\text{pl}^2 e^2}{2f}\varphi^2 }.
	\end{equation}
	The equation of motion for the spatial components $B_i$ is
	\begin{equation}
		\begin{aligned}
			 \frac{\dot a}{a^3}(\partial_ i B_0 - \dot B_i) - \frac{1}{a^2}(\ddot B_i -  \partial_i \dot B_0) + \frac {1}{a^4}(\partial_j^2 B_i - \partial_i\partial_j B_j) - \frac{1}{a^2} \frac{M_\text{pl}^2 e^2}{2f}\varphi^2 B_i =0 \, .
		\end{aligned}
	\end{equation}
Using the constraint Eq.~\eqref{eq:unitaryconstraint}, going to Fourier space and multiplying by $a^2$, the equation of motion becomes
	\begin{equation}
			\ddot B_{i,\mathbf k} + H \dot B_{i,\mathbf k}+ \frac{k^2}{a^2} B_{i,\mathbf k} +2\left(\frac{\dot\varphi}{\varphi}-\frac{\dot f}{2f}+H\right) \frac{k_i k_j \dot B_{j,\mathbf k}}{k^2 +  \frac{M_\text{pl}^2 a^2}{2f} e^2 \varphi^2} + \frac{M_\text{pl}^2 e^2}{2f}\varphi^2 B_{i,\mathbf k}=0.
	\end{equation}
	which is somewhat simplified in conformal time
	\begin{equation}
		 \partial_\tau^2 B_{i,\mathbf k} +  k^2 B_{i,\mathbf k} +2\left(\frac{\partial_\tau \varphi}{\varphi}-\frac{\partial_\tau f}{2f}+\frac{\partial_\tau a}{a}\right) \frac{k_i k_j \partial_\tau B_{j,\mathbf k}}{k^2 +  \frac{M_\text{pl}^2 a^2}{2f} e^2 \varphi^2} + a^2 \frac{M_\text{pl}^2 e^2}{2f}\varphi^2 B_{i, \mathbf k}=0.
	\end{equation}
We now distinguish between transverse ($B^\pm_k$) and longitudinal ($B^L_k$) modes:
	\begin{equation}
		\vec B_ \mathbf k = \hat \epsilon^L_\mathbf k B^L_\mathbf k + \hat \epsilon^+_\mathbf k B^+_\mathbf k + \hat\epsilon^-_\mathbf k B^-_\mathbf k \, ,
	\end{equation}
	with 
	\begin{equation}
		i \mathbf k \cdot \hat \epsilon^L_\mathbf k = |\mathbf k|, \qquad \mathbf k \cdot \hat \epsilon ^\pm_\mathbf k =0 \, .
	\end{equation}
The equations of motion for the transverse and longitudinal modes become:
	\begin{equation}
		\begin{aligned}
			& \partial_\tau^2 B^{\pm}_\mathbf k + (k^2 + a^2 \frac{M_\text{pl}^2e^2}{2f}\varphi^2)B^\pm_\mathbf k =0 \, , \\
			& \partial_\tau^2 B^{L}_\mathbf k    +2\left(\frac{\partial_\tau \varphi}{\varphi}- \frac{\partial_\tau f}{2f}+\frac{\partial_\tau a}{a}\right) \frac{k^2}{k^2 +  \frac{M_\text{pl}^2 a^2}{2f} e^2 \varphi^2}  \partial_\tau B^{L}_\mathbf k + (k^2 + a^2\frac{M_\text{pl}^2e^2}{2f}\varphi^2)B^L_\mathbf k =0 \, .
			& 
		\end{aligned}
	\end{equation}

\subsection{Coulomb gauge}

In Coulomb gauge ($\partial_i B^i =0$), the Goldstone mode $\theta$ remains an explicit dynamical degree of freedom, thus the relevant equations of motion are
	\begin{equation}
		\begin{aligned}
	& \mathcal D_\tau^2 X^\theta +(k^2 + a^2 (\mathcal M^ \theta\,_\theta - \frac 1 6 R \mathcal G^ \theta\,_\theta))X^\theta 
	\\& \qquad +e a^3\frac{M_\text{pl}^2}{2f} \mathcal G^{\theta\theta} (2 B_0 \dot \varphi + ( \dot B_0 + 3H B_0)\varphi -\frac 1 f \varphi \dot f B_0) =0,
			\\& -\frac{1}{a^2} \partial_i^2 B_0 + \frac{M_\text{pl}^2 e^2}{2f}\varphi^2 B_0 -e \frac{M_\text{pl}^2}{2f} (\theta \dot \varphi -\varphi \dot \theta) =0, 
			\\& \frac{\dot a}{a^3}(\partial_ i B_0 - \dot B_i) - \frac{1}{a^2}(\ddot B_i -  \partial_i \dot B_0) + \frac {1}{a^4}\partial_j^2 B_i  - \frac{1}{a^2} \frac{M_\text{pl}^2 e^2}{2f}\varphi^2 B_i - \frac{e}{a^2} \frac{M_\text{pl}^2}{2f}\varphi \partial_i \theta =0. \label{eomCoulomb}
		\end{aligned}
	\end{equation}
	Going to Fourier space, we can solve for $B_{0,\mathbf k}$ in terms of $\theta_\mathbf k$, similarly to the situation in unitary gauge
	\begin{equation}
		B_{0,\mathbf k} = \frac{e \frac{M_\text{pl}^2}{2f}(\theta_ \mathbf k \dot\varphi - \varphi \dot\theta_\mathbf k)}{\frac{k^2}{a^2}+ \frac{M_\text{pl}^2 e^2}{2f}\varphi^2 }.
	\end{equation}
By plugging  the longitudinal mode into the last equation of Eq.~\eqref{eomCoulomb} and demanding that it is zero, we get the additional constraint
	\begin{equation}
		HB_0 + \dot B_0 = \frac{M_\text{pl}^2}{2f} e \varphi \theta.
	\end{equation}
Substituting into the eom for $X^\theta$:
	\begin{equation}
		\begin{aligned}
	& \mathcal D_\tau^2 X^\theta -2 e^2 \frac{M_\text{pl}^4}{4f^2} \mathcal G^{\theta\theta} \frac{\varphi(\partial_\tau \varphi -\frac{\partial_\tau f}{2f}\varphi+ \frac{\partial_\tau a}{a} \varphi)}{\frac{k^2}{a^2} + \frac{M_\text{pl}^2 e^2\varphi^2}{2f}} \mathcal D_\tau X^\theta
	\\&+\Bigg(k^2 + a^2 (\mathcal M^ \theta\,_\theta - \frac 1 6 R \mathcal G^ \theta\,_\theta)  + e^2 \frac{M_\text{pl}^4}{4f^2} \mathcal G^{\theta\theta}  \Big(a^2\varphi^2 + 2\frac{(\partial_\tau \varphi -\frac{\partial_\tau f}{2f}\varphi+ \frac{\partial_\tau a}{a} \varphi)(\partial_\tau\varphi + \frac{\partial_\tau a}{a}\varphi)}{\frac{k^2}{a^2} + \frac{M_\text{pl}^2 e^2}{2f}\varphi^2}
			\\& + 2\frac{\partial_\tau\varphi\varphi(\partial_\tau \varphi -\frac{\partial_\tau f}{2f}\varphi+ \frac{\partial_\tau a}{a} \varphi)}{\frac{k^2}{a^2} + \frac{M_\text{pl}^2 e^2}{2f}\varphi^2} \Gamma^\theta_{h\theta} \Big)\Bigg)X^\theta
			=0 \, 
			.
		\end{aligned}
		\label{eq:XthetaFULL}
	\end{equation}

We must demand that physical observables are identical in the two gauges, and derive a relation between $\theta_\mathbf k$ in Coulomb gauge and $B_ \mathbf k^L$ in unitary gauge. $X_\mathbf k^h$ and $B_\mathbf k^\pm$ are already identical in the two gauges.
The longitudinal component of the electric field\footnote{The gauge field being studied is not the $U(1)$ of the electromagnetic sector. However, we will use the more familiar nomenclature found in electromagnetism. } is given by
	\begin{equation}
		E_\mathbf k^L = \dot B_\mathbf k^L -k B_{0,\mathbf k}.
	\end{equation}
	In unitary and Coulomb gauge we get 
	\begin{equation}
		\text{Unitary:} \, E^L_\mathbf k = \frac{M_\text{pl}^2 e^2}{2f}\varphi^2 \frac{\dot B^L_\mathbf k}{\frac{k^2}{a^2} + \frac{M_\text{pl}^2 e^2}{2f}\varphi^2} \, ,
		\qquad \text{Coulomb:} \, E^L_\mathbf  k = -k\frac{M_\text{pl}^2 e}{2f} \frac{\theta_\mathbf k \dot \varphi - \varphi \dot\theta_\mathbf k}{\frac{k^2}{a^2} + \frac{M_\text{pl}^2 e^2}{2f}\varphi^2}.
	\end{equation}
	Since $E_L$ should not depend on the gauge, we can use these expressions to solve for $B_L$ in terms of $\theta$. We obtain
	\begin{equation}
		B^L_{\mathbf k} = \frac{k}{e\varphi} \theta_\mathbf  k.
		\label{eq:BLtotheta}
	\end{equation}
	It is a straightforward algebraic exercise to show that by using Eq.~\eqref{eq:BLtotheta}, the equation of motion for $B^L_\mathbf k$ and $\theta_\mathbf k$ can be transformed into each other, providing a useful check for our derivation.

	During preheating, when the background inflaton field oscillates, the unitary gauge becomes ill-defined at the times where $\varphi(t)=0$, as can be seen for example in the transformation relation of Eq.~\eqref{eq:BLtotheta}.
	We will perform preheating simulations is the Coulomb gauge, which is always well-defined.

\subsection{Single-field attractor and parameter choices}
\label{sec:parameterchoices}

For Higgs inflation, the function $f(\Phi,\Phi^\dagger)$ is given by \cite{Higgsinfl}:
\beq
	f(\Phi,\Phi^\dagger) = \frac{M_\text{pl}^2}{2}+  \xi \Phi^\dagger \Phi.
\eeq
For typical values of Higgs inflation $\lambda = {\cal O}( 0.01)$ and correspondingly $\xi\sim10^4$. If we consider a different RG flow for the self-coupling $\lambda$, through the introduction of unknown physics before the inflationary scale, $\lambda$ will become smaller or larger at inflationary energies.
Since, as we will show below, the combination ${\lambda/ \xi^2}$  is fixed by the amplitude of the scalar power spectrum, a larger or smaller value of $\lambda$ during inflation will
 lead to a correspondingly larger or smaller value of the nonminimal coupling $\xi$. We will consider values of $\xi$ in the range $10 \le \xi \le10^4$.
The inflationary predictions for the scalar and tensor modes for nonminimally coupled models with $\xi\ge 10$ fall into the large-$\xi$ single-field attractor regime, as described for example in Ref.~\cite{KS}. This results in very simple expressions for the scalar spectral index $n_s$, the tensor-to-scalar ratio $r$ and the running of the spectral index $\alpha$ as a function of the number of $e$-folds at horizon-crossing $N_*$
 \beq
  n_s \simeq 1- {2\over N_*} - {3\over N_*^2} \, ,\quad r\simeq {12\over N_*^2} \, , \quad \alpha = {dn_s\over d\ln k} \simeq -{2\over N_*^2} \left (1+{3\over N_*}\right )\, .
 \label{eq:nsr}
 \eeq
The values for the spectral observables given in Eq.~\eqref{eq:nsr} correspond to single-field background motion.  Multi-field nonminimally coupled models of inflation at large $\xi$ show a very strong single-field attractor behavior. The strength of the attractor was analyzed in 
Ref.~\cite{GKS} for the case of an $SO(N)$-symmetric model, similar to Higgs inflation without gauge fields. The more general case of two-field inflation with generic potential parameters is given in Refs.~\cite{SSK, MultiPreheat1}, showing that the single-field attractor becomes stronger for larger $\xi$ and that it persists not only during inflation but also during the (p)reheating era. For generic initial conditions, the isocurvature fraction $\beta_{\rm iso}$ is exponentially small for random potentials, while for a symmetric potential $\beta_{\rm iso} ={\cal O}(10^{-5})$, as is shown in Ref.~\cite{SSK}.
As discussed in Section \ref{sec:gaugeinflation}, during inflation, the gauge bosons are very massive compared to the Hubble scale, making the single-field attractor behavior of Higgs inflation stronger than the one described in Ref.~\cite{GKS} for the scalar symmetric case. Hence the use of a single-field motion $\varphi(t)$ for the background is well justified during and after Higgs inflation.

The dimensionless power spectrum of the (scalar) density perturbations is measured to be
\beq
A_s \simeq 2\times 10^{-9} \, .
\eeq
Using the tensor-to-scalar ratio from Eq.~\eqref{eq:nsr} with $N_*=55$ yields $r\simeq 3.3\times 10^{-3}$, and hence the tensor power spectrum becomes
\beq
{{\cal P}_T \over  M_{\rm Pl}^2} 
= {2 H^2\over \pi^2 M_{\rm Pl}^2}
= r \times A_s \simeq 6.6 \times 10^{-12} \, .
\eeq
Given that the Hubble scale during inflation is approximately \cite{Higgsinfl}
\beq
H_{\rm infl} ^2 \simeq {\lambda\over 12\xi^2}  M_{\rm Pl}^2\, ,
\label{eq:Hinfl}
\eeq
the Higgs self-coupling and nonminimal coupling must obey the relation
\beq
{\lambda\over \xi^2} \simeq 5\times 10^{-10} \, .
\label{eq:lambdaxi}
\eeq
We keep the value of the Hubble scale fixed and determine the value of $\lambda$ that corresponds to each $\xi$ through Eq.~\eqref{eq:lambdaxi}.

\section{Electroweak Sector}
\label{sec:electroweak}

We now consider the full $SU(2)\times U(1)$ gauge symmetry, as it exists in the electroweak sector of the SM.
The Lagrangian in the Jordan frame is given by 
\begin{equation}
	\begin{aligned}
	S_J = \int d^4x \sqrt{-\tilde g} \Big[&f(\Phi,\Phi^\dagger) \tilde R - \tilde g^{\mu\nu}( \tilde \nabla_\mu \Phi)^\dagger \tilde \nabla_\nu \Phi -\frac{1}{4} \tilde g^{\mu\rho}\tilde g^{\nu\sigma}  B_{\mu\nu}  B_{\rho\sigma} \\& -\frac{1}{4} \tilde g^{\mu\rho}\tilde g^{\nu\sigma} \underline A_{\mu\nu} \cdot \underline A_{\rho\sigma} -\tilde V(\Phi,\Phi^\dagger) \Big],
	\end{aligned}
\end{equation}
with the Higgs doublet
\begin{equation}
	\Phi = \frac{1}{\sqrt 2} \begin{pmatrix}
	\phi^3 + i \phi^4 \\
	\varphi + h + i \theta
	\end{pmatrix}.
\end{equation}
 The covariant derivative $\tilde \nabla_\mu$ is given by:
\begin{equation}
	\tilde \nabla_\mu = \tilde D_\mu + ig' \frac 1 2 Y B_\mu + i g \frac 1 2 \underline A_\mu \cdot \underline\tau ,
\end{equation}
with $Y$ the generator of hypercharge $U(1)$ and $B_\mu$ the corresponding gauge field. The Higgs doublet has hypercharge $+1$. We have also introduced the vector notation
\begin{equation}
	\underline A_\mu \equiv (A_{1,\mu},A_{2,\mu},A_{3,\mu}), \qquad \underline \tau \equiv (\tau_1,\tau_2,\tau_3).
\end{equation}
The $\underline A_\mu$ are the gauge fields corresponding to $SU(2)$ and  $\tau_i$ are the Pauli matrices. The corresponding field strength tensors are:
\begin{equation}
	B_{\mu\nu} = \partial_\mu B_\nu - \partial_\mu B_\nu, \qquad A_{a,\mu\nu} = \partial_\mu A_{a,\nu} - \partial_\nu A_{a,\mu} - g \sum_{b,c =1}^3 \epsilon_{abc} A_{b,\mu} A_{c,\nu}.
\end{equation}
Defining the fields $W_\mu$, $W^\dagger_\mu$, $Z_\mu$ and $A_\mu$ as:
\begin{equation}
	\begin{aligned}
		&W_\mu = \frac{A_{1,\mu}-iA_{2,\mu}}{\sqrt 2} \qquad A_\mu = \sin \theta_W A_{3,\mu} + \cos \theta_W B_\mu\\
		&W^\dagger_\mu = \frac{A_{1,\mu}+iA_{2,\mu}}{\sqrt 2} \qquad Z_\mu =\cos \theta_W A_{3,\mu} - \sin \theta_W B_\mu,
	\end{aligned}
\end{equation}
with 
\begin{equation}
	e = g \sin \theta_W = g' \cos \theta_W,
\end{equation}
the components of the covariant derivative of $\Phi$ are given by:
\begin{equation}
	\tilde \nabla_\mu \Phi = \frac{1}{\sqrt 2} \begin{pmatrix}
		\tilde D_\mu (\phi_3 + i \phi_4) + i\left (e A_\mu + \frac{g \cos2\theta_W}{2 \cos \theta_W}Z_\mu \right)(\phi_3 + i \phi_4) + \frac{ig}{\sqrt 2} W_\mu (\varphi + h + i \theta)\\
		\tilde D_\mu (\varphi +h + i \theta) - \frac{ig}{2 \cos\theta_W} Z_\mu (\varphi +h+ i\theta) + \frac{ig}{\sqrt 2} W^\dagger_\mu (\phi_3 + i \phi_4)
	\end{pmatrix} \, .
	\label{nonabelianDmuPhi}
\end{equation}
The structure of the equations is almost identical to the one studied in the earlier parts of this work, where we focused on the Abelian case.
For $\theta$, the Goldstone mode that becomes the longitudinal polarization of the $Z$ boson, we substitute:
\begin{equation}
	2 eB_\nu \rightarrow -\frac{g}{\cos \theta_W} Z_\nu,
\end{equation}
in our Abelian equation and obtain:
	\begin{equation}
		\mathcal D_\tau^2 X^\theta +(k^2 + a^2 (\mathcal M^ \theta\,_\theta - \frac 1 6 R \mathcal G^ \theta\,_\theta))X^\theta + a^3\frac{M_\text{pl}^2}{2f} \frac{g}{2\cos \theta_W} \mathcal G^{\theta\theta} (2 Z^0 \dot \varphi + (D_\mu Z^\mu)\varphi -\frac{1}{f} Z^0 \varphi \dot f ) =0.
	\end{equation}
The Goldstone bosons $\phi_3$ and $\phi_4$ become the longitudinal modes of the $W^\pm$ bosons. Doing the substitutions $\theta\to \phi_3$ and $\theta\to \phi_4$ in the Abelian equation and
\begin{equation}
	2 eB_\nu \rightarrow i\frac{g}{\sqrt 2} (W_\nu - W_\nu^\dagger), \quad 	2 eB_\nu \rightarrow \frac{g}{\sqrt 2} (W_\nu + W_\nu^\dagger),
\end{equation}
we obtain
\beqn
\nonumber
	\mathcal D_\tau^2 X^{\phi_3} &&+(k^2 + a^2 (\mathcal M^ {\phi_3}\,_{\phi_3} - \frac 1 6 R \mathcal G^ {\phi_3}\,_{\phi_3}))X^{\phi_3} 
	\\&&- a^3\frac{M_\text{pl}^2}{2f} \frac{ig}{2\sqrt 2} \mathcal G^{{\phi_3}{\phi_3}} (2 (W^ 0 - W^{\dagger 0}) \dot \varphi + (D_\mu (W^\mu -W^{\dagger \mu}))\varphi -\frac{1}{f} (W^0 - W^{\dagger 0}) \varphi \dot f ) =0 \, ,
	\\
	\nonumber
	\mathcal D_\tau^2 X^{\phi_4} &&+(k^2 + a^2 (\mathcal M^ {\phi_4}\,_{\phi_4} - \frac 1 6 R \mathcal G^ {\phi_4}\,_{\phi_4}))X^{\phi_4} 
	\\&&- a^3\frac{M_\text{pl}^2}{2f} \frac{g}{2\sqrt 2} \mathcal G^{{\phi_4}{\phi_4}} (2 (W^ 0 + W^{\dagger 0}) \dot \varphi + (D_\mu (W^\mu +W^{\dagger \mu}))\varphi -\frac{1}{f} (W^0 + W^{\dagger 0}) \varphi \dot f ) =0.
\eeqn

At quadratic order, the field strength term for the electroweak case is no more complicated than the Abelian case, it simply contains more fields:
	\begin{equation}
		\mathcal L_\text{gauge} = -\frac 1 2 F_{W \mu\nu}^\dagger F_W^{\mu\nu} - \frac 1 4 F_{Z \mu\nu} F^{\mu\nu}_Z - \frac 1 4 F_{\mu\nu} F^{\mu\nu},\label{FieldStrength}
	\end{equation}
with
	\begin{equation}
		F_{W \mu\nu} = \partial_\mu W_\nu - \partial_\nu W_\mu, \qquad F_{Z \mu\nu} = \partial_\mu Z_\nu - \partial_\nu Z_\mu, \qquad 	F_{ \mu\nu} = \partial_\mu A_\nu - \partial_\nu A_\mu \, .
	\end{equation}
Comparing Eqs.~\eqref{nonabelianDmuPhi} and \eqref{FieldStrength} 
with the Abelian case, we can easily find the equations of motion for the gauge fields.

The photon $A^\mu$ does not couple to the Higgs:
\begin{equation}
	D_\nu F^{\nu \mu} =0 \, .
\end{equation}
The $Z$ boson obeys
\begin{equation}
	D_\nu F^{\nu\mu}_Z - \frac{M_\text{pl}^2}{2f} \frac{g^2}{4 \cos^2 \theta_W}\varphi^2 Z^\mu - \frac{M_\text{pl}^2}{2f} \frac{g}{2 \cos \theta_W}g^{\mu\nu}(\theta \partial_\nu \varphi -\varphi \partial_\nu \theta )=0 \, ,
\end{equation}
and correspondingly for the $W^\pm$ bosons
\begin{equation}
	D_\nu F^{\nu\mu}_W - \frac{M_\text{pl}^2}{2f} \frac{g^2}{2}\varphi^2 W^\mu - \frac{M_\text{pl}^2}{2f} \frac{ig}{2 \sqrt 2}g^{\mu\nu}((\phi_3 + i\phi_4) \partial_\nu \varphi -\varphi \partial_\nu (\phi_3 + i\phi_4) )=0 \, .
\end{equation}

\subsection{Unitary gauge}
In unitary gauge, the equations of motion for the three Goldstone degrees of freedom $\theta$ and $\phi_3,\phi_4$ give the constraints:
\begin{equation}
	D_\mu Z^\mu = \left( -\frac{2\dot\varphi}{\varphi} + \frac{\dot f}{f}\right)Z^0\, , \qquad D_\mu W^\mu = \left( -\frac{2\dot\varphi}{\varphi} + \frac{\dot f}{f}\right)W^0. 
\end{equation}
The  equations of motion  for $Z^\mu$ and $W^\mu$:
\begin{equation}
	D_\nu F^{\nu\mu}_Z -\frac{M_\text{pl}^2}{2f} \frac{g^2}{4 \cos^2 \theta_W} \varphi^2 Z^\mu =0 \, \qquad 	D_\nu F^{\nu\mu}_W -\frac{M_\text{pl}^2}{2f} \frac{g^2}{2} \varphi^2 W^\mu =0.
\end{equation}
These equations are identical to the equations in the Abelian case, albeit with different couplings. The equations for the longitudinal and transverse modes are thus given by:
	\begin{equation}
		\begin{aligned}
			&  \partial_\tau^2 Z^{\pm}_k + (k^2 + a^2 \frac{M_\text{pl}^2}{2f}\frac{g^2}{4 \cos^2 \theta_W}\varphi^2)Z^\pm_k =0 \, , \\
			& \ \partial_\tau^2 Z^{L}_k     +2\left(\frac{\partial _\tau\varphi}{\varphi}- \frac{\partial _\tau f}{2f}+\frac{\partial _\tau a}{a}\right) \frac{k^2}{k^2 +  \frac{M_\text{pl}^2 a^2}{2f} \frac{g^2}{4\cos^2 \theta_W} \varphi^2}  \partial_\tau Z^{L}_k 
			\\ & \qquad \qquad + (k^2 + a^2\frac{M_\text{pl}^2}{2f}\frac{g^2}{4\cos^2 \theta_W}\varphi^2)Z^L_k =0.
			& 
		\end{aligned}
	\end{equation}
	and
	\begin{equation}
	\begin{aligned}
			& \partial_\tau^2 W^{\pm}_k + (k^2 + a^2 \frac{M_\text{pl}^2}{2f}\frac{g^2}{2}\varphi^2)W^\pm_k =0 \, , \\
			& \ \partial_\tau^2 W^{L}_k    +2\left(\frac{\partial _\tau\varphi}{\varphi}- \frac{\partial _\tau f}{2f}+\frac{\partial _\tau a}{a}\right) \frac{k^2}{k^2 +  \frac{M_\text{pl}^2 a^2}{2f} \frac{g^2}{2} \varphi^2}  \partial_\tau W^{L}_k + (k^2 + a^2\frac{M_\text{pl}^2}{2f}\frac{g^2}{2}\varphi^2)W^L_k =0,
			& 
	\end{aligned}
	\end{equation}
	where $W^\pm$ denotes the $\pm$ polarization of the field $W$ (so the $\pm$ does not distinguish $W$ or $W^\dagger$).

\subsection{Coulomb gauge}
The Coulomb gauge for the two types of bosons, $Z$ and $W^\pm$, is defined through the conditions 
\begin{equation}
	\partial_i Z^i =0, \qquad \partial_i W^i=0 \, .
\end{equation}
In Fourier space, we can express $Z_{0,k}$ in terms of $\theta$ and $W_{0,k}$ in terms of $\phi_3$ and $\phi_4$:
	\begin{equation}
		Z_{0,k} = \frac{ \frac{M_\text{pl}^2}{2f} \frac{g}{2 \cos \theta_W} ( \varphi \dot\theta-\theta \dot\varphi )}{\frac{k^2}{a^2}+ \frac{M_\text{pl}^2}{2f} \frac{g^2}{4 \cos^2 \theta_W}\varphi^2 } \, , \qquad 		W_{0,k} = \frac{ \frac{M_\text{pl}^2}{2f} \frac{ig}{2 \sqrt 2} ( \varphi (\dot \phi_3 + i \dot\phi_4)-(\phi_3 + i \phi_4) \dot\varphi )}{\frac{k^2}{a^2}+ \frac{M_\text{pl}^2}{2f} \frac{g^2}{2}\varphi^2 }.
	\end{equation}
	From the decoupling of the longitudinal modes from the equations for the corresponding transverse ones we get the constraints:
	\begin{equation}
		\begin{aligned}
			&HZ_0 + \dot Z_0 = -\frac{M_\text{pl}^2}{2f} \frac{g}{2 \cos \theta_W} \varphi \theta, \\
			&HW_0 + \dot W_0 = -\frac{M_\text{pl}^2}{2f} \frac{ig}{2\sqrt 2} \varphi(\phi_3 + i \phi_4).
		\end{aligned}
	\end{equation}
	Substituting into the equation for $X^\theta$ gives:
	\begin{equation}
		\begin{aligned}
			&\mathcal D_\tau^2 X^\theta -2 \frac{g^2}{4 \cos^2 \theta_W} \frac{M_\text{pl}^4}{4f^2} \mathcal G^{\theta\theta} \frac{\varphi^2\left(\frac{\partial _\tau\varphi}{\varphi}- \frac{\partial _\tau f}{2f}+\frac{\partial _\tau a}{a}\right)}{\frac{k^2}{a^2} + \frac{M_\text{pl}^2 }{2f}\frac{g^2}{4 \cos^2 \theta_W}\varphi^2} \mathcal D_\tau X^\theta +
			\\&\Bigg(k^2 + a^2 (\mathcal M^ \theta\,_\theta - \frac 1 6 R \mathcal G^ \theta\,_\theta) + \frac{g^2}{4 \cos^2 \theta_W} \frac{M_\text{pl}^4}{4f^2} \mathcal G^{\theta\theta} \Big(a^2\varphi^2 + 2\frac{\varphi^2(\partial_\tau\varphi)\left(\frac{\partial _\tau\varphi}{\varphi}- \frac{\partial _\tau f}{2f}+\frac{\partial _\tau a}{a}\right)}{\frac{k^2}{a^2} + \frac{M_\text{pl}^2 }{2f}\frac{g^2}{4 \cos^2 \theta_W}\varphi^2} \Gamma^\theta_{h \theta} 			
			\\& + 2\frac{\varphi^2 \left(\frac{\partial _\tau\varphi}{\varphi}- \frac{\partial _\tau f}{2f}+\frac{\partial _\tau a}{a}\right)( \frac{\partial_\tau \varphi}{\varphi} + \frac{\partial_\tau a}{a})}{\frac{k^2}{a^2} + \frac{M_\text{pl}^2}{2f}\frac{g^2}{4 \cos^2 \theta_W}\varphi^2}  \Big) \Bigg)X^\theta=0 \, ,
		\end{aligned}
	\end{equation}
and substituting into the equation for $X^{\phi_3}$:
	\begin{equation}
			\begin{aligned}
	& \mathcal D_\tau^2 X^{\phi_3} -2 \frac{g^2}{4 } \frac{M_\text{pl}^4}{4f^2} \mathcal G^{\phi_3\phi_3} \frac{\varphi^2\left(\frac{\partial _\tau\varphi}{\varphi}- \frac{\partial _\tau f}{2f}+\frac{\partial _\tau a}{a}\right)}{\frac{k^2}{a^2} + \frac{M_\text{pl}^2 }{2f}\frac{g^2}{2}\varphi^2} \mathcal D_\tau X^{\phi_3} +
	\\& \Bigg(k^2 + a^2 (\mathcal M^ {\phi_3}\,_{\phi_3} - \frac 1 6 R \mathcal G^ {\phi_3}\,_{\phi_3}) 
			+ \frac{g^2}{4 } \frac{M_\text{pl}^4}{4f^2} \mathcal G^{\phi_3\phi_3} \Big(a^2\varphi^2 + 2\frac{\varphi^2(\partial_\tau\varphi)\left(\frac{\partial _\tau\varphi}{\varphi}- \frac{\partial _\tau f}{2f}+\frac{\partial _\tau a}{a}\right)}{\frac{k^2}{a^2} + \frac{M_\text{pl}^2 }{2f}\frac{g^2}{2}\varphi^2} \Gamma^{\phi_3}_{h \phi_3} 
			\\&  + 2\frac{\varphi^2 \left(\frac{\partial _\tau\varphi}{\varphi}- \frac{\partial _\tau f}{2f}+\frac{\partial _\tau a}{a}\right)( \frac{\partial_\tau \varphi}{\varphi} + \frac{\partial_\tau a}{a})}{\frac{k^2}{a^2} + \frac{M_\text{pl}^2}{2f}\frac{g^2}{2 }\varphi^2} \Big)\Bigg)X^{\phi_3}= 0 \, ,
		\end{aligned}
	\end{equation}
and likewise:
	\begin{equation}
		\begin{aligned}
	& \mathcal D_\tau^2 X^{\phi_4} -2 \frac{g^2}{4 } \frac{M_\text{pl}^4}{4f^2} \mathcal G^{\phi_4\phi_4} \frac{\varphi^2\left(\frac{\partial _\tau\varphi}{\varphi}- \frac{\partial _\tau f}{2f}+\frac{\partial _\tau a}{a}\right)}{\frac{k^2}{a^2} + \frac{M_\text{pl}^2 }{2f}\frac{g^2}{2}\varphi^2} \mathcal D_\tau X^{\phi_4}+
	\\& \Bigg(k^2 + a^2 (\mathcal M^ {\phi_4}\,_{\phi_4} - \frac 1 6 R \mathcal G^ {\phi_4}\,_{\phi_4}) + \frac{g^2}{4 } \frac{M_\text{pl}^4}{4f^2} \mathcal G^{\phi_4\phi_4} \Big(a^2\varphi^2 + 2\frac{\varphi^2(\partial_\tau\varphi)\left(\frac{\partial _\tau\varphi}{\varphi}- \frac{\partial _\tau f}{2f}+\frac{\partial _\tau a}{a}\right)}{\frac{k^2}{a^2} + \frac{M_\text{pl}^2 }{2f}\frac{g^2}{2}\varphi^2} \Gamma^{\phi_4}_{h \phi_4}
	\\&+ 2\frac{\varphi^2 \left(\frac{\partial _\tau\varphi}{\varphi}- \frac{\partial _\tau f}{2f}+\frac{\partial _\tau a}{a}\right)( \frac{\partial_\tau \varphi}{\varphi} + \frac{\partial_\tau a}{a})}{\frac{k^2}{a^2} + \frac{M_\text{pl}^2}{2f}\frac{g^2}{2 }\varphi^2}    \Big)\Bigg)X^{\phi_4}=0.
		\end{aligned}
	\end{equation}
	The equations of motion of the transverse modes of the $Z$ and $W$ are:
	\begin{equation}
		\ddot Z_k^\pm + H \dot Z^\pm_k + \frac {1}{a^2}\left(k^2 + \frac{M_\text{Pl}^2}{2f} \frac{g^2}{4 \cos \theta_W}\varphi^2 \right) Z^\pm _k =0,
	\end{equation}
	\begin{equation}
		\ddot W_k^\pm + H \dot W^\pm_k + \frac {1}{a^2}\left(k^2 + \frac{M_\text{Pl}^2}{2f} \frac{g^2}{2}\varphi^2 \right) W^\pm _k =0.
	\end{equation}


\section{Higgs Self-resonance}
\label{sec:Adiabatic}

We now focus on the Higgs fluctuations, neglecting the effects of Goldstone modes and gauge fields. In our linear analysis the Higgs fluctuations do not couple to the gauge field. The equation of motion for the re-scaled fluctuations $X^h (x^\mu)\equiv a(t)  Q^h(x^\mu)$ is
\beq
{\cal D}_\tau^2 X_\mathbf k ^h + \omega^2_h(k,\tau) X_\mathbf k ^h=0 \, ,
\eeq
where the effective frequency is defined as
\beq
{\omega_{h}^2(k,\tau)\over a^2}  
=
{k^2\over a^2} 
+m_{{\rm eff},h}^2 \, .
\label{eq:omegah}
\eeq
For notational simplicity and connection to earlier work \cite{MultiPreheat1, MultiPreheat2, MultiPreheat3} we define the various contributions to the effective mass of the Higgs fluctuations
\beq
m_{{\rm eff},h}^2 \equiv {\cal M}^h\,_{h} -{1\over 6} R = m_{1,h}^2+m_{2,h}^2+m_{3,h}^2+m_{4,h}^2 \,,
\eeq
where $\mathcal M^h \,_h$ was defined in Eq.~\eqref{MIJdef} and
\beqn
m_{1,h}^2 &=& {\cal G}^{hh} ({\cal D}_\varphi {\cal D}_{\varphi} V) \, ,
\\
m_{2,h}^2 &=& -{\cal R}^h_{~LMh}\dot\varphi^L  \dot\varphi^M \, ,
\\
m_{3,h}^2 &=& -{1\over M_{\rm Pl}^2 a^3} {\cal D}_t   \left ({a^3 \over H} \dot \varphi^2   {\cal G}_{hh}\right ) \, ,
\\
m_{4,h}^2 &=& -{1\over 6}R  = (\epsilon -2)H^2 \, .
\eeqn
For the case of fluctuations along the straight background trajectory, as are Higgs fluctuations, the Riemann contribution $m_{2,h}^2$ vanishes identically.
As described in Ref.~\cite{weinberg} and further utilized in Ref.~\cite{MultiPreheat1}, the mode-functions can be decomposed using the vielbeins of the field-space metric. In the  single-field attractor, which exists in the nonminimally coupled models, that include Higgs inflation, both during \cite{KS} and after inflation \cite{MultiPreheat1}, the decomposition of $X_\mathbf k ^h$ into creation and annihilation operators is trivial
\beq
\hat X^h = \int {d^3k\over (2\pi)^{3/2}} 
\left [
 v_k e_1^{~h} \,  \hat a_\mathbf k  e^{i \mathbf k \cdot \mathbf x} 
+
v_k^* e_1^{~h}\, \hat a^\dagger_\mathbf k  e^{-i \mathbf k \cdot \mathbf x} 
\right ] \, ,
\eeq
where $e_1^{~h} = \sqrt{{\cal G}^{hh}}$. Since the vielbeins obey the parallel transport equation ${\cal D}_\tau e_1^{~h} =0$, 
 the equation of motion for the mode-function $v_\mathbf k$ becomes
\beq
{\partial}_\tau^2 v_\mathbf k + \omega_{h}^2(k,\tau) v_\mathbf k=0 \, .
\eeq
We solve the equation in cosmic, rather than conformal time, which is better suited for computations after inflation
\beq
\ddot v_\mathbf k + H \dot v_\mathbf k +{\omega_{h}^2( k,\tau)\over a^2} v_\mathbf k=0 \, ,
\eeq
where the frequency is defined in Eq.~\eqref{eq:omegah}.

We examine the two dominant terms of the effective mass, the one arising from the potential ($m_{1,h}^2$) and the one arising from the coupled metric perturbations ($m_{3,h}^2$). The latter is often overlooked in studies of preheating, perhaps because it is vastly subdominant during inflation. It arises by combining the equation of motion for $\delta\phi$ and the metric perturbation $\psi$, defined through Eq.~\eqref{eq:ds}, in conjunction with the definition of the Mukhanov-Sasaki variables, given in Eq.~\eqref{Qdef}. 

The expression for $m_{1,h}^2$ is
\beq
 m_{1,h}^2  = 
 \frac{\lambda  \varphi^2 \left(\xi  \varphi^2 \left(12 \xi -2 \xi  (6 \xi +1) \varphi^2+1\right)+3\right)}{\left(\xi  \varphi^2+1\right)^2 \left(\xi  (6 \xi +1) \varphi^2+1\right)^2}\simeq 
 -\frac{\lambda }{3 \xi ^3 \varphi^2}
 +
 \frac{\lambda  \left(\varphi^2+18\right)}{18 \xi ^4 \varphi^4} \, ,
\eeq
where we used $\xi\gg 1$ in expressions such as $(6 \xi +1) \simeq 6\xi$. Furthermore, since we are at first interested in studying the behavior during inflation, where analytic progress can be made, we use $\xi\varphi^2 \gg 1$ as an approximation. As we will see, this works reasonably well even close to the end of inflation.
We normalize the effective mass by the Hubble scale 
\beq
 {m_{1,h}^2\over H^2(t)} = \frac{12 \left(\xi  \varphi^2 \left(12 \xi -2 \xi  (6 \xi +1) \varphi^2+1\right)+3\right)}{\varphi^2 \left(\xi  (6 \xi +1) \varphi^2+1\right)^2} \simeq
-\frac{4}{\xi  \varphi^2}
+
\frac{4 }{ \xi ^2 \varphi^4}+ {\cal O}\left ( 1\over \xi^3 \varphi^6 \right ) \, .
\label{eq:m1overH}
\eeq
We can use the single-field slow-roll results
\beq
-N = {3 \over 4}{\xi \varphi^2\over M_{\rm Pl}^2}+ {1\over 8}{ \varphi^2\over M_{\rm Pl}^2} + {\cal O}\left(\log \varphi\over M_{\rm Pl}\right ) \, ,
 \label{eq:Ntoxi}
\eeq
where we went beyond lowest order in $\xi \varphi^2$ and we measure the number of $e$-folds from the end of inflation, meaning that negative values correspond to the inflationary era\footnote{We neglected the contributions coming from the lower end of the integral leading to Eq.~\eqref{eq:Ntoxi}.}.
This leads to
\beq
 {m_{1,h}^2\over H^2(t)} \simeq  
 \frac{3}{N} + {9\over 4N^2} +{\cal O}\left ({1\over N^3} \right ) \, .
 \label{eq:m1hoverH}
\eeq
 If we minimize $m_1^2$  as a function of $\delta = \sqrt \xi \varphi$, the field amplitude that minimizes the mass is
\beq
\delta_{\rm min} =\sqrt{2}+{\cal O}\left ({1\over \xi}\right ) \, ,
\eeq
or equivalently $N_{\rm min} \simeq -1.5$.
 For the minimization we used the full expression for the effective mass and only took the Taylor-expansion for large $\xi$ at the end. We can see that, for $\xi\gg1$ the minimum of $m_1^2$ is independent of $\xi$ and thus occurs at the same value of $\delta$, which will also be the same value of $N$, in the approximation of Eq.~\eqref{eq:Ntoxi}. In general, the function $m_{1,h}^2(N) / H^2$ shows no appreciable difference for different values of $\xi \gg1$ during inflation. This can be easily seen by substituting Eq.~\eqref{eq:Ntoxi} into Eq.~\eqref{eq:m1overH}.
 As shown in Ref.~\cite{MultiPreheat2}, this behavior persists during the time of coherent inflaton oscillations.

The mass component arising from the metric fluctuations is
\beq
m_{3,h}^2 = -\frac{\left(\xi  (6 \xi +1) \varphi^2+1\right) \dot\varphi\left(H(t) (\epsilon (t)+3) \dot\varphi+2 \ddot\varphi\right)}{H(t) \left(\xi  \varphi^2+1\right)^2}
\simeq -\frac{18 \dot\varphi^2}{\varphi^2} \, ,
\eeq
where the last approximation holds during inflation. Using the slow-roll expression for $\dot \varphi$ we get that during inflation
\beq
{m_{3,h}^2\over H^2(t)} \simeq -{9\over 2 N^2} \, .
\label{eq:m3hoverH}
\eeq
This contribution is clearly subdominant to $m_{1,h}^2$, hence it can be safely neglected during inflation.
However, $|m_{3,h}^2|$ grows near the end of inflation, since it is proportional to $\dot\varphi^2$, which at the end of inflation is given by 
\beq
\dot \varphi^2_{\rm end} = {\cal G}^{\varphi\varphi}V = \frac{\lambda \varphi^4}{4 \left(6 \xi ^2\varphi^2+\xi \varphi^2+1\right)} \simeq \frac{\lambda  \varphi^2}{24 \xi ^2} \, .
\eeq
It has been numerically shown in Ref.~\cite{MultiPreheat1} that the field value at the end of inflation is $\sqrt{\xi} \varphi_{\rm end} \simeq 0.8$, leading to
\beq
\dot \varphi^2_{\rm end} \simeq {0.8^2\lambda \over 24 \xi^3} \simeq  {2\lambda \over 75 \xi^3} \, .
\eeq
Numerically we get $m_{3,h}^2 / H^2(t)\simeq -11$ at the end of inflation, in rough agreement with the approximate expressions given above.

\begin{figure} 
\centering
\includegraphics[width=\textwidth]{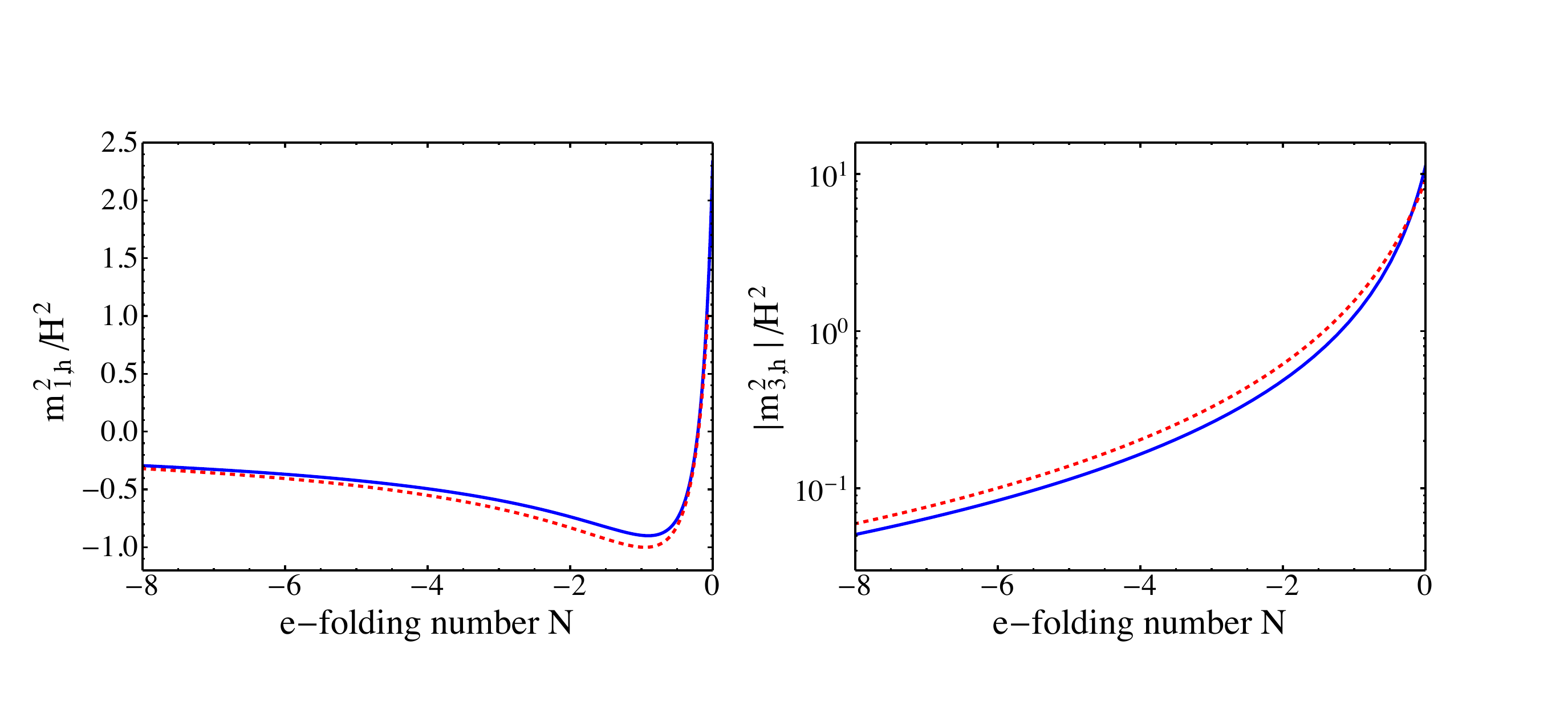}
\caption{
The components of the effective mass of the Higgs fluctuations $m_{1,h}^2$ and $m_{3,h}^2$ rescaled by the Hubble scale. The blue curves show the numerical curves for $\xi=10$ and the red dashed lines the approximate analytic expressions of Eqs.~\eqref{eq:m1hoverH} and \eqref{eq:m3hoverH} respectively.
}
 \label{fig:mhoverH}
\end{figure}

\begin{figure} 
\centering
\includegraphics[width=\textwidth]{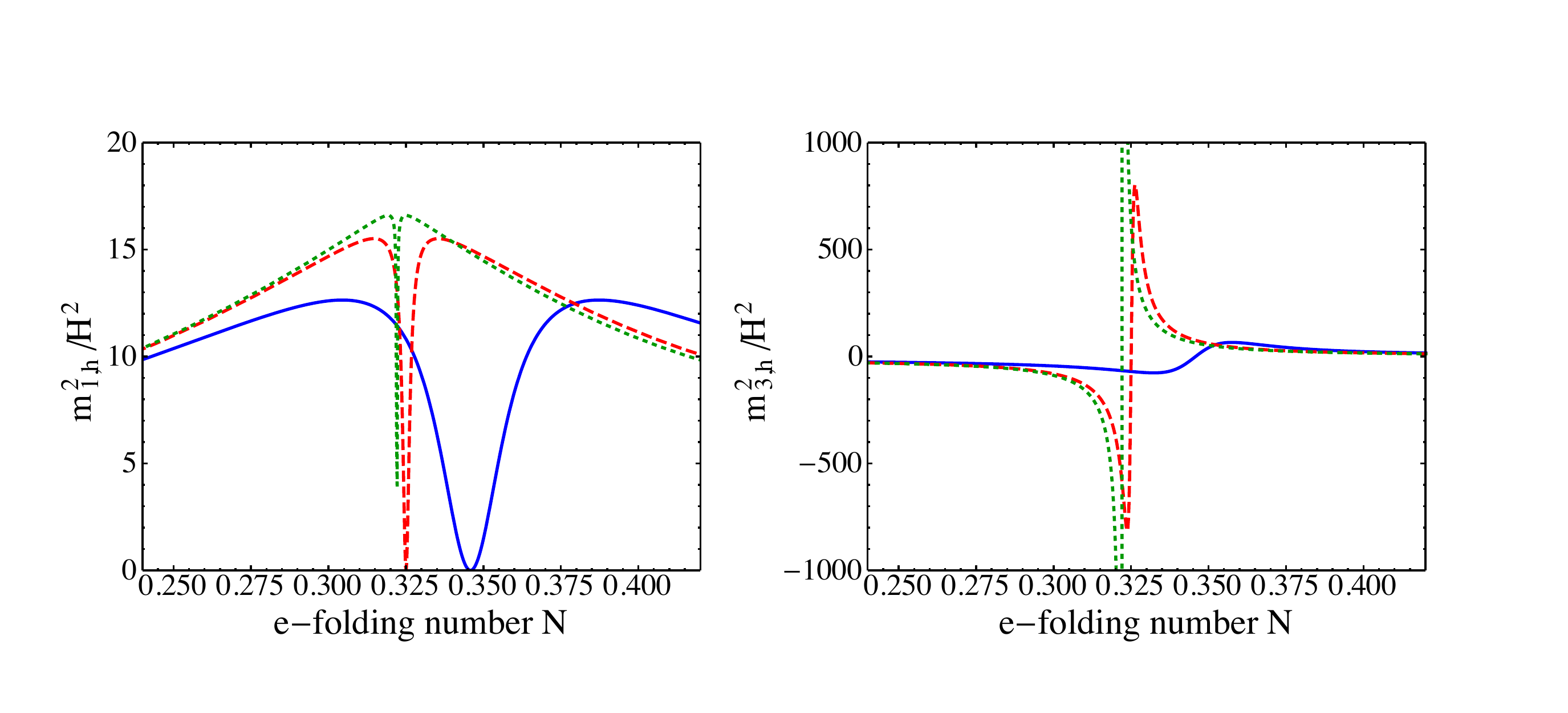}
\caption{
The ratio of the components of the effective mass of the Higgs fluctuations $m_{1,h}^2$ (left) and $m_{3,h}^2$ (right) rescaled by the Hubble scale at the end of inflation. The blue, red dashed and green dotted curves correspond to $\xi=10,10^2,10^3$ respectively.
 }
 \label{fig:mhoverPreg}
\end{figure}

The numerical results for $\xi=10$ are shown in Fig.~\ref{fig:mhoverH}, along with the approximate analytical expressions that we derived. We only show the $\xi=10$ case, since all cases with higher values of the nonminimal coupling exhibit visually identical results.
After the end of inflation the two dominant components of the effective mass of the Higgs fluctuations evolve differently for different values of $\xi$. In Ref.~\cite{MultiPreheat2} the behavior of $m_{1,h}^2$ was analyzed in the static universe approximation. It was shown that for $\xi\gtrsim 100$ the effective mass component $m_{1,h}^2$ quickly approaches a uniform shape regardless of the value of $\xi$. The consequence of that is that the Floquet chart for the inflaton self-resonance also approaches a common form for $\xi\gtrsim 100$. This can be seen in the left panel of Fig.~\ref{fig:mhoverPreg}, where $m_{1,h}^2$ is very similar between $\xi=100$ and $\xi=10^3$, but different for $\xi=10$. The coupled metric fluctuations component of the effective mass has a similar shape for $\xi=100$ and $\xi=10^3$, but for $\xi=10$ it is significantly less pronounced, as seen in the right panel of Fig.~\ref{fig:mhoverPreg}.

\subsubsection{Superhorizon Evolution and Thermalization}
\label{sec:superhorizon}

An important notion when dealing with (p)reheating is the transfer of energy from the inflaton condensate to the radiation degrees of freedom. Naively, one must compute all the power concentrated in the wave-numbers that are excited above to the vacuum energy (different than the adiabatic vacuum at any time) and compare that to the energy density stored in the condensate. However, when dealing with inflationary perturbations, one must keep in mind that computations should refer to modes, whose length-scales are relevant to the dynamics being studied. For curvature perturbations, the use of a finite box was described in Ref.~\cite{Lyth:2007jh}. For preheating, since thermalization proceeds through particle interactions, the relevant length-scales are those that allow for particle interactions, hence sub-horizon scales, or short wavelengths.

 The parametric excitation of long-wavelength modes has been extensively studied
\cite{Traschen:1990sw, Finelli:1998bu, Bassett:1998wg, Parry:1998pn,Bassett:1999mt, Bassett:1999ta, Felder:1999wt,  Easther:1999ws, Afshordi:2000nr, Tsujikawa:2002nf}. It has been demonstrated that the coupled metric fluctuations lead to an enhancement of --particularly-- long wavelength modes \cite{Bassett:1998wg, Bassett:1999mt, Bassett:1999ta,Easther:1999ws, Tsujikawa:2002nf},  which is larger than the one computed using a rigid background. Furthermore, the amplification of long-wavelength modes, even on super Hubble scales, does not violate causality, as discussed for example in Ref.~\cite{Finelli:1998bu,Bassett:1998wg, Bassett:1999mt, Bassett:1999ta, Tsujikawa:2002nf}. Intuitively, the inflaton condensate has a super-Hubble correlation length and can thus consistently affect super-Hubble modes. 

While UV modes encounter the complication of possibly being excited for wavenumbers that exceed the unitarity bound (this doesn't occur for Higgs modes),  the IR modes have a different conceptual difficulty: since thermalization occurs when particles interact and exchange energy, in order to lead to a thermal distribution, modes that are super-horizon are ``frozen-in" and hence cannot take part in such processes\footnote{Generically in multifield models, one would not expect the curvature perturbations to remain ``frozen in" when stretched outside the Hubble radius, since multifield interactions can generate non-adiabatic pressure, which in turn will source changes in the gauge-invariant curvature perturbations on arbitrarily long length-scales. However, in models like Higgs inflation that feature strong single-field attractor dynamics during inflation, the non-adiabatic pressure effectively vanishes and the long-wavelength modes remain ``frozen in," akin to the expected behavior in simple single-field models. Details on the  single-field attractor in such models can be found in Ref.~\cite{KS, GKS, MultiPreheat1}.}. Hence, it is normal to only consider modes that have large enough physical wave-numbers, that place them inside the horizon at the instant in time that we are considering. Modes that have longer wavelengths are frozen outside the horizon and do not contribute to the thermalization process. They should be summed over and added to the local background energy density. We will skip this last step, as their contribution is subdominant, compared to the energy density stored in the inflaton condensate.
\begin{figure}
\centering
\includegraphics[width=0.8\textwidth]{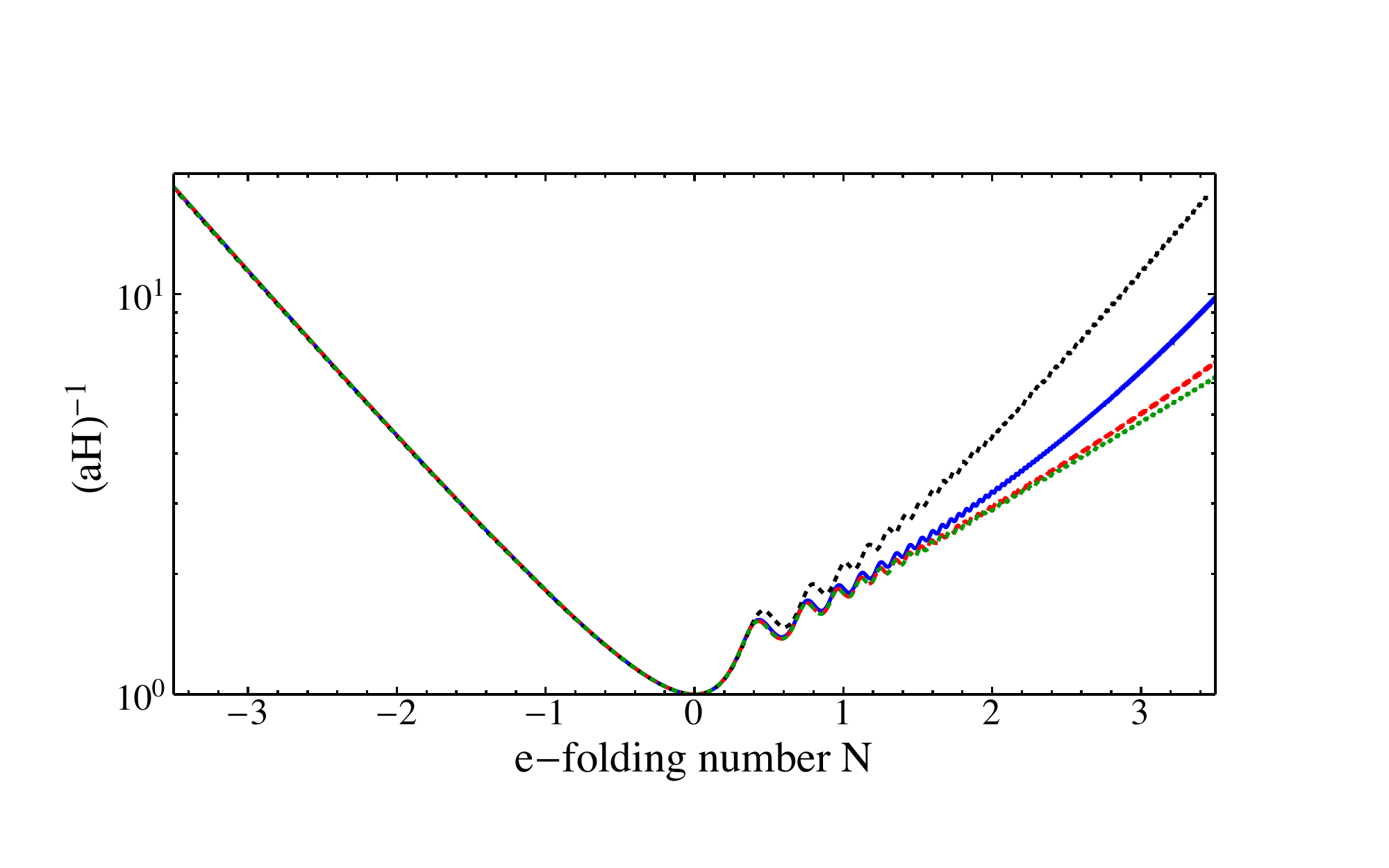}
\caption{
The size of the comoving Hubble radius during and after inflation for $\xi=10,10^2,10^3,10^4$ (black, blue, red and green respectively).
 }
 \label{fig:NvsaH}
\end{figure}
In Figure \ref{fig:NvsaH} we see the evolution of the comoving Hubble radius, shrinking during inflation and growing after that. We also see that different values of $\xi$ lead to different post-inflationary evolution, which is expected, since the effective equation of state of the background dynamics after inflation depends strongly on $\xi$, as shown in Ref.~\cite{MultiPreheat1}. More specifically, large nonminimal couplings $\xi \gtrsim100$ lead to a prolonged period of matter-domination-like expansion, which can last  for several $e$-folds in the absence of back-reaction.
As we will see in the next sections, the majority of the parametric resonance effects occur for $N\lesssim 3$ $e$-folds, placing the entirety of the reheating dynamics inside the matter-dominated background era for large values of $\xi$.
In order to take into account the relevant wavenumbers consistently, we use an adaptive code, that only sums up the contribution of modes that are inside the horizon at the point in time when computing the energy-density of the Higgs field fluctuations.
 
\subsection{Preheating}

We now move to the computation of the energy density in the Higgs particles that are produced during preheating. A detailed analysis was performed in Ref.~\cite{MultiPreheat3}. However, all computations were initialized at the end of inflation, thereby neglecting the amplification of small wavelength modes during the last e-folds of inflation.
We initialize all computations at $4.5$ $e$-folds before the end of inflation, in order to ensure  that all relevant modes are well described by the Bunch-Davies (BD) vacuum solution
\beq
v_{k,h}\simeq {1\over \sqrt {2k}} e^{-i{k \tau}} \, .
\eeq

\begin{figure}
\centering
\includegraphics[width=\textwidth]{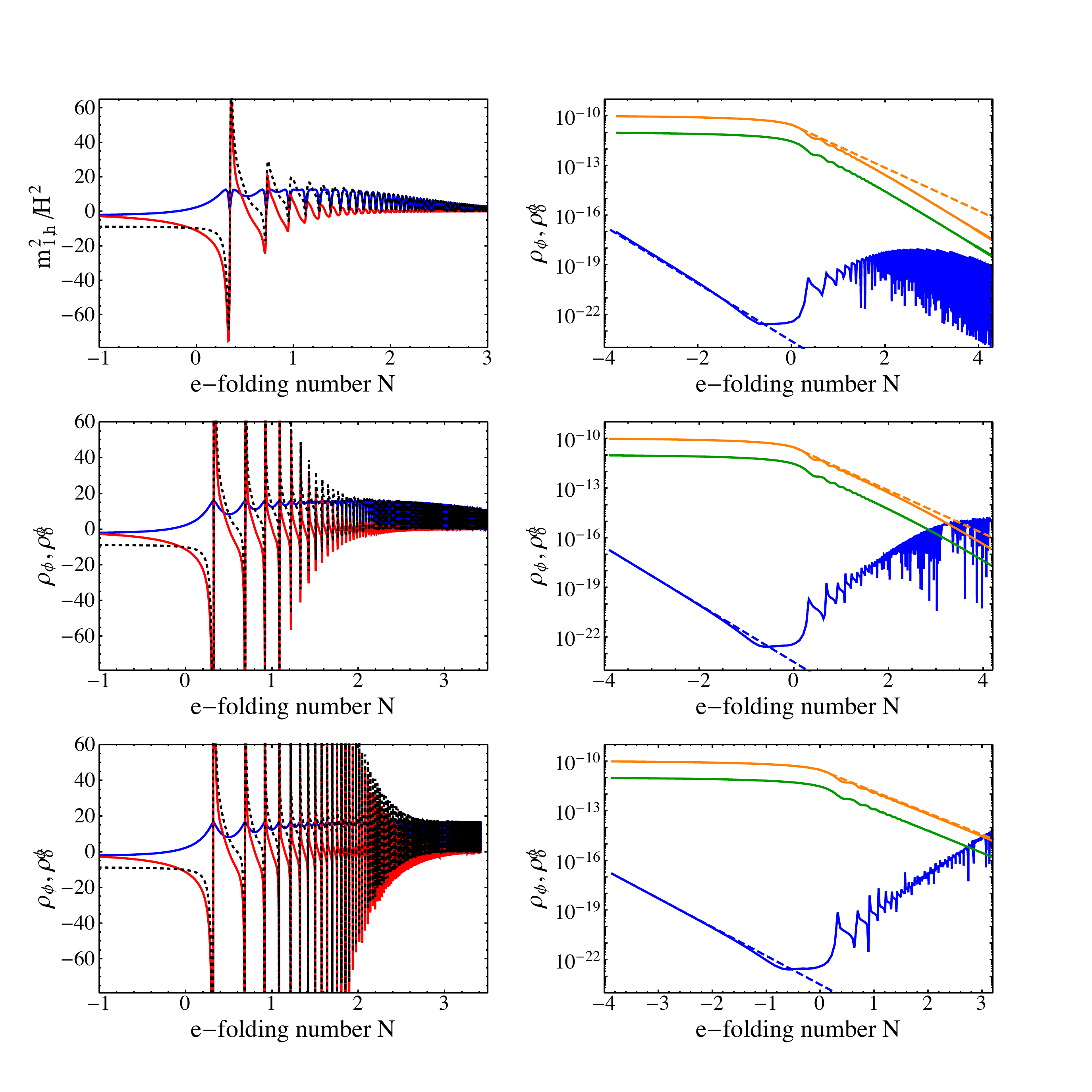} 
\caption{
{\it Left:} The  effective mass-squared (black-dotted), along with the contributions from the potential (blue) and the coupled metric perturbations (red). 
\\
{\it Right:} The energy density in the background Higgs condensate (orange) and the Higgs fluctuations (blue) for $\xi= 10, 10^2, 10^3$ (top to bottom). The green line shows $10\%$ of the background energy density, which is used as a proxy for the limit of our linear analysis. The orange-dashed line is $\rho_0 a^{-4}$, corresponding to the red-shifting of the background energy density during radiation-dominated expansion.
 }
 \label{fig:selfresonance}
\end{figure}

We see in the right panel of Fig.~\ref{fig:selfresonance} that at early times (before the end of inflation), the energy density in Higgs modes (indicated by the solid blue line) decays as $a^{-4}$ (indicated by the dotted line), in keeping with the expectation for modes in the BD state. However, approximately one e-fold before the end of inflation, the evolution of the energy density in Higgs modes departs from $a^{-4}$, because the low $k$-modes are enhanced with respect to the BD spectrum. This enhancement occurs because $m_{{ \rm eff}, h}^2 < 0$, an early tachyonic amplification phase driven largely by the effect of coupled metric perturbations. An immediate consequence of this fact is that one would underestimate the true amount of growth by starting the computation in a BD-like vacuum state at the end of inflation.

The right panels of fig.~\ref{fig:selfresonance} present the results for the energy transfer into Higgs particles for $\xi=10,10^2, 10^3$. Preheating completes when the energy density in the Higgs fluctuations (blue line) becomes equal to the energy density of the background field (orange line). However, the linear analysis is expected to break down when the energy density of the Higgs fluctuations becomes comparable to that of the inflaton field. As an indicator of the validity of the linear theory, which neglects backreaction of the excited modes onto the background, the green line shows $10 \%$ of the energy density of the inflation field. 

For all values of $\xi$ studied, the system exhibits an amplification of inflaton (Higgs) fluctuations. This is mainly caused by the periodic negative contribution of $m_{3,h}^2$ to the effective mass-squared $m_{{\rm eff},h}^2$, which is plotted in the left panel of Fig.~\ref{fig:selfresonance}. This is the term arising from  considering the effect of the coupled metric perturbations at linear order.
 As shown in Ref.~\cite{MultiPreheat3} and further reiterated in Fig.~\ref{fig:selfresonance}, the amplification driven by $m_{3,h}^2$ lasts longer for larger values of $\xi$. 
Specifically, the time at which the  tachyonic resonance regime stops scales as $ t \sim \sqrt{\xi} H_{\rm end}^{-1}$, as shown in Ref.~\cite{MultiPreheat3}.
However, for $\xi> 100$ the differences are irrelevant (in the simplified linear treatment), since the universe will have preheated already by $N\simeq 3$ $e$-folds. Hence for $\xi>100$, self-resonance of the Higgs field leads to predictions for the duration of preheating that are almost independent of the exact value of $\xi$. 

After the  tachyonic resonance has shut off (and if preheating has not completed yet), the modes undergo parametric resonance, driven by the oscillating effective mass term $m_{1,h}^2$. However, for very long-wavelength modes $k\simeq 0$, the Floquet exponent vanishes \cite{MultiPreheat2}, and the amplification is polynomial in time rather than exponential, hence significantly weaker. 
As shown in Ref.~\cite{MultiPreheat2} the maximum Floquet exponent in the static universe approximation is $\mu_{k,{\rm max}} T \approx 0.3$, where $T$ is the background period. Using the relation $\omega / H \simeq 4$, which was derived in Ref.~\cite{MultiPreheat1} for $\omega = 2\pi /T$, the maximum Floquet exponent is experssed as $\mu_k \sim 0.5 H$. Hence the Floquet exponent is too small to lead to  an efficient amplification of Higgs fluctuations in an expanding universe.
Thus the early time tachyonic resonance, driven by the coupled metric fluctuation is crucial for preheating the universe through Higgs particle production. 

For $\xi=10$ the situation is significantly different. Both tachyonic resonance, due to the coupled metric fluctuations encoded in $m_{3,h}^2$, as well as parametric resonance due to the potential term $m_{1,h}^2$ become inefficient earlier, leading to a slower growth of the fluctuations and the energy density that they carry and an incomplete preheating.
\begin{figure}
\centering
\includegraphics[width=0.7\textwidth]{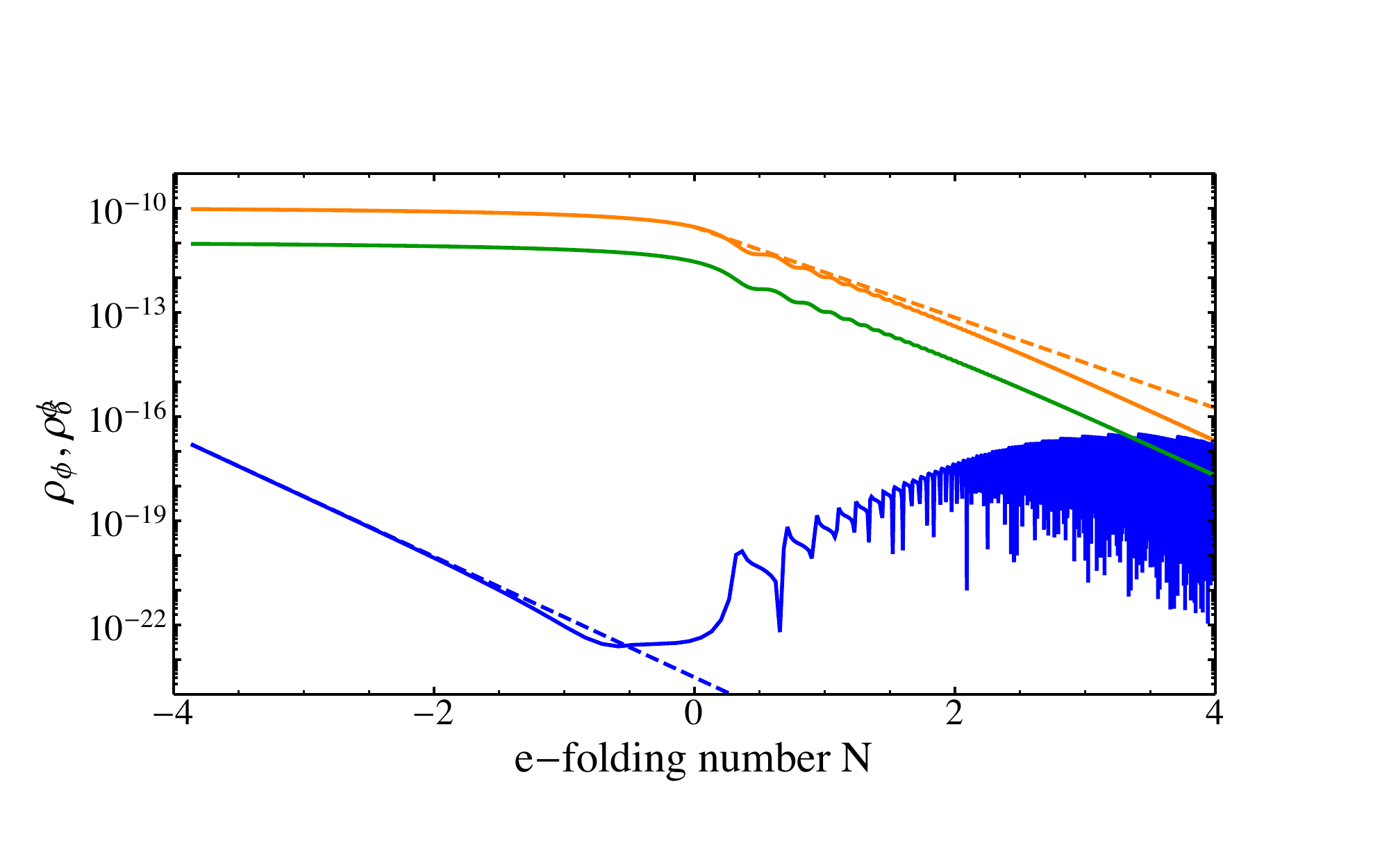}
\caption{
The energy density in the background Higgs condensate (orange) and the Higgs fluctuations (blue) for the marginal case of $\xi= 30$ (top to bottom). The green line shows $10\%$ of the background energy density, which is used as a proxy for the limit of our linear analysis. The orange-dashed line is $\rho_0 a^{-4}$, corresponding to the red-shifting of the background energy density during radiation-dominated expansion.
 }
 \label{fig:selfresonancexi30}
\end{figure}
However, for smaller values of the nonminimal coupling $\xi = {\cal O}(10)$ one must take into account
another important feature, namely the evolution of the background. As shown in Ref.~\cite{MultiPreheat1}, larger values of $\xi$ put the universe into a prolonged matter-dominated state ($w=0$). This means that the energy density of the background condensate redshifts as $a^{-3} = e^{-3N}$. For small values of $\xi$, however, the universe passes briefly through the background (average) equation of state $w=0$ and after the first $e$-fold approaches $w\simeq 1/3$. Fig.~\ref{fig:selfresonancexi30} shows the evolution of the energy density in Higgs modes for the marginal case of $\xi=30$.  We see that the fluctuation energy density in the Higgs modes would be always smaller than the background, if the background evolved with $w\simeq 0$, as indicated by the orange dashed line. However, the fact that the background energy density redshifts faster ($w\simeq 1/3$) allows for complete preheating. Simply put, nonminimal couplings in the ``intermediate" regime of $\xi= {\cal O}(10)$ exhibit a shorter period of tachyonic-parametric amplification, while at the same time following a background evolution of $\rho_{\rm \phi} \sim e^{-4N}$. 

\begin{figure}
\centering
\includegraphics[width=0.9\textwidth]{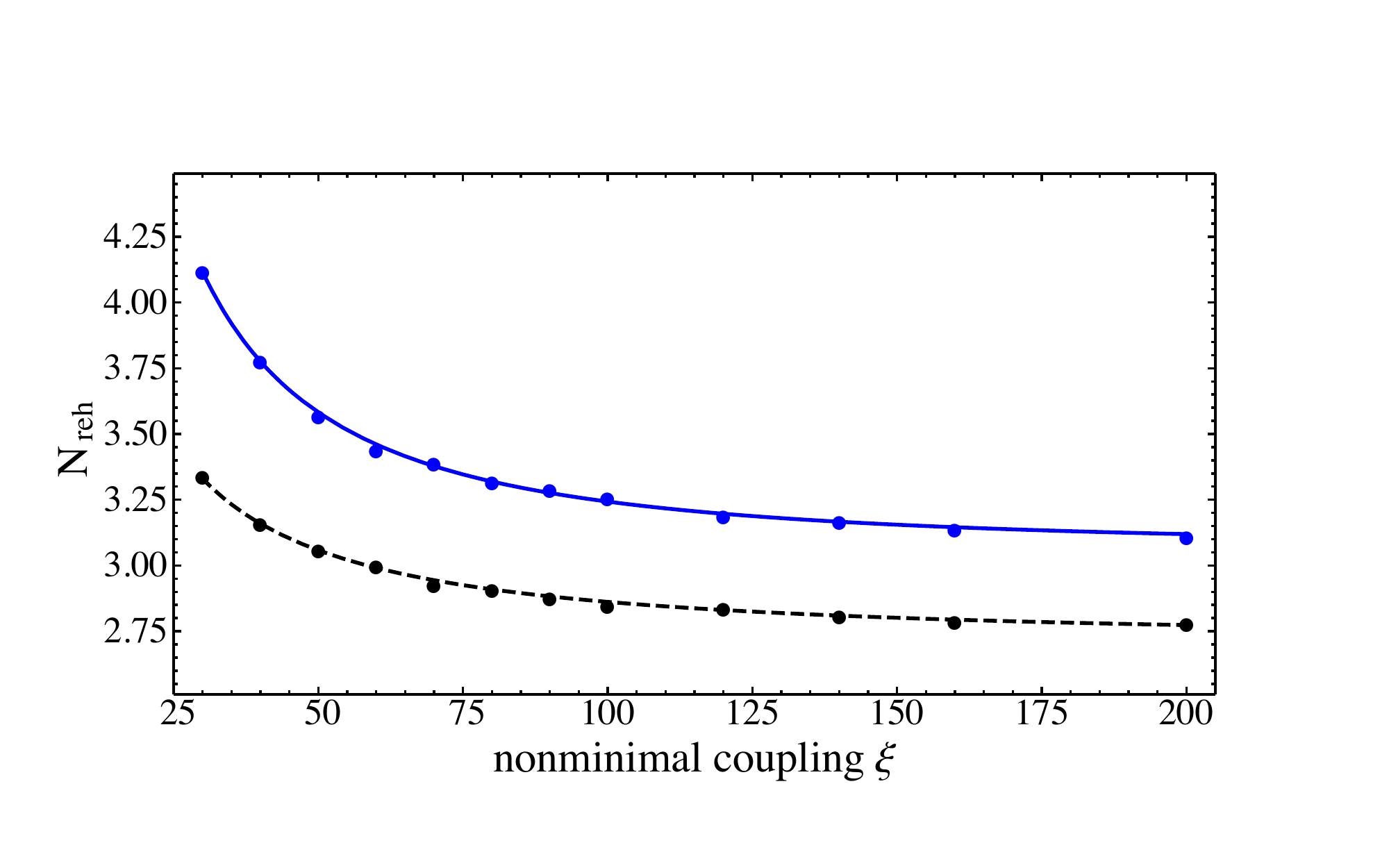} 
\caption{
The number of $e$-folds after inflation when the energy density in the Higgs fluctuations equals the background energy density $N_{\rm reh}$ (blue solid) or $10\%$ of the background energy density $N_{\rm br}$ (black dashed). 
 }
 \label{fig:RhoIntersect}
\end{figure}

We distinguish two time points relevant for preheating: $N_{\rm reh}$ is the time at which the energy density in the linear fluctuations equals the background energy density, which we take as the time of complete preheating and $N_{\rm br}$ is the time at which the energy density in the linear fluctuations equals $10\%$ of the background energy density, which is the point at which back-reaction effects may become important. We have numerically found that self-resonance of the Higgs field becomes insufficient to preheat the universe at $\xi < 30$. In particular, the results for $N_{\rm reh}(\xi)$ can be fitted by a simple analytical function, as shown in Fig.~\ref{fig:RhoIntersect}:
\begin{equation}
N_{\rm reh}(\xi) \simeq \frac{21}{ \xi(1+ 0.016\xi)} + 3 \, ,
\label{eq:Nrehapprox}
\end{equation}
for $ \xi \gtrsim 30$, where complete preheating is possible, at least in the linear approximation that we used. For $\xi>100$, $N_{\rm reh}$ becomes largely independent of $\xi$, as expected from the results of Fig.~\ref{fig:selfresonance}. 

As a final note, we must say that the results were insensitive to the exact value of the maximum wave-number considered. This is due to the fact that the small (but sub-horizon) wavenumbers $k= {\cal O}(H_{\rm end})$ are exponentially amplified and dominate the fluctuation energy density shortly after the end of inflation. Hence we do not need to implement any scheme to subtract the vacuum contribution from large-$k$ modes, since it is vastly subdominant for any reasonable UV cutoff. 


\section{Gauge / Goldstone boson production}
\label{sec:Isocurvature}

\subsection{Evolution during Inflation and Initial Conditions for Preheating}
\label{sec:gaugeinflation}

We  use the equations of motion derived in the Abelian model in unitary gauge, in order to study the evolution of gauge fields during inflation. The unitary gauge is well defined in this period, since $\varphi(t)$ does not vanish. The values of $B^{\pm,L}_\mathbf k$ at the end of inflation serve as initial conditions for preheating. Especially for initializing lattice simulations, which are increasingly expensive to start deeper within inflation, accurate knowledge of the spectrum of gauge fields at the end of inflation is essential.
During preheating, unitary gauge is not well-defined at moments when $\varphi(t)=0$, so we use Coulomb gauge. In order to determine the initial condition for $\theta_\mathbf k$, we will use Eq.~\eqref{eq:BLtotheta}, which relates $B^L_{\mathbf k}$ in unitary gauge to $\theta_\mathbf k$ in Coulomb gauge.

The equations of motion for the longitudinal and transverse modes in unitary gauge in conformal time $\tau$ are
\beqn
&&\partial_\tau^2B^L_{\mathbf k} + 2 \left ( {\partial_\tau \varphi\over \varphi}  - {\partial_\tau f \over 2f} + {\partial_\tau a\over a}\right ) {k^2 \over k^2 +{M_{\rm Pl}^2 a^2 \over 2f} e^2\varphi^2 } \partial_\tau B^L_{\mathbf k} + 
\left (   k^2 + a^2 {M_{\rm Pl}^2 \over 2f } e^2\varphi^2\right )B^L_{\mathbf  k}=0 \, ,
\label{eq:BLeomtau}
\\
&& \partial^2_\tau B^{\pm}_\mathbf k + \left(k^2 + a^2 \frac{M_\text{pl}^2e^2}{2f}\varphi^2\right)B^\pm_\mathbf k =0 \, , 
 \label{eq:Bpmeomtau}
\eeqn
where $e$ is the $U(1)$ gauge coupling.
These equations are of the form
\beq
\partial^2_\tau B^{I}_{\mathbf k} + \left ({\partial\over \partial\tau}\log(b_I) \right )\partial_\tau B^{I}_{\mathbf k} +\omega_{I}^2 (k,\tau)B^I_{\mathbf k}=0  \, ,
\eeq
with $I$ denoting either $L$ or $\pm$ polarization and the $b_I$ and $\omega^2_I$ given by:
\beq
	\begin{aligned}
		& b_{L} (k,\tau) = \left (1+{k^2  2f\over M_{\rm Pl}^2 a^2 e^2 \varphi^2}\right)^{-1} \, , \qquad 
&&\omega_{L}^2(k,\tau)= k^2 + a^2 {M_{\rm Pl}^2 \over 2f } e^2\varphi^2 \, , \\
		& b_{\pm}(k,\tau) = 1\, , \qquad && \omega_{\pm}^2(k,\tau) =  k^2 + a^2 {M_{\rm Pl}^2 \over 2f } e^2\varphi^2 \, .
	\end{aligned}
\eeq
After integrating by parts, we rewrite the quadratic action in Fourier space as
\beq
S^I = \int d\tau {\cal L}_{I}(\tau)=\int d\tau \int d^3k \, b_{I} (k,\tau) \left [    {1\over 2} \left |  \partial_\tau B^I_{\mathbf k}  \right |^2-{1\over 2} \omega_{I}^2 (k,\tau) \left | B^I_{\mathbf k}\right|^2
\right ] \, ,
\label{eq:actionbI}
\eeq
and  follow the same quantization procedure as the one appearing in Ref.~\cite{Lozanov:2016pac}. This is the standard method used to quantize models with noncanonical kinetic terms, 
which include nonminimally coupled models in the Einstein frame. The canonical momentum is
\beq
\pi_{I,\mathbf k}(\tau ) = {\delta {\cal L}(\tau) \over \delta (\partial_\tau B^I_{-\mathbf k}(\tau)) } = b_{I}\partial_\tau B^I_\mathbf k(\tau) \, ,
\eeq
and the commutator relation of the operator $\hat B^I_{\mathbf k}(\tau)$ is
\beq
[\hat B^I_{\mathbf k}(\tau) , \partial_\tau \hat B^J_{\mathbf q}(\tau)] = i  {1\over b_{I}(k, \tau) } \delta^{IJ}\delta(\mathbf k+ \mathbf q) \, .
\eeq
We decompose the field operator $\hat B^I_\mathbf k(\tau) $ in terms of creation and annihilation operators
\beq
\hat B^I_{\mathbf k}(\tau)  = \hat a^I_\mathbf k u^I_k(\tau) +\hat a^{I\dagger} _{-\mathbf k} u^{I*}_k (\tau) \, ,
\eeq
where the mode-function $u^I_k(\tau)$ satisfies the same equation of motion as the field operator $\hat B^I_{\mathbf k}(\tau)$, Eq.~\eqref{eq:BLeomtau}. 

As long as the adiabaticity condition $\left| \frac{\partial_\tau\omega}{\omega^2} \right|  \ll 1$ holds \cite{Lozanov:2016pac}, the modes  can be described by the WKB-approximation 
\beqn
\nonumber
u^I_k(\tau ) &=& {\alpha^I \over \sqrt 2} {1\over \sqrt{b_{I}(k,\tau)} \sqrt{\omega_{I}(k,\tau)} }\exp \left ( - i \int d\tau' \omega_{I}(k,\tau') \right ) 
\\
&+&
 {\beta^I \over \sqrt 2} {1\over \sqrt{b_{I}(k,\tau)} \sqrt{\omega_{I}(k,\tau)} }\exp \left ( + i \int d\tau' \omega_{I}(k,\tau') \right ) 
\, .
\label{eq:ukLfull}
\eeqn

The behavior is different for modes with $|k\tau| > x_c$ (early times/short wavelengths) and $|k\tau| < x_c$ (late times/long wavelengths), with $x_c$ given by:
\beq
	x_c = \sqrt{\frac{12 \xi}{\lambda}}\,e \, ,
	\label{eq:defxc}
\eeq
corresponding to the ratio of the gauge boson mass to the Hubble scale during inflation. Both cases can be described by the WKB, but they exhibit different behavior that we describe hereafter.

At early times and for large values of the wavenumber $k$ the wavefunction $u_k^I(\tau)$ must match onto the Bunch-Davies vacuum solution. We focus on the longitudinal mode first. We can take the limit of early times (or sub-horizon modes) analytically, when $1\ll |k\tau | \simeq k/ (aH)$, resulting in
\beq
\omega_{L}(k,\tau) =\sqrt{ k^2 + a^2 {M_{\rm Pl}^2 \over 2f } e^2\varphi^2} \to k \, ,
\eeq
and
\beq
b_{L}(k,\tau)  \to  
{ M_{\rm Pl}^2 a^2 e^2 \varphi^2 \over k^2  2f} \, .
\eeq
Putting everything together, the mode function for $|k\tau| >x_c$ becomes
\beq
u_k^L(\tau ) \to    {1\over \sqrt{2k}}{{k \tau }\over x_c} 
e^{-ik\tau} \, ,
\label{eq:ukLapprox}
\eeq
where we used $2f \simeq \xi\varphi^2$ and $x_c^2 ={ M_{\rm Pl}^2 e^2 /( \xi H^2)} $.

The transverse modes are canonically normalized and furthermore conformally coupled at early times, hence their mode function becomes
\beq
u_k^\pm(\tau) \to  {1\over \sqrt{2k}} e^{-ik\tau} \, .
\eeq
Overall $\alpha^{L,\pm}=1$ and $\beta^{L,\pm}=0$ in Eq.~\eqref{eq:ukLfull}.

\subsubsection{Single field attractor strength from gauge  interactions}

The super-horizon evolution ($k\ll aH$) of isocurvature fluctuations is an indicator of the (in)stability of the classical background trajectory (see for example Ref.~\cite{Renaux-Petel:2015mga}). We will analyze the behavior of the gauge fields and the possible effects on the stability of the single field attractor. We will mainly focus on the longitudinal mode, since it will be amplified most efficiently during preheating.

During inflation we can rewrite the equations of motion using $x=-k\tau$ as the time variable. If we further make use of  the de-Sitter approximation ($\tau = -1/aH$) and take $\varphi(t)$ as a constant, the equations of motion become\footnote{It is worth noting that the equations of motion for the gauge fields during inflation look very similar in structure to the ones derived for a minimally coupled charged inflaton in Ref. \cite{Lozanov:2016pac}. As shown in Ref. \cite{KMS}, the field-space is asymptotically flat for large field values, hence all covariant derivatives can be substituted for partial derivatives during inflation, at lowest order in $1/\xi$.}
\beqn
&&\partial_x^2 B_\mathbf k^\pm + \left ( 1 + {x_c^2\over x^2}  \right )B_\mathbf k^\pm=0 \, ,
\label{eq:BpmdS}
\\
&&\partial_x^2 B_\mathbf k^L - {2\over x} {1 \over 1 +    {x_c^2\over x^2}}\partial_x B_\mathbf k^L+ \left ( 1 + {x_c^2\over x^2}\right )B_\mathbf k^L=0 \, ,
\label{eq:BLdS}
\eeqn
with $x_c$ as defined in eq. (\ref{eq:defxc}). As expected, we recover the solution of Eq.~\eqref{eq:ukLapprox} in the limit $|k\tau| \gg x_c$.

By using the relation between $\xi$ and $\lambda$ given in Eq.~\eqref{eq:lambdaxi} which is required by the normalization of the power spectrum,
Eq.~\eqref{eq:defxc} gives
\beq
x_c =  {{\cal O}(10^{5} )\over \sqrt{\xi}} \, ,
\label{eq:xcnumber}
\eeq
where we took $e\simeq 1$ for Standard Model gauge couplings during inflation.
For $\xi\gtrsim 1$, where the CMB observables and the inflationary dynamics fall into the ``large $\xi$" attractor, $x_c\gg 1$ for all values of interest. Thus, the first of the two cases that were examined in Ref. \cite{Lozanov:2016pac}, $x_c<1$ and $x_c>1$, does not apply for nonminimally coupled models of inflation with large $\xi$, unless one takes a very weakly coupled gauge field $e\ll1$, making such a value very different to gauge couplings found in the SM.

 For the longitudinal mode $B^L$ the presence of a first-derivative term is important for $x<x_c$, leading to
 \beq
 u_L(k,\tau) = {1\over  \sqrt{2k}} \left ({k|\tau|\over x_c}\right)^{1/2} (k|\tau|)^{-i \, x_c} ~,~ k|\tau| <x_c  \, .
 \label{eq:BLxlessxc}
 \eeq
The details of the derivation are given in Appendix A.

\begin{figure}[h]
\centering
\includegraphics[width=\textwidth]{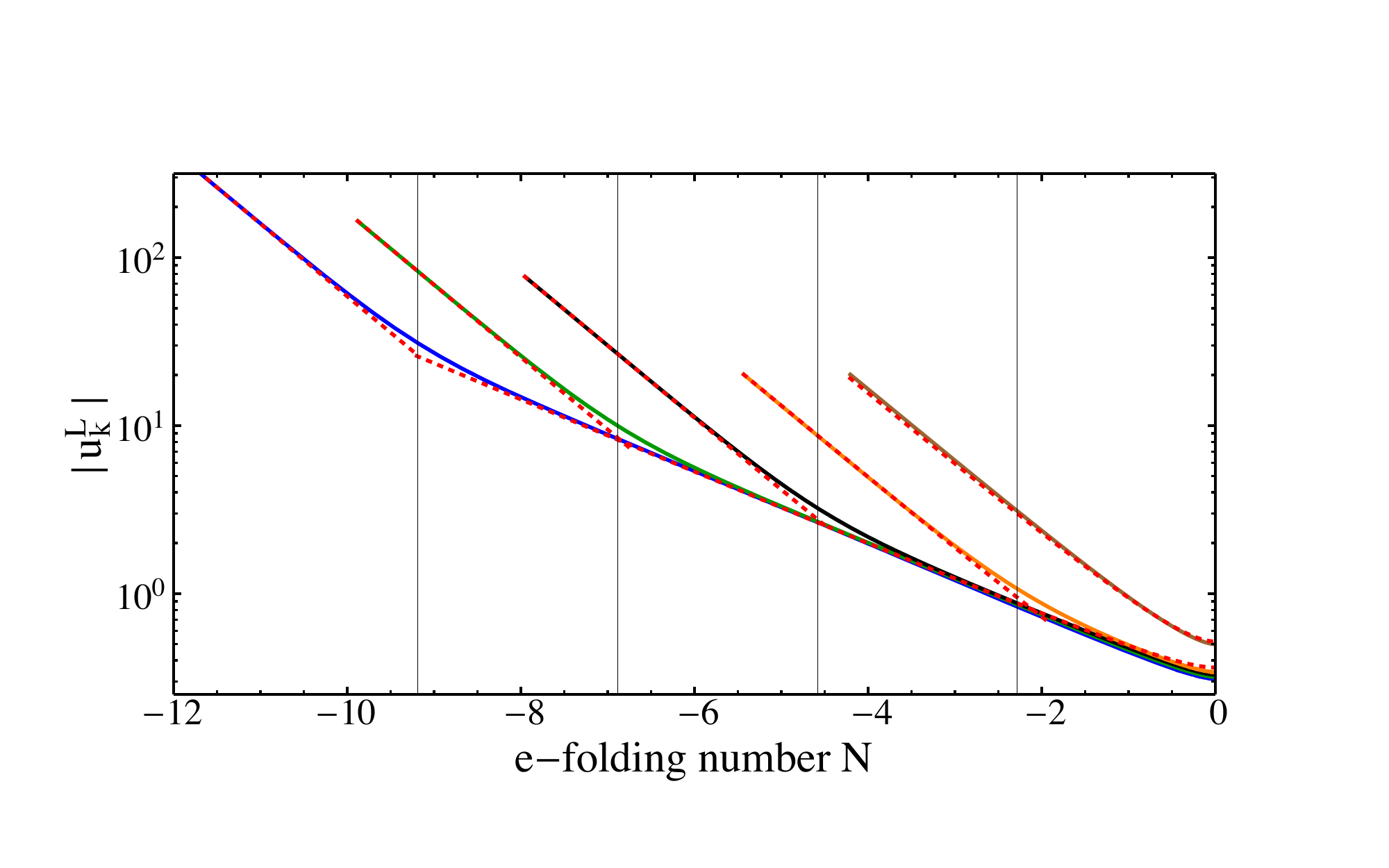}
\caption{
The evolution of the longitudinal gauge field mode during inflation for $\xi=10^3$. The solid lines correspond to the numerical solution for $k / H_{\rm end} = 1,10,10^2,10^3, 10^4$
(blue, green, black, orange and brown respectively), along with the approximate solutions of Eqs.~\eqref{eq:ukLapprox} and \eqref{eq:BLxlessxc} (red-dotted). The vertical lines correspond to the points $k|\tau| = x_c$, where the matching between the two asymptotic regimes is performed. For the brown curve this does not occur during inflation.
}
 \label{fig:BLkvsN}
\end{figure}

Fig.~\ref{fig:BLkvsN} shows the evolution of certain wavenumbers from $k|\tau|\gg x_c\gg 1$ until the end of inflation. It is evident that the simple scalings of 
$\left|u_L(k|\tau|\gg x_c)\right|\propto |\tau|$ and $\left |u_L(k|\tau|\ll x_c)\right|\propto \sqrt{|\tau|}$ agree very well with the full numerical evolution across a wide range of wavenumbers.
While $\xi=1000$ was chosen for Fig.~\ref{fig:BLkvsN}, different values of a nonminimal coupling $\xi\gg 1$ lead to similar results.

Following Eq.~\eqref{eq:xcnumber}, gauge fields during Higgs inflation become very massive, when compared to the Hubble scale. 
This further reinforces the single-field description of the background trajectory, discussed in Section \ref{sec:parameterchoices}, since the orthogonal direction(s), described equally well through the scalar degree of freedom $\theta$ or through the longitudinal polarization of the gauge boson $B^L$, are very massive, making the background single-field trajectory a stable one\footnote{We must note here, that our analysis only shows the linearized stability of the single field trajectory, not the approach towards it from generic initial conditions $\{\Phi, \partial_t \Phi\}$. The latter was performed in Ref. \cite{GKS} for an $SO(4)$ symmetric model, meant to describe Higgs inflation without gauge couplings.}.
It is interesting to note, that due to the relation between $\xi$ and $\lambda$, arising from the normalization of the power spectrum, the gauge fields become less massive for larger $\xi$, meaning that the ratio of the gauge field mass to the Hubble scale becomes smaller. Hence, the single field attractor, at least in the linearized analysis, becomes weaker for larger $\xi$. This is opposite to the case of a scalar-only multi-field model with a nonsymmetric potential, where the attractor strength increases with $\xi$, as shown in Ref.~\cite{MultiPreheat1}. While for the SM the gauge couplings are large enough to make the gauge field much heavier than the Hubble scale, 
one can construct more general inflationary models, involving a Higgs-like field and the associated gauge sector.
In this case weakly coupled gauge sectors might leave observational imprints through oscillations of the background during inflation. A search for ``primordial clocks" \cite{Chen:2014cwa, Chen:2015lza} in these models is beyond the scope of the present work, because they do not arise in SM Higgs inflation, but could provide a useful tool for exploring gauge field phenomena in broader classes of nonminimally coupled inflation. 

The transverse modes are significantly easier to analyze, since 
Eq.~\eqref{eq:BpmdS} makes clear that the $B^\pm$ are conformally coupled at early times and will become massive (and thus be suppressed) for $x< x_c$. In the de-Sitter approximation, Eq.~\eqref{eq:BLdS} can be solved exactly using Hankel functions, resulting in
\beq
u_k^\pm(\tau) = {\sqrt{-k\tau}}
\sqrt{\pi\over 4k} H_z^{(1)} (-k\tau) e^{i z {\pi\over 2} + i{\pi\over 4}}\, ,
\eeq
where $z = \sqrt{{1\over 4} - x_c^2}$, as described for example in Ref.~\cite{Lozanov:2016pac}. The analysis of the transverse modes is essentially identical to the minimally coupled case of  Ref.~\cite{Lozanov:2016pac}. Since they will not be significantly amplified during preheating, we will not discuss them further.

\subsubsection{Initial conditions in Coulomb gauge}

Having explored in detail the behavior of the longitudinal gauge fields during inflation, we focus on their form close to the end of inflation and the start of the (p)reheating era. Since we are interested in the details of the vacuum (as will be evident later), we compare the adiabatic vacuum during inflation, given in Eq.~\eqref{eq:ukLfull} to the approximate analytic expressions derived for $u_L(k,\tau)$, as well as to the numerically derived values. 
It is a straightforward exercise to expand $b_L(k,\tau)$ and $\omega_L(k,\tau)$ in the two limiting cases of $k|\tau|$ to see that the WKB expression given in Eq.~\eqref{eq:ukLfull} with $\alpha=1$ and $\beta=0$
 matches Eqs.~\eqref{eq:ukLapprox} and \eqref{eq:BLxlessxc} in the appropriate limits.

\begin{figure}
\centering
\includegraphics[width=\textwidth]{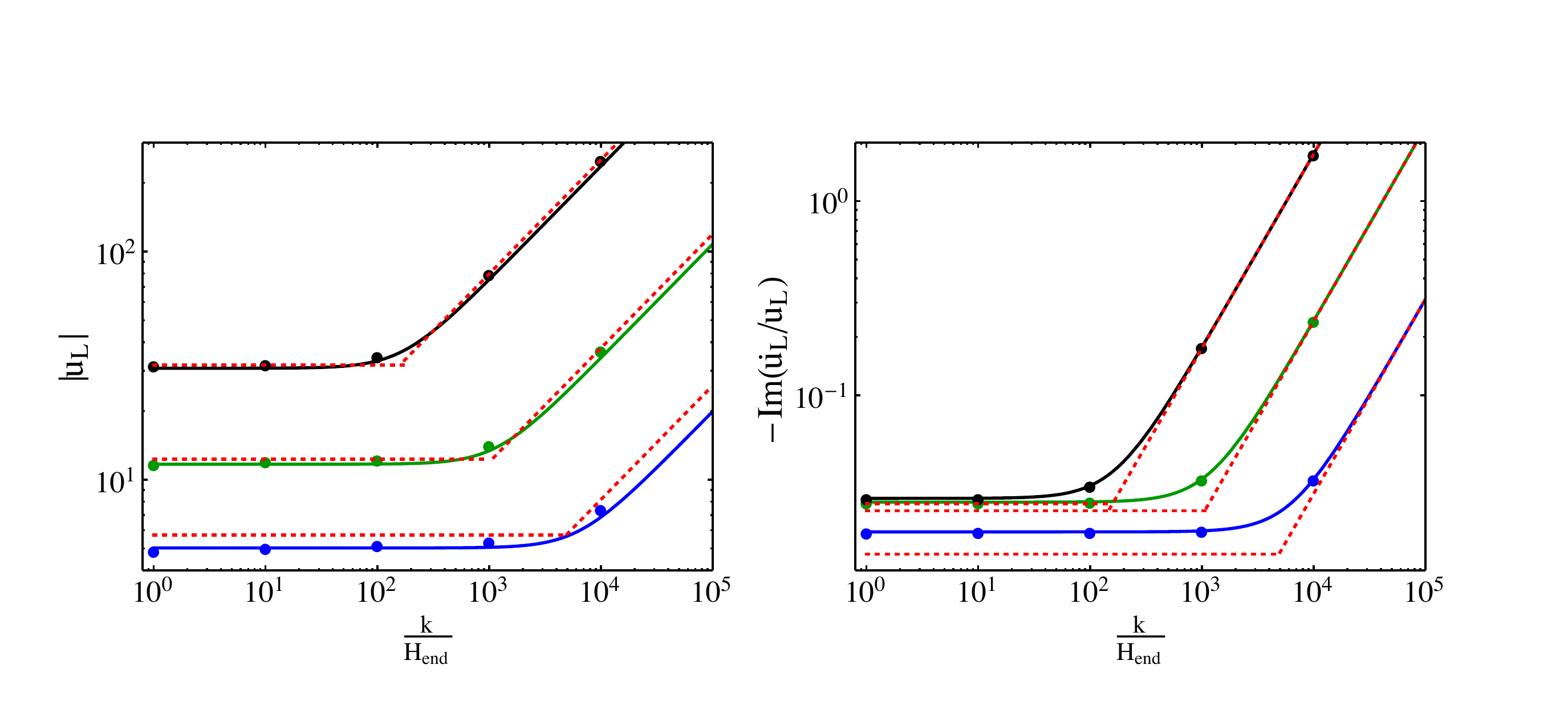}
\caption{
The mode-function amplitude (left) and frequency (right) for $N=0,2,4$ $e$-folds before the end of inflation (blue, green and black respectively). Solid lines correspond to the WKB expression of Eq.~\eqref{eq:ukLfull}  and red-dotted lines correspond to the approximate solutions for $x\gg x_c$ and $x\ll x_c$. The dots show the full numerical results.
}
 \label{fig:uLvsk}
\end{figure}

Fig.~\ref{fig:uLvsk} shows the comparison of the WKB solution of Eq.~\eqref{eq:ukLfull}, the approximate expressions of Eqs.~\eqref{eq:ukLapprox} and \eqref{eq:ukLapproxlate}, as well as the numerical results from the modes shown in Fig.~\ref{fig:BLkvsN}. We see an excellent agreement between all three, with the exception of the modes around $k|\tau| \sim x_c$, where the approximate expressions fail, since they were derived using the limits $k|\tau|\gg x_c$ or $k|\tau|\ll x_c$. We must also note that we used the approximation $\tau = -1/aH$ for the analytically derived expressions, hence we expect some discrepancy close to the end of inflation. 
This agreement has a significant physical meaning: since the adiabatic vacuum follows the evolution of the mode-functions, there is no particle production during inflation. We can thus begin our numerical computations at the end of inflation, unlike the case of Higgs self-resonance, where we needed to initialize our simulations several $e$-folds before the end of inflation, in order to capture nontrivial dynamics that took place during the last stages of inflation itself.

The initial conditions in Coulomb gauge can be easily read off from the unitary gauge solutions, using Eq.~\eqref{eq:BLtotheta}. It is interesting to note that there is no $\xi$-dependent term in the relation of $\theta_{\mathbf k}$ to $B^L_{\mathbf k}$. 
The initial conditions that we will use for the computations in Coulomb gauge are:
\beqn
\theta_{\mathbf k}(t_{\rm in}) &=&{1\over \sqrt 2} {e\varphi(t_{\rm in})\over k} {1\over \sqrt{b_{L}(k,t_{\rm in})} \sqrt{\omega_{L}(k,t_{\rm in})} } \, ,
\label{eq:thetainit1}
\\
\dot \theta_{\mathbf k}(t_{\rm in}) &=&- i \, {\omega_L(k,t_{\rm in}) \over a(t_{\rm in})} \times \theta_k(t_{\rm in}) \, .
\label{eq:thetainit2}
\eeqn

Before we conclude the analysis of the gauge field evolution during inflation, let us focus on the case of $k|\tau|\gg x_c$, where the initial conditions for preheating are
\beqn
\theta_\mathbf k(\tau_{\rm in})  &  \approx&   {e\phi \over x_c}{{  \tau_{\rm in}  }\over \sqrt {2k}} \, ,
\\
\dot \theta_\mathbf k(\tau_{\rm in})   &\approx& 
\theta_\mathbf k(\tau_{\rm in}) \times \left ( 
{i \, k \over a(\tau_{\rm in})}\right ) \, .
\eeqn
It is reassuring that for large wavenumbers the coupling constant $e$ drops out of the initial conditions for the $\theta$ field (since $x_c\propto e$), hence the decoupling limit is trivially obtained. For $k|\tau|<x_c$ it is slightly more complicated to see that, since for $e\to0$ we get $x_c\to0$, hence that region shrinks into nonexistence as we take the decoupling limit. Also, we would have to compute the expressions for $x_c\ll 1$ before we send $e\to 0$ in that case. Since the case of $e\ll 1$ does not apply to Higgs inflation, we will not pursue it further.

\subsection{Preheating}
\label{sec:gaugepreheat}

We start by rewriting Eq.~\eqref{eq:XthetaFULL} in a somewhat more compact way
\beqn
\nonumber
{\cal D}^2_\tau X^\theta 
&&- \partial_\tau \log\left(1+{\tilde m_B^2 \over k^2} \right ) {\cal D}_\tau X^\theta
\\
+&&
\left [
k^2 +a^2 m_{{\rm eff},\theta}^2
+ \tilde m_B^2 + \left ({\partial_\tau \varphi \over \varphi} + \frac{\partial_\tau a}{a}  -  {\partial_\tau f \over 2f}\right )  \partial_\tau \log \left ( 1+ {\tilde m_B^2 \over k^2}\right) 
\right ] X^\theta=0 \, ,
\label{eq:Xtheta_eta}
\eeqn
where we defined the gauge field mass
\beq
\tilde m_B^2 \equiv e^2 \varphi^2\frac{M_\text{pl}^2}{2f} a^2 \, ,
\eeq
and $X^\theta = a(t)\cdot \theta$. We normalize the scale-factor as $a\equiv1$ at the end of inflation. The effective mass of the Goldstone mode $\theta$ in the absence of gauge fields is
\beq
m_{{\rm eff},\theta}^2 \equiv {\cal M}^\theta\,_{\theta} -{1\over 6} R = m_{1,\theta}^2+m_{2,\theta}^2+m_{3,\theta}^2+m_{4,\theta}^2 \, ,
\eeq
with
\beqn
m_{1,\theta}^2 &=& {\cal G}^{\theta\theta} ({\cal D}_\theta {\cal D}_{\theta} V) \, ,
\\
m_{2,\theta}^2 &=& -{\cal R}^\theta_{~hh\theta} \, \dot\varphi^2 \, ,
\\
m_{3,\theta}^2 &=& 0 ,
\\
m_{4,\theta}^2 &=& -{1\over 6}R  = (\epsilon -2)H^2 \, .
\eeqn

The numerical solution of Eq.~\eqref{eq:Xtheta_eta} was performed in cosmic rather than conformal time, since this is more convenient for numerical simulations after the end of inflation. The computations were initialized at the end of inflation, according to Eqs.~\eqref{eq:thetainit1} and \eqref{eq:thetainit2}. 

We can follow the quantization method described in \cite{MultiPreheat1} and utilized in Section \ref{sec:Adiabatic} for the study of Higgs self-resonance
\beq
\hat X^\theta = \int {d^3k\over (2\pi)^{3/2}} 
\left [
 z_k e_2^{~\theta} \,  \hat a_\mathbf k  e^{i \mathbf k \cdot \mathbf x} 
+
z_k^* e_2^{~\theta}\, \hat a^\dagger_\mathbf k  e^{-i \mathbf k \cdot \mathbf x} 
\right ] \, ,
\eeq
where $e_2^{~\theta} = \sqrt{{\cal G}^{\theta\theta}}$. 
Using the vielbein decomposition, the covariant derivatives are effectively substituted by partial ones 
\beqn
\nonumber
\partial^2_\tau z_k &&- \partial_\tau \log (1+\tilde m_B^2/k^2)\cdot \partial_\tau z_k 
\\
+&&
\left (k^2 + a^2 m_{{\rm eff},\theta}^2
+
\tilde m_B^2 + {1\over 2}
\partial_\tau \log \left (\tilde m_B^2  \sqrt{2f\over M_{\rm Pl}^2}\right )
  \partial_\tau \log \left ( 1+ {\tilde m_B^2 \over k^2}\right)
\right )z_k
=0 \, .
\eeqn
In order to eliminate the first-derivative term we can use the rescaled variable $\tilde z_k$, defined as
\beq
z_k = \sqrt{1+{\tilde m^2_B\over k^2}}\, \tilde z_k \equiv T \cdot \tilde z_k \, ,
\eeq
leading to 
\beq
 \partial^2_\tau \tilde z_k 
 +
\omega^2_z
\, \tilde  z_k=0 \, ,
\eeq
where
\beq
\omega^2_z=
k^2 + a^2 m_{{\rm eff},\theta}^2
+
\tilde m_B^2 + {1\over 2}
\partial_\tau \log \left (\tilde m_B^2  \sqrt{2f\over M_{\rm Pl}^2}\right )
  \partial_\tau \log (T^2)
  +{\partial_\tau^2 (\tilde m_B^2) \over 2k^2 T^2} - {3\over 4} {(\partial_\tau \tilde m_B^2)^2 \over k^4 T^4} \, ,
  \label{eq:omegaz2}
\eeq
where $\tilde m_B^2$ is larger than $m^2_{1,\theta}$ and $m^2_{4,\theta}$.
As discussed extensively in Refs.~\cite{MultiPreheat1, MultiPreheat2, MultiPreheat3} for the case of a purely scalar multi-field model with large nonminimal couplings to gravity, the field-space manifold is asymptotically flat for large field values and exhibits a curvature ``spike" at the origin $\varphi(t) \simeq0$.  This ``Riemann spike" is exhibited in the effective mass of the isocurvature modes $m^2_{{\rm eff},\theta}$, more specifically in the $m_{2,\theta}^2 $ component, which is subdominant for all times away from the zero-crossings of the background value of the inflaton field $\varphi(t)$. We will not reproduce the entirety of the Floquet structure of this model, both because we do not wish to repeat the analysis of \cite{MultiPreheat2}, and because, as we will see in the subsequent section, the first zero-crossing of $\varphi(t)$ is the only relevant one for preheating through gauge modes.

In order to estimate the maximum excited wavenumber $k_{\rm max}$, we consider the following approximation, containing only the dominant terms
\beq
\omega_{z, \, {\rm approx}}^2 \equiv k^2 + a^2 m^2_{2,\theta} + \tilde m_B^2 \, ,
\eeq
where $\tilde m_B^2$ dominates over all subsequent terms in Eq.~\eqref{eq:omegaz2} for large $k$.
\begin{figure}
\centering
\includegraphics[width=\textwidth]{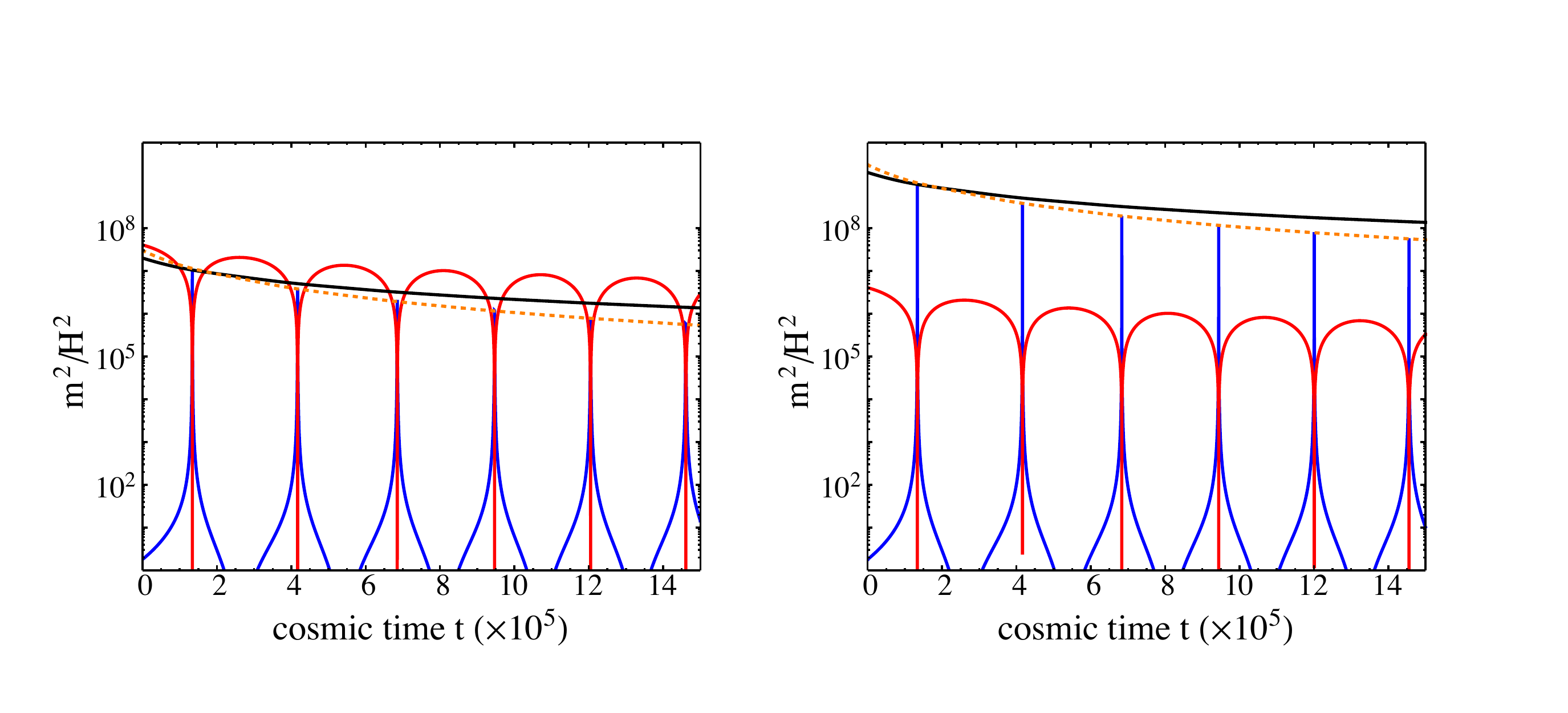}
\caption{Dominant components of the effective frequency-squared for $\xi=10^3$ (left) and $\xi=10^4$ (right). Color coding is as follows: 
$  \tilde m_B^2/ a^2$ (red), 
$m^2_{2,\theta}$ (blue) and 
$k^2/a^2$ (black) for the maximum excited wavenumber $k_{\rm max}$. The orange-dotted curve shows the scaling $a^{-3}$. 
}
 \label{fig:meffchiplot}
\end{figure}
Fig. \ref{fig:meffchiplot} shows the three contributions to $\omega_{z, \, {\rm approx}}^2$ for $\xi=10^3 ,10^4$. As shown in Ref.~\cite{MultiPreheat3}, the scaling of the spike in the effective mass is
\beq
{\left . m^2_{2,\theta}\right |_{\rm max}\over \langle H(t) \rangle^2} ={\cal O}(10) \xi^2 \, ,
\eeq
 where $\langle H(t) \rangle$ is a time-averaged version of the Hubble scale over the early oscillatory behavior.
The range of excited wavenumbers is given by the relation
\beq
 k^2 \lesssim a^2 \left . m^2_{2,\theta}\right |_{\rm max}  \, ,
\eeq
assuming that the spike of $m^2_{2,\theta} $ dominates over $\tilde m_B^2$ near $\varphi(t)=0$. Each subsequent inflaton zero-crossing affects a smaller range of wavenumbers, since $m^2_{2,\theta} \propto H^2 \propto \rho_{\rm infl.} \propto a^{-3}$, where we assumed $w=0$ for the averaged background evolution. Altogether $k_{\rm max}^2 \propto a^{-1}$, hence the maximum excited wavenumber shrinks for every subsequent inflaton oscillation.
The maximum comoving wavenumber after the first inflaton zero-crossing, where $a(t) \approx 1$, is
\beq
k_{\rm max}^2 ={\cal O}(10) \xi^2 H_{\rm end}^2 = {\cal O}(1) \lambda \, M_{\rm Pl}^2 \, ,
\label{eq:kmax}
\eeq
where we used Eq.~\eqref{eq:Hinfl} and $H_{\rm end} \approx 0.5 H_{\rm infl}$. This is in agreement with Ref.~\cite{Ema:2016dny}.
We focus primarily on the first inflaton zero-crossing, since the produced gauge bosons will decay into fermions between two subsequent background zero-crossings, hence Bose enhancement is lost. This was shown in Refs.~\cite{GarciaBellido:2008ab,Bezrukov:2008ut} 
 and will be discussed in detail in Section~\ref{sec:decays}.

The second dominant component of the gauge field effective frequency-squared is $\tilde m_B^2$, which scales simply as
\beq
{\tilde m_B^2/  a^2 \over H_{\rm end}^2} ={M_{\rm Pl}^2 e^2\over 2f} { \varphi^2 } {1\over H_{\rm end}^2} =  {\cal O}(1) {\xi\over \lambda} = {\cal O}(1) {10^{10}\over \xi} \, ,
\label{eq:mAoverH}
\eeq
where the $\lambda-\xi$ relation given in Eq.~\eqref{eq:lambdaxi} was used at the last step.  We can see that for $\xi=10^3$ the maxima of the two contributions $\tilde m_B^2 $ and $  m^3_{2,\theta}$ are similar, as shown in Fig.~\ref{fig:meffchiplot} .

Computing the energy density transferred from the inflaton condensate into the gauge field modes requires more attention than the corresponding computation of Section \ref{sec:Adiabatic} for the Higgs self-resonance. In the case of Higgs self-resonance, the range of excited wavenumbers is $k_{\rm max}^h \sim H$. A naive computation of the energy density in the local adiabatic (WKB) vacuum for the same modes gives $\rho_{BD} \sim k_{\rm max}^4 \sim H^4$ which is $10$ orders of magnitude smaller than the background energy density\footnote{Any computation that does not involve vacuum subtraction, including lattice simulations such as Ref.~\cite{Adshead:2015pva, Adshead:2016iae}, deals with classical quantities and computes the energy density of the vacuum modes as if they were physical. Such a computation is valid as long as the unphysical energy density of the vacuum modes is vastly subdominant.}. In that case we do not need to subtract this unphysical vacuum contribution from the energy density of the Higgs modes, since the energy density in the parametrically amplified modes is exponentially larger. 

For the case of gauge fields the maximum wavenumber up to which modes can be excited is given in Eq.~\eqref{eq:kmax}. The vacuum energy density in these modes, naively computed, is $\rho_{\rm BD} \sim k_{\rm max}^4 \sim \lambda^2 M_{\rm Pl}^4$. The total energy density in the inflaton field is $\rho_{\rm infl} = 3 H^2 M_{\rm Pl}^2$ leading to $\rho_{\rm BD}  / \rho_{\rm infl} \sim \lambda \, \xi^2\sim 10^{-10}\xi^4$. This is much greater than unity for large values of the nonminimal coupling. We thus need to remove the unphysical vacuum contribution to the energy density using the adiabatic subtraction scheme \cite{BirrellDavies}. In this scheme we compare the wave-function of the gauge fields to the instantaneous adiabatic vacuum, computed in the WKB approximation, isolating the particle number for each wavenumber $k$. The particle number corresponding to a mode $v_k$ is given by:
\beq
	n_k = \frac{\omega_k}{2} \left(\frac{|\dot v_k|^2}{\omega_k^2} + |v_k|^2 \right) -\frac{1}{2} \, .
\eeq

A drawback of this method is that the particle number is only well defined when the adiabaticity condition holds $\dot \omega_k / \omega_k^2\ll1$, thus we cannot define the particle number in the vicinity of the ``Riemann spike", when $\varphi(t)=0$\footnote{Ref.~\cite{Ema:2016dny} computed the particle number, working in the Jordan frame, arriving at similar results. The energy of the gauge fields was subsequently computed using the value of the gauge field mass directly on the ``Riemann spike".
We refrain from using $\left .m_{2,\theta}^2\right|_{\rm max}$ as an indicator of the gauge field mass, since the particle number is not a well defined quantity there. For $\xi\approx 10^3$, the two contributions to the gauge field mass, $m_{2,\theta}^2$ and $\tilde m_B^2$ are comparable, as shown in Fig.~\ref{fig:meffchiplot}, which does not hold for other values of $\xi$.
}.
The energy density is easily computed through the particle number as
\beq
\rho^{L,\theta} = \int {d^3k\over (2\pi)^3 }n_k \omega_k \, .
\eeq

\begin{figure} 
\centering
\includegraphics[width=\textwidth]{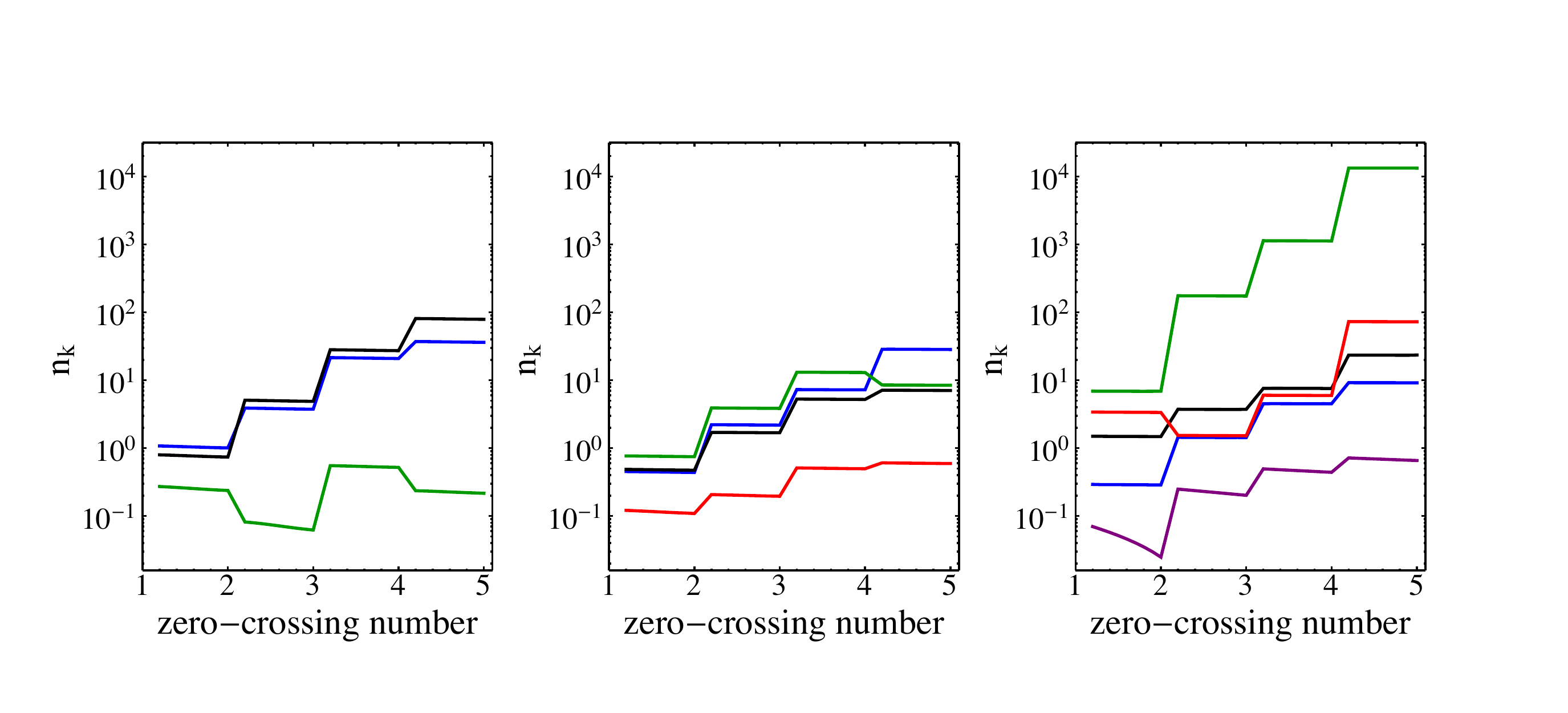}
\caption{
The particle number density for $k/ H_{\rm end} = 1, 150, 550, 2600, 28000$ (blue, black, green, red and purple respectively). From left to right: $\xi = 10^2, 10^3, 10^4$. If a colored curve is missing from a panel, the corresponding wavenumber is not excited.
 }
 \label{fig:nkvstime}
\end{figure}

\begin{figure} 
\centering
\includegraphics[width=\textwidth]{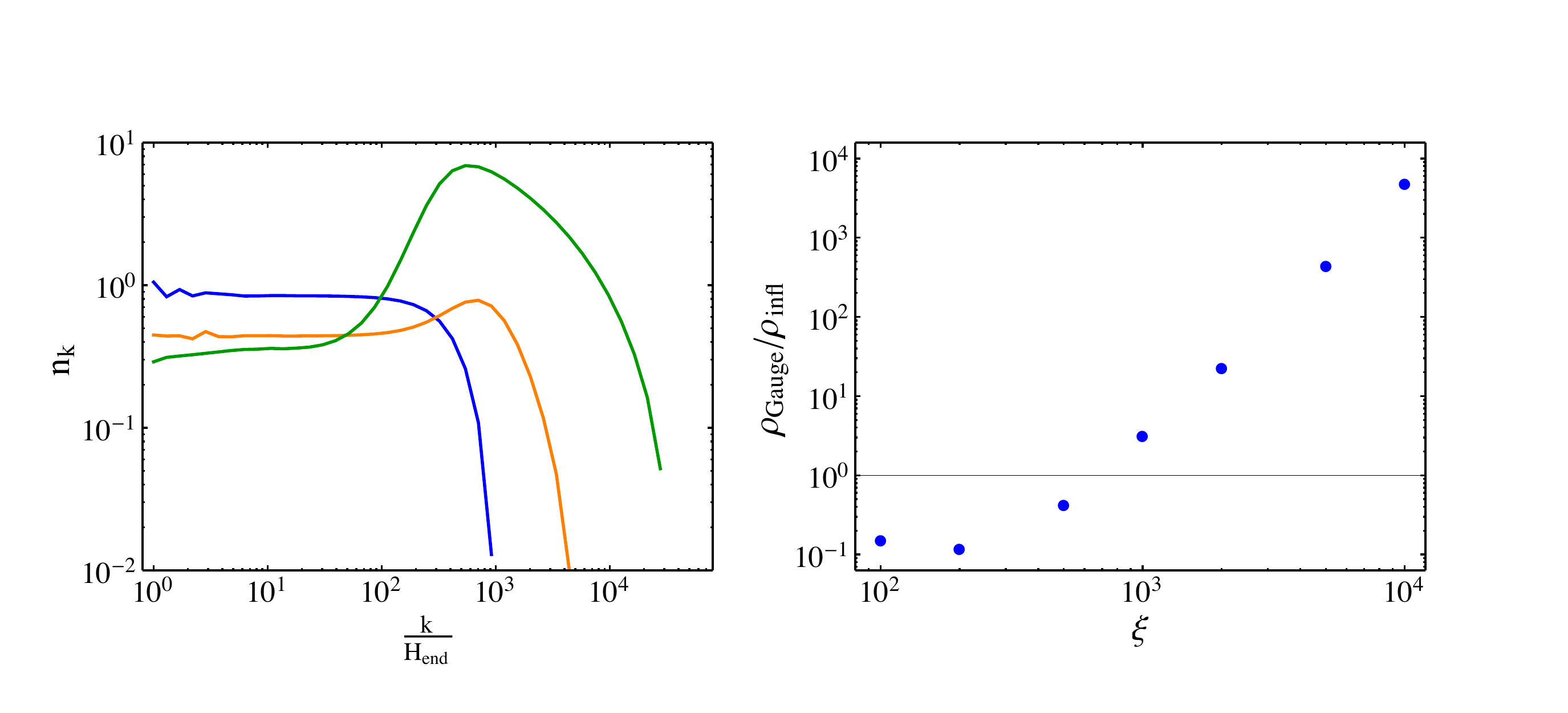}
\caption{
Left: The particle number density after the first inflaton zero-crossing for $\xi=10^2, 10^3, 10^4$ (blue, orange and green respectively)
\\
Right: The ratio of the energy density in gauge fields to the background inflaton energy density as a function of the nonminimal coupling $\xi$ after the first zero-crossing. We see that for $\xi\gtrsim 10^3$ gauge boson production can preheat the universse after one background inflaton zero-crossing, hence it is much more efficient than Higgs self-resonance. }
 \label{fig:nvsxiANDrhovsxi}
\end{figure}

Both the particle number and the energy density can be computed  equally well using the field $\theta_\mathbf k$ or $B^L_\mathbf k$, since the only moment for which the longitudinal gauge fields are not defined is when $\varphi(t)=0$. At this instant we cannot define the particle number either way, since there is no well-defined adiabatic vacuum. Fig.~\ref{fig:nkvstime} shows the evolution of the particle number density for a few values of the comoving wavenumber after the first few inflaton zero-crossings, neglecting the effect of particle decays, as described in Section \ref{sec:decays}. The left panel of Fig.~\ref{fig:nvsxiANDrhovsxi} shows the particle number density per $k$-mode for $\xi=10, 10^2, 10^3$ after the first inflaton zero-crossing. The condition of Eq.~ \eqref{eq:kmax} for the maximum excited wavenumber $k_{\rm max}$ is evident.

At this point, it is worth performing a simple estimate of the energy density that can be transferred to the gauge field modes away from the first point $\varphi(t)=0$. 
\beq
\rho = \int {d^3k\over (2\pi)^3} n_k \omega_k \sim \langle n\rangle \,  \tilde m_B  \, k_{\rm max}^3
\sim
\langle n\rangle \,
\left ( {10^5 \over \sqrt{\xi} } H_{\rm end} \right)
\,
\left (
 \lambda^{3/2} M_{\rm Pl}^3
 \right )
 \sim \langle n\rangle M_{\rm Pl}^4 10^{-15} \xi^{5/2} \, ,
\eeq
where $\langle n\rangle$ is the average occupation number. The background inflaton energy density is $\rho_{\rm infl} = H^2 M_{\rm Pl}^2 \sim 10^{-11}M_{\rm Pl}^2$, hence for $\xi\gtrsim 10^3$ the transfer of energy is enough to completely drain the inflaton condensate within one zero-crossing of $\varphi(t)$, if we take the particle number shown in Fig.~\ref{fig:nvsxiANDrhovsxi} into account. The right panel of  Fig.~\ref{fig:nvsxiANDrhovsxi} shows the ratio of the energy density in gauge fields to the background energy density of the inflaton after the first zero-crossing. Obviously, values of $\rho_{\rm gauge} / \rho_{\rm infl} >1$ are not physical but signal the possibility of complete preheating.

\subsection{Unitarity scale cut-off}

So far we have computed the excitation of gauge field modes of arbitrary wavenumber $k<M_{\rm Pl}$.
However the unitarity scale sets a limit above which no analytical (perturbative) treatment can be trusted. The unitarity scale for Higgs inflation and more generally for nonminimally coupled models, has received extensive attention in the literature. We will follow the analysis of Ref.~\cite{BezrukovInflaton}, where a field-dependent unitarity scale was derived in both the Jordan and Einstein frames. 

The unitarity scale at the end of inflation is $k_{{\rm UV},1} \equiv M_{\rm Pl}/\sqrt{\xi}$, which becomes $k_{{\rm UV},2} \equiv M_{\rm Pl}/{\xi}$ for even smaller values of the background Higgs field. It is straightforward to estimate the relation of the unitarity scale to the maximum excited wavenumber
\beqn
{k_{{\rm UV},1} \over k_{\rm max}}  
=
 {1\over \sqrt{ \xi  \lambda}} \sim{ 5\times 10^4\over \xi^{3/2}} \, ,
 \\
 {k_{{\rm UV},2} \over k_{\rm max}}  
=
 {1\over \xi \sqrt{\lambda }} \sim{ 5\times 10^4\over \xi^{2}}\, .
\eeqn
We see that, depending on the value of the non-minimal coupling $\xi$, the wavenumber of the produced gauge bosons can exceed the field-dependent unitarity scale. 
New physics is needed above the unitarity scale and it is not clear how this new physics will change particle production for such large wavenumbers.
We do not wish to propose any UV completion of the Standard Model in order to address the dynamics above the unitarity scale. We will instead provide a conservative estimate of 
 the energy density in gauge bosons in the presence of unknown UV physics that suppresses particle production with large wavenumbers (above the unitarity scale). Simply put, we will compute the energy density by introducing a UV cut-off at  $k_{{\rm UV},1}$ or $k_{{\rm UV},2}$.

If we consider the UV cut-off at $k_{{\rm UV},1}$, both $\xi=10^3$ and $\xi=10^4$ preheat entirely after one inflaton zero-crossing, since  $k_{{\rm UV},1} \gtrsim k_{\rm max}(\xi=1000)$, as can be seen from Fig.~\ref{fig:nvsxiANDrhovsxi}. If instead we place the UV cut-off at $k_{{\rm UV},2}$, the gauge fields do not carry enough energy to completely preheat the universe after one inflaton zero-crossing, regardless of the value of the nonminimal coupling $\xi$. We thus conclude that preheating into gauge fields is very sensitive to unknown UV physics, since the majority of the energy density is carried by high-$k$ modes, whose number density in a UV-complete model can be much different than the one computed here. 
It is worth noting that the excitation of Higgs fluctuations occurs entirely below the unitarity scale, hence it is not UV sensitive. We will not consider any UV cut-off for the remainder of this work, unless explicitly stated.

\section{Scattering, Decay and backreaction}
\label{sec:decays}

So far we have computed the parametric excitation of particles, either Higgs or gauge bosons, from the oscillating Higgs condensate during preheating. With the exception of the brief discussion in Section \ref{sec:superhorizon}, the interactions of the resulting particles have been completely ignored. However, as discussed in Refs.~\cite{Bezrukov:2008ut, GarciaBellido:2008ab}, certain types of decays of the produced particles can suppress Bose enhancement and thus effectively shut off preheating.
We will discuss in turn 
\begin{enumerate}
\item[A.] the decay of Higgs particles into gauge bosons and fermions,
\item[B.] the scattering of Higgs particles into gauge bosons and fermions,
\item[C.] the decay of parametrically produced gauge bosons, 
\item[D.] the scattering of gauge bosons into fermions and Higgs bosons and
\item[E.] possible effects arising from non-Abelian interactions of the produced $W$ and $Z$ bosons.
\end{enumerate}
Any of the above mentioned processes can suppress or shut off the resonances. Due to their inherent differences, we will explore them separately

\subsection{Higgs decay}
\label{sec:higgsdecay}

In the Standard Model, Higgs particles can decay into pairs of fermions or gauge bosons. The fermion masses are
\beq
m_f^2 = {y_f^2\over 2} {\varphi^2 \over 2f}
\label{eq:mf}
\eeq
while the gauge boson masses were extensively studied in Section~\ref{sec:gaugepreheat}. For now, it is enough to consider the part of the gauge field mass analogous to $m_f$ in Eq.~\eqref{eq:mf} with the Yukawa coupling substituted by the gauge coupling.

We start with the process of a Higgs particle decaying into two gauge bosons. In order for this to be kinematically allowed, the following relation must hold: $m_h >2m_{\rm gauge}$. It is straightforward to see that $m_h \ll 2m_{\rm gauge}$, at least for $\varphi(t)\neq0$. When $\varphi(t)=0$, the Riemann contribution to the gauge field mass (the ``Riemann spike") dominates, keeping the relation $m_h \ll 2m_{\rm gauge}$ valid at all times. Hence the Higgs field cannot decay into gauge bosons, as long as the background Higgs condensate follows the evolution that is derived neglecting back-reaction.

The decays of Higgs bosons to fermions deserve closer attention, due to the fact that small Yukawa couplings for some fermions (like electrons and positrons) can make them much lighter than the Higgs particles, hence kinematically open the decay channel. Furthermore, fermion masses do not have a Riemann component, hence when $\varphi(t)$ crosses zero, fermions become instantaneously massless, making the decay even easier. A similar analysis of kinematical blocking of perturbative decays during reheating was performed in \cite{Freese:2017ace}, when the Higgs field was a light spectator field during inflation, rather than playing the role of the inflaton itself. 

We will compute each component of the Higgs field $m_{h,1}$ and $m_{h,3}$ separately.
We begin with the potential contribution 
\beq
m_{h,1}^2 = {\lambda M_{\rm Pl}^2\over \xi}{\delta^2 (\delta^2 (12\xi-12\xi\delta^2) +3)\over 
(1+\delta^2)^2 (1+6\xi\delta^2)^2} 
\sim 
 {\lambda\over 3\xi^2}M_{\rm Pl}^2 \sim H_{\rm end}^2
 \label{eq:mh1Decays1}
 \eeq
where $\delta = \xi\varphi^2$ and $\delta \simeq 0.8$ at the end of inflation, as discussed in Refs.~\cite{MultiPreheat1, MultiPreheat2, MultiPreheat3}. The value in Eq.~\eqref{eq:mh1Decays1} holds at the start of preheating and until the cross-over time $t_{\rm cross} \sim \sqrt{\xi} H_{\rm end}^{-1}$. The expression for $t_{\rm cross}$  was derived in Ref.~\cite{MultiPreheat3}. For $t<t_{\rm cross}$ metric perturbations dominate the effective mass, resulting in tachyonic amplification. For $t>t_{\rm cross}$ the Higgs particle mass $m_{h,1}^2$ decreases slowly with time. For $\xi=10,10^2,10^3$ the cross-over time occurs at $N_{\rm cross}\simeq  1.5, 2.5, 3.2$ respectively.

\begin{figure}[h]
\centering
\includegraphics[width=\textwidth]{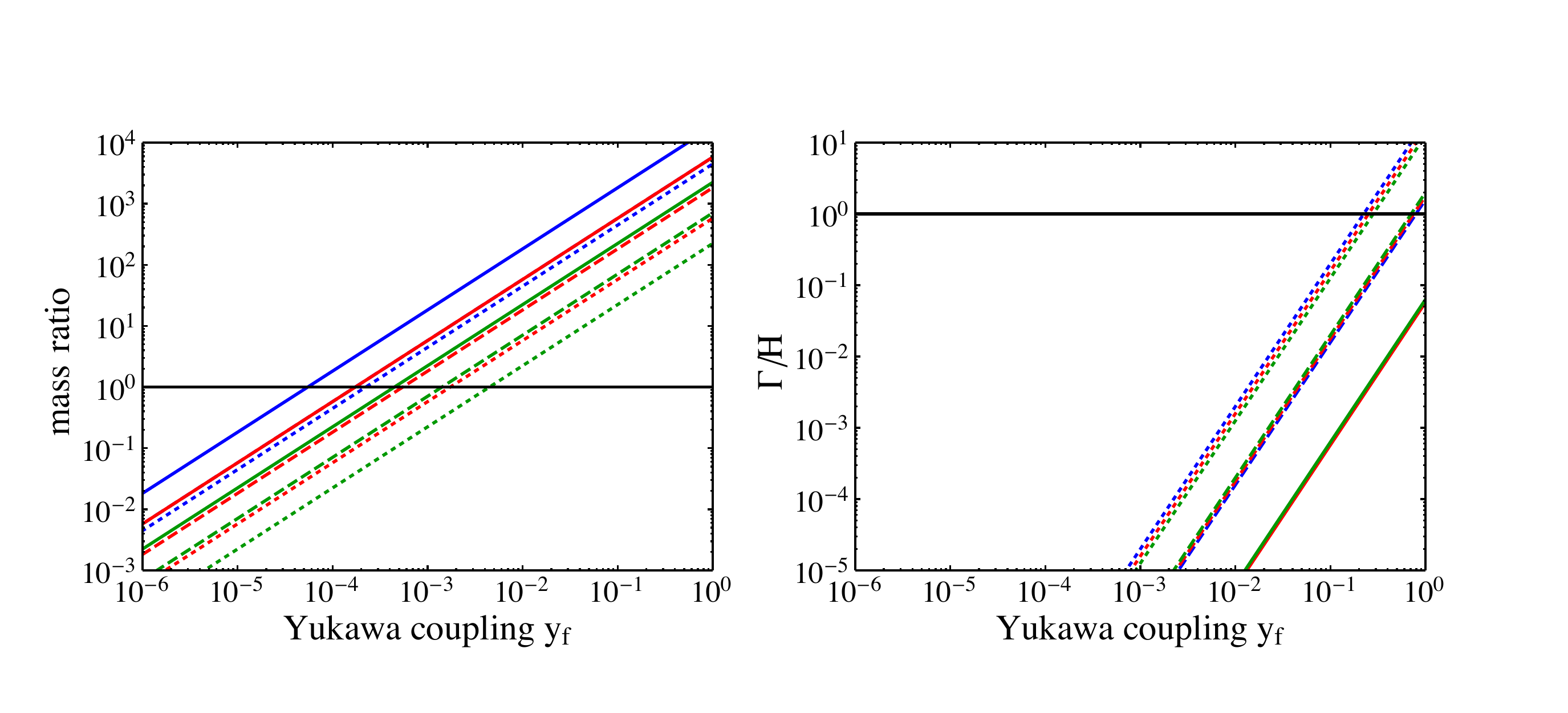}
\caption{
Left: Fermion to Higgs mass ratio as a function of the Yukawa coupling for $\xi=10,100,10^3$ (blue, red and green respectively) at $N=0,1.5,3$ (solid, dashed, dotted respectively).
\\
Right: Ratio of the decay rate to the Hubble rate. The color-coding is the same.
}
 \label{fig:mh1decay}
\end{figure}

Fig.~\ref{fig:mh1decay} shows the ratio of the fermion to the Higgs mass at the end of inflation as a function of the Yukawa coupling, for different values of the nonminimal coupling. We see that the decay is kinematically possible for small Yukawa couplings. Furthermore, the decay channel is less constrained for later times and larger nonminimal coupling.
The perturbative decay rate of Higgs particles to fermions is given by
\beq
\Gamma = {y_f^2 \over 8\pi} m_{h} \, .
\eeq
Fig.~\ref{fig:mh1decay} shows the ratio $\Gamma / H $, which must be greater than unity in order for the decay to be efficient. It is clear that, in the parameter range where the decay is kinematically allowed, it is very inefficient. This can be intuitively understood since $m_h \sim H$, $m_f$ is proportional to $y_f$ and $\Gamma$ is proportional to $y_f^2$, hence $\Gamma/H$ is suppressed by an extra factor of the Yukawa coupling compared to $m_f/ m_h$. This conclusion does not change, even if one considers the short increase in the mass of the Higgs modes due to the coupled metric fluctuations term $m_{3,h}^2$. Even though $m_{3,h}^2$ has a large positive spike, its duration is too small to allow for a significant decay of Higgs particles into fermions.

Before we conclude this section, we will make one further note, regarding the evolution of fermion masses. Eq.~\eqref{eq:mf} shows that fermions become massless when $\varphi(t) =0$. The distinction between computing the fermion mass during reheating by either using an averaged quantity for the Higgs VEV or by using the full time-dependence was explored in \cite{Freese:2017ace}. In order to explore possible effects of the time-dependence of the fermion mass, we focus on the case of $\xi=10^3$ and choose a large Yukawa coupling $y_w=1$, since that provides the largest decay rate to Hubble scale ratio $\Gamma / H \simeq 10$, as shown in Fig.~\ref{fig:mh1decay}. The time per oscillation that $m_f < m_h$ is $\Delta t * H\simeq 10^{-3}$. Hence $\Gamma / \Delta t \ll 1$, meaning that the time when fermions are massless is too small to significantly deplete the Higgs boson population.

\subsection{Higgs scattering}

While we saw that Higgs decays to both gauge bosons and fermions are either kinematically blocked or extremely weak during preheating, the same might not be true for Higgs scatterings, due to the large occupation number, close to the time of complete preheating.
 The kinematical blocking arguments still apply, since the relation $m_h > 2m_{f,A}$ is replaced by $m_h > m_{f,A}$, hence is weakened only by a factor of two. As we saw, the kinematical constraints are significant, hence we will only consider scattering of
 Higgs particles into the lightest fermions (electron-positron pairs), with $y_e  = {\cal O}(10^{-6})$. The relevant rate is
\beq
\Gamma = n \sigma v \, .
\eeq 
The Higgs particles are heavy $m_h>H$ and have small wavenumbers $ k/a \lesssim H$, hence will be nonrelativistic. We will take $v=c\equiv 1$ as an upper limit. The number density of Higgs particles is approximately
\beq
n \approx {\rho_h\over m_h} \le {\rho_{\rm {\rm infl}}\over m_h} \, ,
\eeq
where $\rho_h = \rho_{\rm infl}$ at the point of complete preheating. The cross-section is
\beq
\sigma \approx {y_e^4 \over 8\pi m_h^2}\, .
\eeq
Putting everything together we arrive at
\beq
{\Gamma\over H} \le y_e^4 {1\over 8\pi} { \rho_{\rm tot}\over m_h^3 H}
= y_e^4 {3\over 8\pi} {M_{\rm Pl}^2 H\over m_h^3} \, .
\eeq
It is easy to see that $\Gamma / H\ll1 $ since $y_e^4\simeq 10^{-24}$, $M_{\rm Pl}^2 / H^2 \simeq 10^{10}$ and $H/ m_h<1$.

\begin{figure}[h]
\centering
\includegraphics[width=0.75\textwidth]{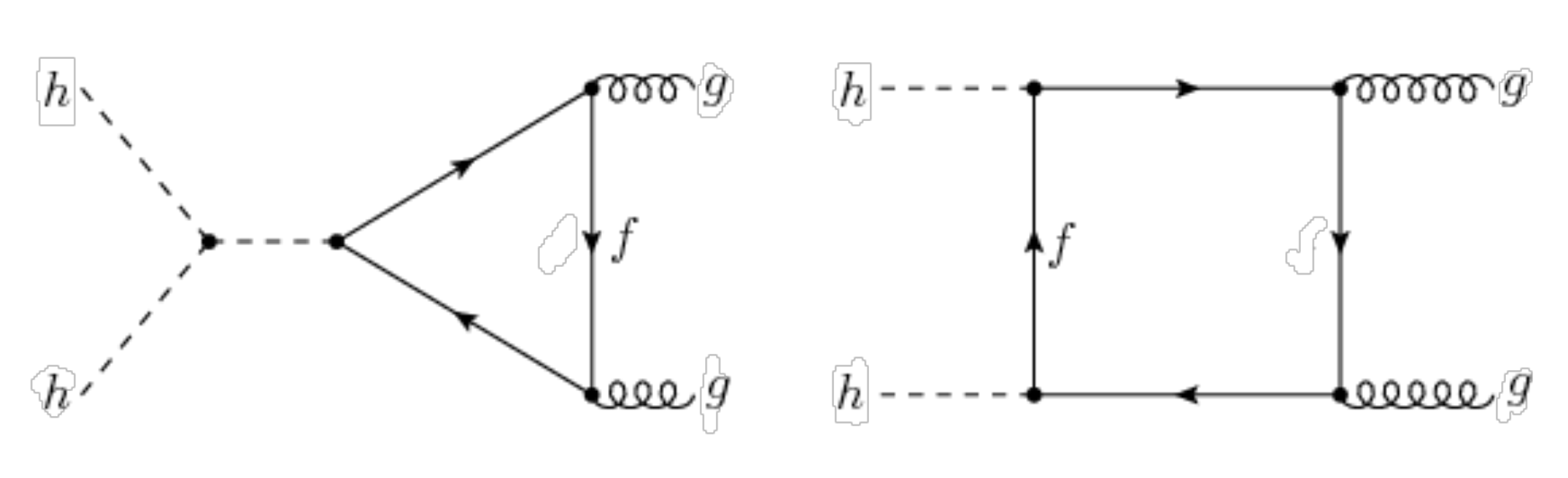}
\caption{
Loop diagrams that contribute to the scattering of Higgs bosons to gluon pairs.
}
 \label{fig:higgstogluons}
\end{figure}

It is also worth briefly noting other scattering diagrams leading to the depletion of the Higgs population. Two examples are shown in Fig.~\ref{fig:higgstogluons}, which are the inverse of gluon fusion processes considered for the LHC.
In general they suffer from the same suppression factors as the tree-level scattering: light fermions come with small Yukawa couplings, while heavy ones will lead to suppression factors from the fermion loops. We will not discuss these processes further.

\subsection{Gauge decay}

Following Refs.~\cite{GarciaBellido:2008ab,Bezrukov:2008ut} the decay width of the W and Z bosons to fermions is given by
\beqn
\Gamma_W &=& {3g^2\over 16\pi} m_W \, ,
\label{eq:gammaW}
\\
\Gamma_Z &=& {g_2^2\over 8\pi^2 \cos^2\theta_W} m_Z \left (  
{7\over 2} - {11\over 3} \sin^2\theta_W + {49\over 9} \sin^4\theta_W
\right) \, ,
\label{eq:gammaZ}
\eeqn
where the decay widths are obtained by summing over all allowed decay channels into SM fermions.
The decay of the Z boson to a pair of Higgs particles proceeds similarly. Using the gauge boson mass given in Eq.~\eqref{eq:mAoverH}, we see that $\Gamma_{W,Z}  / H \gg1$, hence the produced gauge bosons population is depleted within a Hubble time, or between two consecutive inflaton background zero-crossings $\varphi(t)=0$. There are two issues that need to be addressed: the possible decay of particles during their production close to the Riemann spike at $\varphi(t)=0$ and the decay away from $\varphi(t)=0$, when the $\tilde m_B^2 = e^2 \varphi^2 (M_{\rm Pl}^2/2f)$ component dominates the gauge field mass. In both cases, we will approximate the total decay of the particle number as
\beq
n (t) = n_0 \, e^{-\int_{t_0}^t \Gamma(t') dt'}
\label{eq:noft}
\eeq
where $\Gamma(t)$ is defined through Eqs.~\eqref{eq:gammaW} and \eqref{eq:gammaZ} by considering the time-dependent mass of the gauge bosons. We will focus only on the cases of $\xi=10^3$ and $\xi=10^4$. During the spike, the particle number is not a well defined quantity, since an adiabatic vacuum cannot be constructed, due to the violation of the adiabaticity condition. We will however compute the exponential decay factor of Eq.~(\ref{eq:noft}) as an estimate of possible particle decays. We choose the limits of integration to correspond to the times for which adiabaticity is violated, hence particle production occurs. This is also the time at which the Riemann spike is pronounced. For all cases we get $e^{-\int_{t_0}^t \Gamma(t') dt'}>0.5$, hence there is no significant particle decay. We will thus neglect this altogether.

\begin{figure} 
\centering
\includegraphics[width=\textwidth]{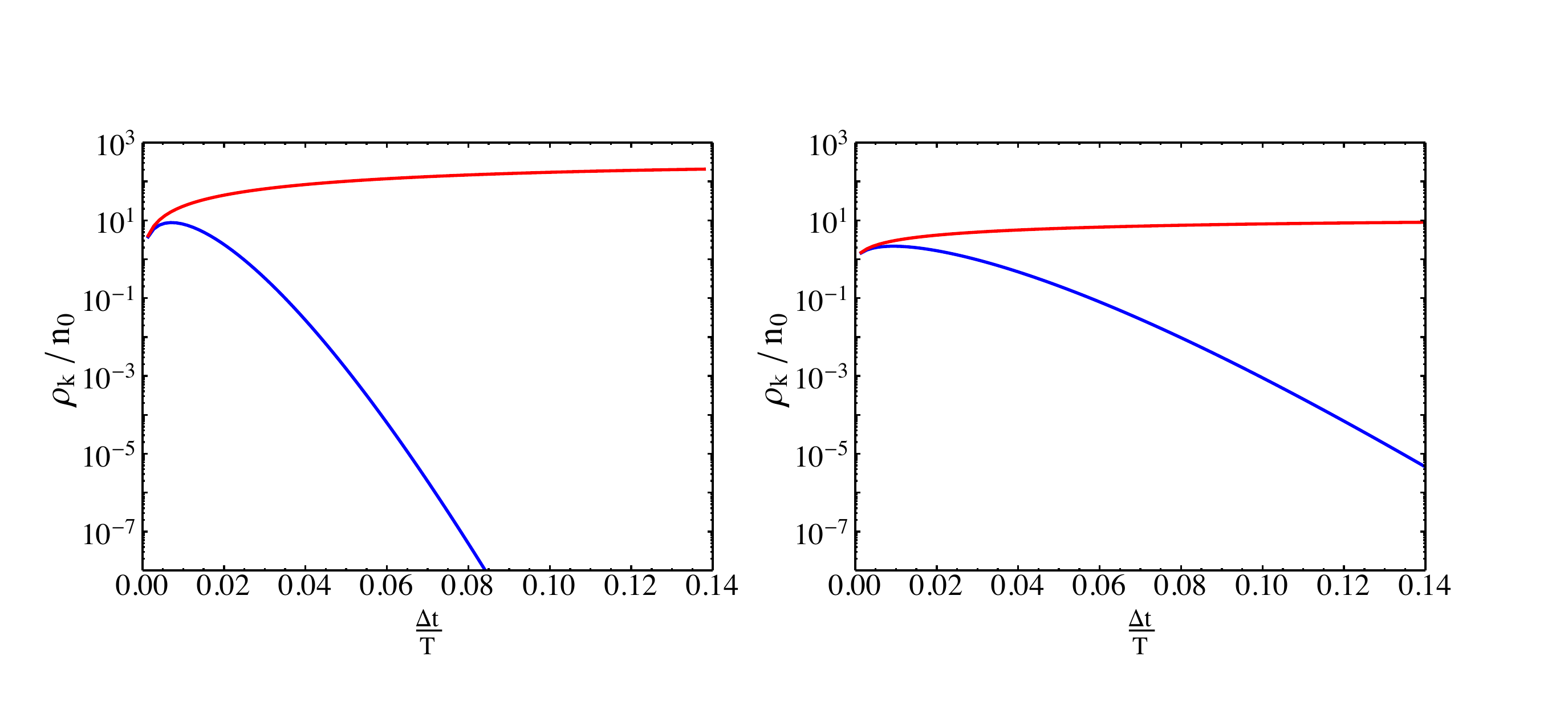}
\caption{
Energy density per mode (in arbitrary units) with (blue) and without (red) considering particle decays for $\xi = 10^3$ (left) and $\xi = 10^4$ (right). The time is rescaled by the period of background oscillations.
}
 \label{fig:rhodecay}
\end{figure}

However, after the particle production has taken place at $\varphi=0$, the particle number is a constant, if one neglects decays, and the particle mass is growing sharply as $m_{W,Z}^2 \sim \varphi^2$. We rewrite the equation for the energy density in the gauge sector as
\beq
\rho^{L,\theta} = \int {d^3k\over (2\pi)^3 }n_0 e^{-\int_{t_0}^t \Gamma(t') dt'} \omega_k \, .
\eeq
Fig.~\ref{fig:rhodecay} shows the energy density per particle number of a random excited $k$-mode as $\rho \simeq n(t) m_A $, with $A$ denoting any gauge field. We see that decays into fermions completely deplete the produced gauge boson population within far less than a period of background oscillations. Hence, in order for the energy transfer to be able to preheat the universe, the energy density in the gauge fields must be equal to the energy density in the inflaton condensate as soon as the adiabaticity condition is restored. The fact that the particle decays during the ``Riemann spike" are insufficient to suppress gauge boson production shows that this is indeed possible.

\subsection{Gauge scattering}

Instead of decaying into fermions, gauge bosons can also scatter into Higgs particles or fermion-antifermion pairs. We will estimate the rate of the Higgs scattering to Higgs bosons. The scattering rate is
$
\Gamma = n \sigma v
$
where we take $v=c$ and
\beqn
&&m \simeq {H M_{\rm Pl}^2 \over m_A} \simeq H M_{\rm Pl}^2  \sqrt{\xi} 10^{-5} \, ,
\\
&&\sigma \simeq {\alpha^2 \over s} \simeq {1\over (\sqrt \lambda M_{\rm Pl})^2} \, ,
\eeqn
where we computed the number density using the condition of complete preheating and we took the Mandelstam variable $s \simeq k_{\rm max}^2$. Altogether
\beq
{\Gamma\over H} \simeq {10^{5} \over \xi^{3/2}}
\eeq
where we used the relation between $\lambda$ and $\xi$ given in Eq.~\eqref{eq:lambdaxi}. Since $\Gamma / H \lesssim 1$ for $\xi\gtrsim 10^3$, gauge field scatterings are not important. This is different from other cases of preheating into gauge bosons, such as \cite{Adshead:2016iae}, where gauge boson scattering is extremely efficient. The difference is that in the present case the number density is not large, but the average energy carried by each gauge boson is, due to the large range of excited wavenumbers.

\subsection{Non-Abelian effects}

Since we are using an Abelian $U(1)$  gauge field as a proxy for preheating into SM $W$ and $Z$ bosons, we must estimate the possible non-Abelian effects. 
As long as the linear analysis holds, the electroweak sector can be decomposed into $3$ almost identical Abelian copies. A numerical example of the relation between an $SU(2)$ gauge field and its $3$ Abelian copies at low field values is shown in Ref.~\cite{Adshead:2017xll}.

However, once the the gauge field modes become sufficiently populated, their true non-Abelian nature cannot be neglected. The relevant term in the non-Abelian Lagrangian is
\begin{equation}
{\cal L}_{\rm non-Abelian} \subset -{1\over 4} f_{abc}f_{ade} A^{b\mu} A_{\mu}^d A^{c\nu}A_{\nu}^e
\end{equation}
where $f_{abc}$ and $f_{ade}$ are $SU(2)$ structure constants.
In the equation of motion for the gauge field strength $A_i$, this term in the Lagrangian will induce a term of the form $g^2 A_j A^j A_i$, which has the form of an effective non-Abelian mass term. Using a Hartree-type approximation we can define the non-Abelian contribution to the gauge field mass-squared as $m^2_{\rm non-Abelian} \sim g^2 \langle A A\rangle $.
 We estimate $\langle A A\rangle$ through the energy density of the gauge fields as $\rho \simeq m_A^2 \langle A ^2\rangle$. Taking as a maximum value $\rho=\rho_{\rm infl} = H^2 M_{\rm Pl}^2$ we estimate
\beq
\langle A^2 \rangle \simeq 10^{-10} \xi M_{\rm Pl}^2 \, .
\eeq
In order for the non-Abelian mass contribution to suppress particle production, it must dominate over $m_{\theta,2}^2$. However, we know that $m_{\theta,2}^2 \simeq \xi ^2 H^2 \simeq \xi^2 10^{-12} M_{\rm Pl}^2$, meaning that for $\xi \gtrsim10^3$ the ``Riemann spike" dominates over the possible non-Abelian mass contribution. Hence, we expect the explosive tranfer of energy from the inflaton to the gauge fields to persist even in the full $SU(2)\times U(1)$ sector.

A further phenomenon that has been observed during simulations of preheating of a non-Abelian Higgsed sector is described in Ref.~\cite{Enqvist:2015sua}. There, the decay of the Higgs condensate through resonant decay of electroweak bosons is simulated. Non-Abelian gauge boson interactions led to an extended momentum distribution. Particles with such high momenta are energetic enough to scatter off the Higgs condensate and fragment it, thereby shutting off any further parametric resonance. In the case of Higgs inflation the gauge fields produced do not survive long before decaying into fermions, due to their large masses. Hence this is unlikely to be an issue in the present case.

\section{Observational consequences}
\label{sec:observables}

Observing reheating is difficult due to the inherently small length scales involved. However, there are two important quantities that can be used to connect reheating to particle physics processes or CMB observables: the reheat temperature $T_{\rm reh}$ and the number of $e$-folds of an early matter dominated epoch in the expansion history of the universe $N_{\rm matter}$.

\begin{figure} 
\centering
\includegraphics[width=0.7\textwidth]{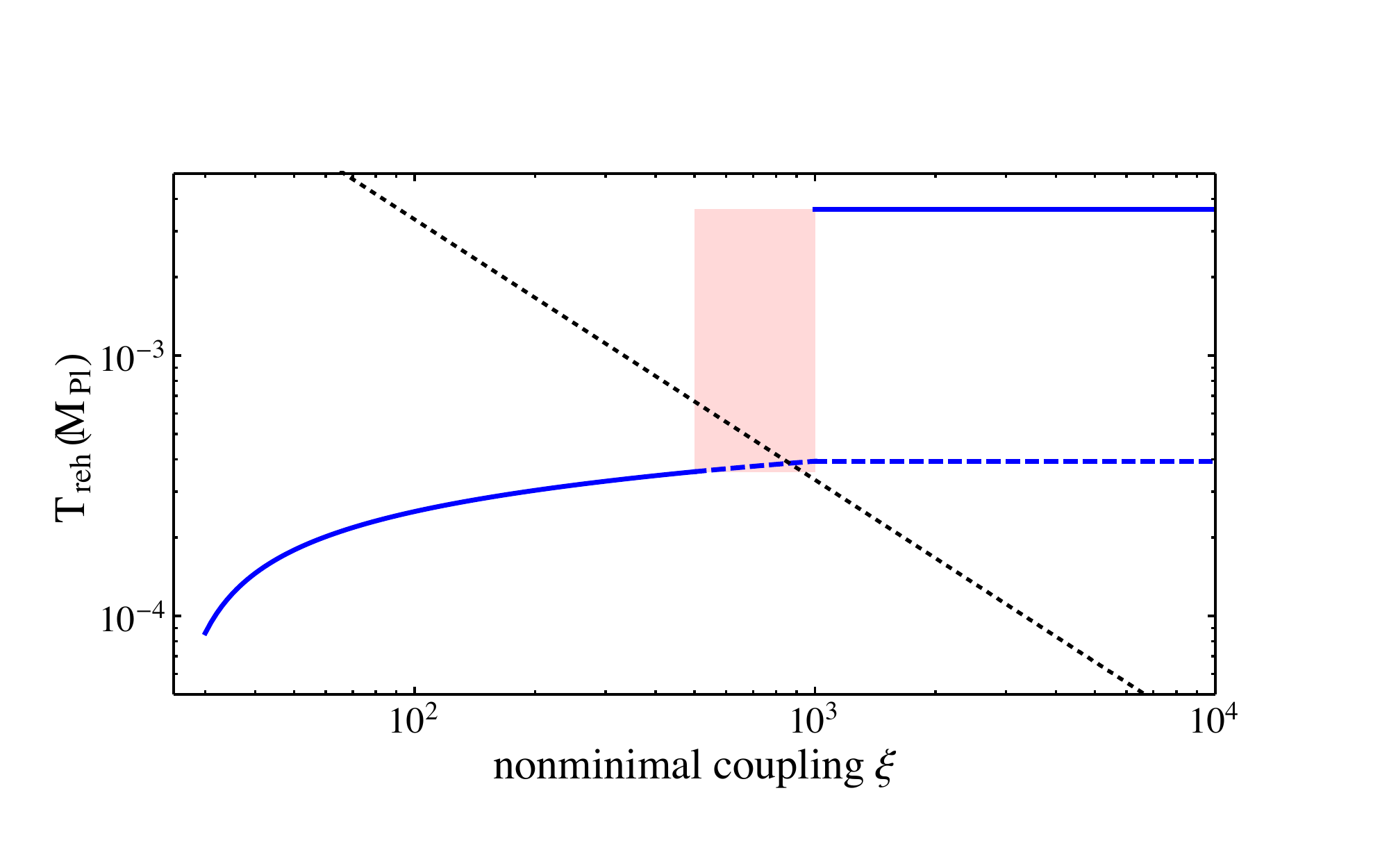}
\caption{
Reheat temperature in units of $M_{\rm Pl}$ as a function of the nonminimal coupling $\xi$. The discontinuity at $\xi\simeq 10^3$ occurs due to the instantaneous preheating to gauge fields. The light red region represents the uncertainty of the exact threshold of instantaneous preheating to gauge fields. The black-dotted line corresponds to the unitarity scale constraint.
The blue-dashed line shows the reheat temperature due entirely to Higgs self-resonance, assuming gauge boson production above the unitarity scale is suppressed due to unknown UV physics.
}
 \label{fig:Treh}
\end{figure}

The reheat temperature is computed using the Hubble scale at the instant when $\rho_{\rm infl} = \rho_{\rm rad}$ as
\beq
3M_{\rm Pl}^2 H^2 = \rho = \sigma_{SB} T_{\rm reh}^4 \, ,
\label{eq:Treh}
\eeq
where $\sigma_{SB} = \pi^2 / 60$ is the Stefan-Boltzman constant. 
For instantaneous reheating from gauge field production, which happens for $\xi \gtrsim1000$, the Hubble scale is $H \simeq H_{\rm end}$. For $\xi \lesssim 1000$ preheating proceeds through Higgs self-resonance, leading to a smaller value of the energy density as shown in Fig.~\ref{fig:selfresonance}. The monotonic increase of the reheat temperature $T_{\rm reh}$ as a function of the nonminimal coupling $\xi$ is shown in Fig.~\ref{fig:Treh}. It must be noted that Eq.~\eqref{eq:Treh} assumes the immediate transition to a thermal state after preheating has ended. For the case of Higgs self-resonance, this will occur through efficient scattering of Higgs bosons to the rest of the SM. For the case of instantaneous preheating to gauge fields, the situation is more complicated. In that case the number density of gauge bosons is not exponentially large, as is usually the case in preheating. On the contrary, the transfer of energy to gauge fields is done primarily through the production of fewer high-momentum modes $k_{\rm max} \sim \sqrt{\lambda}M_{\rm Pl}$. A fraction of the produced $W$ and $Z$ bosons will decay to leptons, while another fraction will decay into quark and antiquarks that will eventually hadronize. The approach to thermal equilibrium will thus be more complicated. We leave the study of the thermalization process for future work and we use Eq.~\eqref{eq:Treh} as an estimate of the reheat temperature, under the assumption of efficient thermalization.

However, a high reheat temperature may pose a challenge for any computation that goes beyond the linearized analysis that we presented, due to possible conflicts with the unitarity scale. Since thermalization of the reheating products will result in a blackbody spectrum, we can take the typical momentum involved to be $k \sim 3T_{\rm reh}$, which is thus the typical momentum exchange in particle scatterings inside the plasma. Since complete reheating means that the inflaton condensate will have completely decayed, the unitarity scale is $k_{{\rm UV},2} \equiv M_{\rm Pl}/{\xi}$. The typical particle momenta are below the unitarity scale for $3T<k_{{\rm UV},2}$. As shown in Fig.~\ref{fig:Treh}, for $\xi\lesssim 10^3$, the resulting plasma has a low enough temperature to avoid processes that exceed the unitarity scale, at least neglecting the tail of the thermal spectrum. For $\xi \gtrsim 10^3$, the unitarity scale  $k_{{\rm UV},2}$ will be exceeded by the typical wavenumbers in the system. Even if one constructs a model that suppresses gauge field excitations with $k>k_{{\rm UV},2}$, Higgs self-resonance will preheat the universe within $3$ $e$-folds, leading to $T_{\rm reh} \sim 5 \times 10^{-4} M_{\rm Pl}$, which is larger than the unitarity scale for $\xi \gtrsim 10^3$.

The number of matter-dominated $e$-folds of post-inflationary expansion is a non-monotonic  function of the nonminimal coupling. For $\xi\gtrsim10^3$, instantaneous reheating leads to a universe filled with gauge field modes of high wave-numbers, hence the universe transitions immediately to radiation domination (assuming no UV suppression). We must note that the decay of the inflaton condensate makes the gauge fields light, hence relativistic.
 For small values of the nonminimal coupling $\xi= {\cal O}(10)$, the background evolves as $w\approx 1/3$, hence the evolution of the universe is that of radiation domination soon after the end of inflaton, even if preheating is not  efficient. Hence $N_{\rm matter}=0$ for both large and ${\cal O}(10)$ values of the nonminimal coupling. There is an intermediate region of $\xi = {\cal O}(100)$, where preheating happens through self-resonance and the background evolves following an average equation of state of $w\approx 0$ \cite{MultiPreheat1} before preheating completes. In that regime of nonminimal couplings $N_{\rm matter} \approx N_{\rm reh} \approx 3$, slightly shifting  the predictions of the CMB compared to the approximation of instantaneous reheating \cite{AHKK}, where the equation of state is assumed to transition from $w=-1/3$ at the end of inflation to $w=1/3$ immediately afterwards.

\section{Conclusions}
\label{sec:Conclusions}

Higgs inflation is an appealing way to realize inflation within the particle content of the  Standard Model, by coupling the Higgs field nonminimally to the gravity sector with a large value of the nonminimal coupling. We analyzed the nonperturbative decay of the Higgs condensate into Higgs bosons and electroweak gauge fields, finding distinct behavior for different ranges of values of the nonminimal coupling $\xi$.

The self-resonance of the Higgs field leads to preheating after $N_{\rm reh}\simeq 4$ $e$-folds  for values of the nonminimal coupling $\xi \gtrsim 30 $. For large values $\xi >100$ the inflaton can transfer all of its energy into nonrelativistic Higgs modes within $N_{\rm reh}\approx 3$, independent of the exact value of the nonminimal coupling. The dominant contribution to the parametric excitation of Higgs modes is the effect of coupled metric fluctuations. In order to accurately capture the amplitude of the Higgs wavefunction, the computation must be initiated before the end of inflation.  

The excitation of gauge bosons is much more dramatic, reminiscent of the purely scalar case of preheating in multi-field inflation with nonminimal couplings \cite{MultiPreheat1,  MultiPreheat2, MultiPreheat3}. Gauge fields are excited after the first zero-crossing of the inflaton field, up to wavenumbers $k_{\rm max} \sim \sqrt{\lambda} M_{\rm Pl}$. This leads to the possibility of the inflaton condensate transferring the entirety of its energy density to $W$ and $Z$ bosons immediately after the end of inflation, leading to instantaneous preheating. $W$ and $Z$ bosons will efficiently decay into SM fermions, ultimately filling the universe with a thermal plasma. Estimates of perturbative decay and non-Abelian effects show that gauge field production is robust against both.

The efficiency of the reheating stage can have observational consequences. The values of the spectral observables $n_s$ and $r$ are related to the time $N_*$ when the CMB-relevant modes exited the horizon during inflation. For Higgs inflation and related models
the CMB observables are given by $n_s \simeq 1 - 2/N_* - 3/N_*^2$ and $r \simeq 12/N_*^2$. Depending on the speed of the transition from the end of inflation to radiation-dominated expansion of the universe, the observationally relevant $N_*$ may vary, shifting the predictions for $n_s$ and $r$.

The use of Coulomb, rather than unitary gauge for our computations  allows us to tie the results to the purely scalar case studied in Refs.~\cite{MultiPreheat1, MultiPreheat2, MultiPreheat3}, as well as apply the results to other models with curved field-space manifolds. One such example is another version of Higgs and Higgs-like inflation, proposed in Ref.~\cite{Lindealpha1}. In that model, the necessary nonminimal coupling is small and negative, accompanied by a minimum of the Higgs potential at a large vacuum expectation value during inflation. The analysis of this model is left for future work and can provide a possible method for probing the Higgs potential during inflation, through its effect on the preheating behavior and the reheat temperature, rather than the CMB observables alone.

Another modification of Higgs inflation is based on the assumption of the existence of an inflection point in the Higgs inflation potential \cite{Bezrukov:2014bra}. This can have interesting consequences, such as primordial black hole production \cite{Ezquiaga:2017fvi}, even though the robustness of a critical point in the Higgs potential is debated \cite{Masina:2018ejw}.
Recent studies of Higgs inflation involving nonminimal couplings in the Palatini formulation of gravity \cite{Rasanen:2017ivk, Enckell:2018kkc} can also have different preheating phenomenology. Exploring the preheating phenomenology of these models is interesting and can be performed using the techniques applied here. Such analyses can provide unique handles in order to probe the Higgs potential at energy scales that are out of reach for the LHC and any future accelerator.

\section*{Appendix A: Gauge field modes during inflation for $|k\tau| < x_c$}

Significant analytical progress can be made in computing the spectrum of the various modes close to the end of inflation, at which point we start the preheating computation. 
As shown also in Ref.~\cite{Lozanov:2016pac}, for the case of $x_c\gg 1$, which is where Higgs inflation falls, the spectrum at the end of inflation is indistinguishable from the de-Sitter results (at least in the case of quadratic inflation, which was the example used in Ref.~\cite{Lozanov:2016pac}).

 Fortunately, as pointed out in \cite{Lozanov:2016pac} but not further pursued there, the equation of motion for $x<x_c$ can be analytically solved using hypergeometric functions
\beqn
u^L(k,\tau)&=&
\nonumber
c_1 (-1)^{\frac{1}{4} \left(1-\nu\right)} x_c^{\frac{1}{2} \left(\nu-1\right)} (k \tau )^{\frac{1}{2} \left(1-\nu\right)} \, _1F_1\left(\frac{1}{4}-\frac{1}{4} \nu;1-\frac{1}{2}\nu;\frac{k^2 \tau ^2}{x_c^2}\right)
\\
&+&c_2 (-1)^{\frac{1}{4} \left(\nu+1\right)} x_c^{\frac{1}{2} \left(-\nu-1\right)} (k \tau )^{\frac{1}{2} \left(\nu+1\right)} \, _1F_1\left(\frac{1}{4} \nu+\frac{1}{4};\frac{1}{2} \nu+1;\frac{k^2 \tau ^2}{x_c^2}\right) \, ,
\eeqn
where $\nu = \sqrt{1-4 x_c^2}$ and $c_1, c_2$ are integration constants.
This is a rather cumbersome formula that doesn't provide a lot of insight. By Taylor expanding it for values of $k|\tau| \ll x_c$ we get a rather simple expression
\beqn
u^L(k,\tau)&\simeq &c_1 \times(-1)^{\frac{1}{4} \left(1-\nu\right)} \left(\frac{k\tau}{x_c}\right)^{\frac{1}{2} \left(1-\nu\right)}+
c_2 \times(-1)^{\frac{1}{4} \left(\nu+1\right)} \left(\frac{k\tau}{x_c}\right)^{\frac{1}{2} \left(1+\nu\right)}\eeqn
We require $c_2=0$, in order for the phases to match the Bunch Davies form $e^{-i k \tau}$ in the past. We must also set $c_1$ such that the norm matches to Eq.~\eqref{eq:ukLapprox} for $k|\tau|=x_c$.
\beq
\left | u_L(k,\tau)\right | \simeq \left |c_1\right | e^{\frac{1}{4} \pi  \Im\left(\sqrt{1-4 x_c^2}\right)}e^{\frac{1}{2} \Re\left(\left(1-\sqrt{1-4 x_c^2}\right) \log \left(\frac{x}{x_c}\right)\right)}
\eeq
If we work in the regime $x_c\gg 1$, which is true for Higgs inflation, we approximate $\sqrt{1-4 x_c^2} \simeq i \, 2x_c$, and the above expression simplifies to
\beq
\left | u_L(k,\tau)\right | \simeq \left |c_1\right | e^{\frac{1}{2} \pi x_c} \left ({x\over x_c}\right )^{1/2} \, ,
\eeq
hence equating this to Eq.~\eqref{eq:ukLapprox} for $k|\tau|=x_c$ reveals the value of the integration constant $c_1$
\beq
|c_1| = e^{-{1\over 2}\pi x_c}{1\over \sqrt{2k}} \, .
\eeq
Altogether, the evolution of the mode-function is
\beq
 u_L(k,\tau) = {1\over  \sqrt{2k}} \left ({k|\tau|\over x_c}\right)^{1/2} (k|\tau|)^{-i \, x_c} ~,~ k|\tau| <x_c  ,
 \eeq
where we dropped an arbitrary pure phase term.
The derivative is
\beq
{\partial_\tau u_L(k,\tau) \over u_L(k,\tau)}  ={-1\over \tau}\left ( {1\over 2} - i \, x_c\right)  \, .
\eeq
As a side-note, the fact that the term proportional to $i$ is negative, shows that we rightly chose the right-moving wave\footnote{In reality, solving the full equation of motion in cosmic time with all factors included, the result is not a perfect right-moving wave. However, this is still a very good approximation to use as an initial condition both for the current linear computation of fluctuations as well as for future lattice simulations.}.
Again, dropping an arbitrary phase, the initial conditions for preheating computations are
\beqn
 u_L(k,\tau_{\rm in}) &=& {1\over \sqrt{2k}} \left ({k\, \tau_{\rm in}\over x_c}\right)^{1/2} \, ,
 \label{eq:ukLapproxlate}
\\
\dot u_L(k,\tau_{\rm in})   &=&   B_L(k,\tau_{\rm in}) 
  {1\over a(\tau_{\rm in} )\tau_{\rm in}}\left ({1\over 2} - i \, x_c \right ) \simeq 
 B_L(k,\tau_{\rm in}) 
  { H(\tau_{\rm in})  }\left ({1\over 2} - i \, x_c \right ) \, ,
\eeqn
for wave-numbers such that $|k\tau_{\rm in}| < x_c$. Since $x_c\gg 1$, we can drop the $1/2$ factor in the above equation.

\acknowledgements{We are indebted to David Kaiser and Marieke Postma for invaluable discussions and comments on the manuscript. 
The authors gratefully acknowledge support from the Dutch Organisation for Scientific Research (NWO).
}


\begin{thebibliography}{999}



  \bibitem{Higgsdiscovery} ATLAS Collaboration, ``Observation of a new particle in the search for the Standard Model Higgs boson with the ATLAS detector at the LHC," Phys Lett. B716 (2012): 1-29 [arXiv:1207.7214 [hep-ex]]; CMS Collaboration, ``Observation of a new boson at a mass of 125 GeV with the CMS experiment at the LHC," Phys. Lett. B716 (2012): 30 [arXiv:1207.7235 [hep-ex]].


\bibitem{LythRiotto} D. H. Lyth and A. Riotto, ``Particle physics models of inflation and the cosmological density perturbation," Phys. Rept. {\bf 314}, 1 (1999), arXiv:hep-ph/9807278.


\bibitem{GuthKaiser} A. H. Guth and D. I. Kaiser, ``Inflationary cosmology: Exporing the universe from the smallest to the largest scales," Science {\bf 307},884 (2005)  [arXiv:astro-ph/0502328].

\bibitem{Mazumdar} A. Mazumdar and J. Rocher, ``Particle physics models of inflation and curvaton scenarios," Phys. Rept. {\bf 497}, 85 (2011) [arXiv:1001.0993 [hep-ph]].

\bibitem{Guth:1982ec} 
  A.~H.~Guth and S.~Y.~Pi,
``Fluctuations in the New Inflationary Universe,''
  Phys.\ Rev.\ Lett.\  {\bf 49}, 1110 (1982).
  
      \bibitem{Higgsinfl} F. L. Bezrukov and M. E. Shaposhnikov, ``The Standard Model Higgs boson as the inflaton," Phys. Lett. {\bf B659}, 703 (2008) [arXiv:0710.3755 [hep-th]].

  
\bibitem{Callan} C. G. Callan, Jr., S. R. Coleman, and R. Jackiw, ``A new improved energy-momentum tensor," Ann. Phys. (N.Y.) {\bf 59}, 42 (1970).

\bibitem{Bunch} T. S. Bunch, P. Panangaden, and L. Parker, ``On renormalization of $\lambda \phi^4$ field theory in curved space-time, I," J. Phys. A {\bf 13}, 901 (1980); T. S. Bunch and P. Panangaden, ``On renormalization of $\lambda \phi^4$ field theory in curved space-time, II," J. Phys. A {\bf 13}, 919 (1980).

\bibitem{BirrellDavies} N. D. Birrell and P. C. W. Davies, {\it Quantum Fields in Curved Space} (New York: Cambridge University Press, 1982).

\bibitem{Buchbinder} I. L. Buchbinder, S. D. Odintsov, and I. L. Shapiro, {\it Effective Action in Quantum Gravity} (New York: Taylor and Francis, 1992).

\bibitem{ParkerToms} L. E. Parker and D. J. Toms, {\it Quantum Field Theory in Curved Spacetime} (New York: Cambridge University Press, 2009).

\bibitem{Odintsov1991} S. D. Odintsov, ``Renormalization group, effective action, and Grand Unification Theories in curved spacetime," Fortsch. Phys. {\bf 39}, 621 (1991).

\bibitem{Bounakis:2017fkv} 
  M.~Bounakis and I.~G.~Moss,
``Gravitational corrections to Higgs potentials,''
  JHEP {\bf 1804}, 071 (2018)
  [arXiv:1710.02987 [hep-th]].

\bibitem{Markkanen2013} T. Markkanen and A. Tranberg, ``A simple method for one-loop renormalization in curved spacetime," JCAP 08 (2013): 045, arXiv:1303.0180 [hep-th].


  \bibitem{Fujii} Y. Fujii and K. Maeda, {\it The Scalar-Tensor Theory of Gravitation} (New York: Cambridge University Press, 2003).

\bibitem{Faraoni2004} V. Faraoni, {\it Cosmology in Scalar-Tensor Gravity} (Boston: Kluwer, 2004).




\bibitem{SimoneHertzbergWilczek} A. de Simone, M. P. Hertzberg, and F. Wilczek, ``Running inflation in the Standard Model," Phys. Lett. B678 (2009): 1 [arXiv:0812.4946 [hep-ph]].
  
  
  \bibitem{BezrukovMass} F. L. Bezrukov, A. Magnin, and M. E. Shaposhnikov, ``Standard Model Higgs boson mass from inflation," Phys. Lett. B675 (2009): 88 [arXiv:0812.4950 [hep-ph]]; F. L. Bezrukov and M. E. Shaposhnikov, ``Standard Model Higgs boson mass from inflation: two loop analysis," JHEP 0907 (2009): 089 [arXiv:0904.1537 [hep-ph]].


\bibitem{Barvinsky} A. O. Barvinsky, A. Yu. Kamenshchik, C. Kiefer, A. A. Starobinsky, and C. F. Steinwachs, ``Asymptotic freedom in inflationary cosmology with a nonminimally coupled Higgs field," JCAP 0912 (2009): 003, arXiv:0904.1698 [hep-ph]; A. O. Barvinsky, A. Yu. Kamenshchik, C. Kiefer, A. A. Starobinsky, and C. F. Steinwachs, ``Higgs boson, renormalization group, and naturalness in cosmology," arXiv:0910.1041 [hep-ph].


\bibitem{Allison:2013uaa} K. Allison, ``Higgs $\xi$-inflation for the 125-126 FeV Higgs: A two-loop analysis," JHEP 02 (2014): 040, arXiv:1306.6931 [hep-ph].


\bibitem{KS} D. I. Kaiser and E. I. Sfakianakis, ``Multifield inflation after Planck: The case for nonminimal couplings," Phys. Rev. Lett. {\bf 112}, 011302 (2014), arXiv:1304.0363 [astro-ph.CO].

\bibitem{LindeRoest} R. Kallosh, A. Linde, and D. Roest, ``Universal attractor for inflation at strong coupling," Phys. Rev. Lett. {\bf 112}, 011303 (2014), arXiv:1310.3950 [hep-th].


\bibitem{Starobinsky:1980te} 
  A.~A.~Starobinsky,
 ``A New Type of Isotropic Cosmological Models Without Singularity,''
  Phys.\ Lett.\ B {\bf 91}, 99 (1980)
  [Phys.\ Lett.\  {\bf 91B}, 99 (1980)].


\bibitem{Lindealpha}  R. Kallosh and A. Linde, ``Multi-field conformal cosmological attractors," JCAP 1312 (2013): 006, arXiv:1309.2015 [hep-th];  R. Kallosh, A. Linde, and D. Roest, ``Superconformal inflationary $\alpha$-attractors," JHEP 1311 (2013): 198, arXiv:1311.0472 [hep-th]; M. Galante, R. Kallosh, A. Linde, and D. Roest, ``The unity of cosmological attractors," Phys. Rev. Lett. {\bf 114}, 141302 (2015), arXiv:1412.3797 [hep-th]; R. Kallosh and A. Linde, ``Planck, LHC, and $\alpha$-attractors," Phys. Rev. D {\bf 91}, 083528 (2015), arXiv:1502.07733 [astro-ph.CO]; J. J. M. Carrasco, R. Kallosh, and A. Linde, ``Cosmological attractors and initial conditions for inflation," Phys. Rev. D {\bf 92}, 063519 (2015), arXiv:1506.00936 [hep-th].

 \bibitem{Christodoulidis:2018qdw} 
  P.~Christodoulidis, D.~Roest and E.~I.~Sfakianakis,
  ``Angular inflation in multi-field ${\alpha}$-attractors,''
  arXiv:1803.09841 [hep-th].
  
  
\bibitem{Higgs2018} 
  Y.~Akrami {\it et al.} [Planck Collaboration],
  ``Planck 2018 results. X. Constraints on inflation,''
  arXiv:1807.06211 [astro-ph.CO].
  

  \bibitem{BTW} B. A. Bassett, S. Tsujikawa, and D. Wands, ``Inflation dynamics and reheating," Rev. Mod. Phys. {\bf 78}, 537 (2006), arXiv:astro-ph/0507632.

  
\bibitem{LythLiddle} D. H. Lyth and A. R. Liddle, {\it The Primordial Density Perturbation: Cosmology, Inflation, and the Origin of Structure} (New York: Cambridge University Press, 2009).

   \bibitem{Baumann} D. Baumann, ``TASI Lectures on inflation," arXiv:0907.5424 [hep-th].

  
\bibitem{MartinRingeval} J. Martin, C. Ringeval, and V. Vennin, ``Encyclopedia inflationaris," Phys. Dark Univ. {\bf 5-6}, 75 (2014), arXiv:1303.3787 [astro-ph.CO].

\bibitem{GKN} A. H. Guth, D. I. Kaiser, and Y. Nomura, ``Infltationary paradigm after Planck 2013," Phys. Lett. B {\bf 733}, 112 (2014), arXiv:1312.7619 [astro-ph.CO].

\bibitem{LindePlanck} A. D. Linde, ``Inflationary cosmology after Planck 2013," arXiv:1402.0526 [hep-th].

  
\bibitem{MartinRev} J. Martin, ``The observational status of cosmic inflation after Planck," arXiv:1502.05733 [astro-ph.CO].


\bibitem{Steigman} G. Steigman, ``Primordial nucleosynthesis in the precision cosmology era," Ann. Rev. Nucl. Part. Sci. {\bf 57}, 463 (2007), arXiv:0712.1100 [astro-ph].

\bibitem{FieldsBBN} B. D. Fields, P. Molaro, and S. Sarkar, ``Big-bang nucleosynthesis," Chin. Phys. C {\bf 38}, 339 (2014), arXiv:1412.1408 [astro-ph.CO].

\bibitem{Cyburt} R. H. Cyburt, B. D. Fields, K. A. Olive, and T.-H. Yeh, ``Big bang nucleosynthesis: 2015," Rev. Mod. Phys. {\bf 88}, 015004 (2016), arXiv:1505.01076 [astro-ph.CO].



\bibitem{AdsheadEasther} P. Adshead, R. Easther, J. Pritchard, and A. Loeb, ``Inflation and the scale dependent spectral index: Prospects and strategies," JCAP 1102 (2011): 021, arXiv:1007.3748 [astro-ph.CO].

\bibitem{Dai} L. Dai, M. Kamionkowski, and J. Wang, ``Reheating constraints to inflationary models," Phys. Rev. Lett. {\bf 113}, 041302 (2014), arXiv:1404.6704 [astro-ph.CO].

\bibitem{Creminelli} P. Creminelli, D. L. Nacir, M. Simonovi, G. Trevisan, and M. Zaldarriaga, ``$\varphi^2$ inflation at its endpoint," Phys. Rev. D {\bf 90}, 083513 (2014), arXiv:1405.6264 [astro-ph.CO].

\bibitem{MartinReheat} J. Martin, C. Ringeval, and V. Vennin, ``Observing the inflationary reheating," Phys. Rev. Lett. {\bf 114}, 081303 (2015), arXiv:1410.7958 [astro-ph.CO].

\bibitem{GongLeungPi} J.-O. Gong, G. Leung, and S. Pi, ``Probing reheating with primordial spectrum," JCAP 05 (2015): 027, arXiv:1501.03604 [hep-ph].

\bibitem{CaiGuoWang} R.-G. Cai, Z.-K. Guo, and S.-J. Wang, ``Reheating phase diagram for single-field slow-roll inflationary models," Phys. Rev. D {\bf 92}, 063506 (2015), arXiv:1501.07743 [gr-qc].


\bibitem{Cook} J. L. Cook, E. Dimastrogiovanni, D. A. Easson, and L. M. Krauss, ``Reheating predictions in single field inflation," JCAP 04 (2015): 004, arXiv:1502.04673 [astro-ph.CO].

\bibitem{Heisig} V. Domcke and J. Heisig, ``Constraints on the reheating temperature from sizable tensor modes," Phys. Rev. D {\bf 92}, 103515 (2015), arXiv:1504.00345 [astro-ph.CO].

\bibitem{AHKK} M. A. Amin, M. P. Hertzberg, D. I. Kaiser, and J. Karouby, ``Nonperturbative dynamics of reheating after inflation: A review," Int. J. Mod. Phys. D {\bf 24}, 1530003 (2015), arXiv:1410.3808 [hep-ph].


\bibitem{MPHBaryogenesis} M. P. Hertzberg and J. Karouby, ``Baryogenesis from the inflaton field," Phys. Lett. B {\bf 737}, 34 (2014), arXiv:1309.0007 [hep-ph]; M. P. Hertzberg and J. Karouby, ``Generating the observed baryon asymmetry from the inflaton field," Phys. Rev. D {\bf 89}, 063523 (2014), arXiv:1309.0010 [hep-ph].

\bibitem{Adshead:2015jza} 
P.~Adshead and E.~I.~Sfakianakis, ``Leptogenesis from left-handed neutrino production during axion inflation,'' Phys. Rev. Lett. {\bf 116}, 091301 (2016), arXiv:1508.00881 [hep-ph].

\bibitem{Adshead:2015kza}  P.~Adshead and E.~I.~Sfakianakis, ``Fermion production during and after axion inflation,'' JCAP 1511 (2015): 021, arXiv:1508.00891 [hep-ph].
  
\bibitem{Adshead:2017znw} 
  P.~Adshead, A.~J.~Long and E.~I.~Sfakianakis,
``Gravitational Leptogenesis, Reheating, and Models of Neutrino Mass,''
  Phys.\ Rev.\ D {\bf 97}, no. 4, 043511 (2018)
  [arXiv:1711.04800 [hep-ph]].

\bibitem{Mustafa} K. D. Lozanov and M. A. Amin, ``End of inflation, oscillons, and matter-antimatter asymmetry," Phys. Rev. D {\bf 90}, 083528 (2014), arXiv:1408.1811 [hep-ph].



\bibitem{Adshead:2015pva} 
  P.~Adshead, J.~T.~Giblin, T.~R.~Scully and E.~I.~Sfakianakis,
 ``Gauge-preheating and the end of axion inflation,''
  JCAP {\bf 1512}, no. 12, 034 (2015)
  [arXiv:1502.06506 [astro-ph.CO]].
  
\bibitem{Adshead:2016iae} 
  P.~Adshead, J.~T.~Giblin, T.~R.~Scully and E.~I.~Sfakianakis,
``Magnetogenesis from axion inflation,''
  JCAP {\bf 1610}, 039 (2016)
  [arXiv:1606.08474 [astro-ph.CO]].





\bibitem{Georg:2017mqk} 
  J.~Georg and S.~Watson,
``A Preferred Mass Range for Primordial Black Hole Formation and Black Holes as Dark Matter Revisited,''
  JHEP {\bf 1709}, 138 (2017)
  [arXiv:1703.04825 [astro-ph.CO]].


\bibitem{Carr:2018nkm} 
  B.~Carr, K.~Dimopoulos, C.~Owen and T.~Tenkanen,
``Primordial Black Hole Formation During Slow Reheating After Inflation,''
  Phys.\ Rev.\ D {\bf 97}, no. 12, 123535 (2018)
  [arXiv:1804.08639 [astro-ph.CO]].

\bibitem{Cai:2018rqf} 
  R.~G.~Cai, T.~B.~Liu and S.~J.~Wang,
``Reheating sensitivity on primordial black holes,''
  arXiv:1806.05390 [astro-ph.CO].


\bibitem{GarciaBellido:2008ab} 
  J.~Garcia-Bellido, D.~G.~Figueroa and J.~Rubio,
``Preheating in the Standard Model with the Higgs-Inflaton coupled to gravity,''
  Phys.\ Rev.\ D {\bf 79}, 063531 (2009)
  doi:10.1103/PhysRevD.79.063531
  [arXiv:0812.4624 [hep-ph]].
  
  \bibitem{Bezrukov:2008ut} 
  F.~Bezrukov, D.~Gorbunov and M.~Shaposhnikov,
 ``On initial conditions for the Hot Big Bang,''
  JCAP {\bf 0906}, 029 (2009)
  doi:10.1088/1475-7516/2009/06/029
  [arXiv:0812.3622 [hep-ph]].



\bibitem{MultiPreheat1} 
  M.~P.~DeCross, D.~I.~Kaiser, A.~Prabhu, C.~Prescod-Weinstein and E.~I.~Sfakianakis,
``Preheating after Multifield Inflation with Nonminimal Couplings, I: Covariant Formalism and Attractor Behavior,''
  Phys.\ Rev.\ D {\bf 97}, no. 2, 023526 (2018)
  [arXiv:1510.08553 [astro-ph.CO]].

\bibitem{MultiPreheat2}  
  M.~P.~DeCross, D.~I.~Kaiser, A.~Prabhu, C.~Prescod-Weinstein and E.~I.~Sfakianakis,
``Preheating after multifield inflation with nonminimal couplings, II: Resonance Structure,''
  Phys.\ Rev.\ D {\bf 97}, no. 2, 023527 (2018)
  [arXiv:1610.08868 [astro-ph.CO]].

\bibitem{MultiPreheat3} 
  M.~P.~DeCross, D.~I.~Kaiser, A.~Prabhu, C.~Prescod-Weinstein and E.~I.~Sfakianakis,
 ``Preheating after multifield inflation with nonminimal couplings, III: Dynamical spacetime results,''
  Phys.\ Rev.\ D {\bf 97}, no. 2, 023528 (2018)
  [arXiv:1610.08916 [astro-ph.CO]].

\bibitem{Ema:2016dny} 
  Y.~Ema, R.~Jinno, K.~Mukaida and K.~Nakayama,
``Violent Preheating in Inflation with Nonminimal Coupling,''
  JCAP {\bf 1702}, no. 02, 045 (2017)
  doi:10.1088/1475-7516/2017/02/045
  [arXiv:1609.05209 [hep-ph]].

   
\bibitem{Hertzberg} M. P. Hertzberg, ``On inflation with non-minimal coupling," JHEP {\bf 1011} (2010): 023 [arXiv:1002.2995 [hep-ph]].
  

\bibitem{unitarity} 
R. N. Lerner and J. McDonald, ``A unitarity-conserving Higgs inflation model," Phys. Rev. D82 (2010): 103525 [arXiv:1005.2978 [hep-ph]]; 
S. Ferrara, R. Kallosh, A. Linde, A. Marrani, and A. Van Proeyen, ``Superconformal symmetry, NMSSM, and inflation," Phys. Rev. D83 (2011): 025008 [arXiv:1008.2942 [hep-th]]; 
F. Bezrukov, A. Magnin, M. Shaposhnikov, and S. Sibiryakov, ``Higgs inflation: Consistency and generalizations," JHEP 1101 (2011): 016 [arXiv:1008.5157 [hep-ph]]; 
G. F. Giudice and H. M. Lee, ``Unitarizing Higgs inflation," Phys. Lett. B694 (2011): 294 [arXiv:1010.1417 [hep-ph]]; 
F. Bezrukov, D. Gorbunov, and M. Shaposhnikov, ``Late and early time phenomenology of Higgs-dependent cutoff," arXiv:1106.5019 [hep-ph]]; 
R. N. Lerner and J. McDonald, ``Unitarity-violation in generalized Higgs inflation models," arXiv:1112.0954 [hep-ph]; 
%
D. A. Demir, ``Gravi-Natural Higgs and Conformal New Physics," arXiv:1207.4584 [hep-ph];
M.~Atkins and X.~Calmet,
``Remarks on Higgs Inflation,''
  Phys.\ Lett.\ B {\bf 697}, 37 (2011)
  [arXiv:1011.4179 [hep-ph]];
  %
  X.~Calmet and R.~Casadio,
  ``Self-healing of unitarity in Higgs inflation,''
  Phys.\ Lett.\ B {\bf 734}, 17 (2014)
  [arXiv:1310.7410 [hep-ph]].

\bibitem{BezrukovInflaton} F. Bezrukov, ``The Higgs field as an inflaton," Class. Quant. Grav. {\bf 30}, 214001 (2013), arXiv:1307.0708 [hep-ph].

\bibitem{Burgess} C. P. Burgess, H. M. Lee, and M. Trott, ``Power-counting and the validity of the classical approximation during inflation," JHEP {\bf 0909} (2009): 103 [arXiv:0902.4465 [hep-ph]]; C. P. Burgess, H. M. Lee, and M. Trott, ``Comment on Higgs inflation and naturalness," JHEP {\bf 1007} (2010): 007 [arXiv:1002.2730 [hep-ph]].


\bibitem{Rubio:2018ogq} 
  J.~Rubio,
``Higgs inflation,''
  arXiv:1807.02376 [hep-ph].

  
\bibitem{KMS} D. I. Kaiser, E. A. Mazenc, and E. I. Sfakianakis, ``Primordial bispectrum from multifield inflation with nonminimal couplings," Phys. Rev. D {\bf 87}, 064004 (2013), arXiv:1210.7487 [astro-ph.CO].


\bibitem{GKS} R. N. Greenwood, D. I. Kaiser, and E. I. Sfakianakis, ``Multifield dynamics of Higgs inflation," Phys. Rev. D {\bf 87}, 044038 (2013), arXiv:1210.8190 [hep-ph].



\bibitem{SSK} K. Schutz, E. I. Sfakianakis, and D. I. Kaiser, ``Multifield inflation after Planck: Isocurvature modes from nonminimal couplings," Phys. Rev. D {\bf 89}, 064044 (2014), arXiv:1310.8285 [astro-ph.CO].



  
  \bibitem{HGF} 
  P.~Adshead, E.~Martinec, E.~I.~Sfakianakis and M.~Wyman,
  ``Higgsed Chromo-Natural Inflation,''
  JHEP {\bf 1612}, 137 (2016)
  [arXiv:1609.04025 [hep-th]].
  ;
   P.~Adshead and E.~I.~Sfakianakis,
``Higgsed Gauge-flation,''
  JHEP {\bf 1708}, 130 (2017)
  [arXiv:1705.03024 [hep-th]].
  
  
  
  \bibitem{tensorNG} 
  A.~Agrawal, T.~Fujita and E.~Komatsu,
``Large tensor non-Gaussianity from axion-gauge field dynamics,''
  Phys.\ Rev.\ D {\bf 97}, no. 10, 103526 (2018)
  [arXiv:1707.03023 [astro-ph.CO]]
  ;
   B.~Thorne, T.~Fujita, M.~Hazumi, N.~Katayama, E.~Komatsu and M.~Shiraishi,
``Finding the chiral gravitational wave background of an axion-SU(2) inflationary model using CMB observations and laser interferometers,''
  Phys.\ Rev.\ D {\bf 97}, no. 4, 043506 (2018)
  [arXiv:1707.03240 [astro-ph.CO]]
  ;
  A.~Agrawal, T.~Fujita and E.~Komatsu,
 ``Tensor Non-Gaussianity from Axion-Gauge-Fields Dynamics : Parameter Search,''
  JCAP {\bf 1806}, no. 06, 027 (2018)
  [arXiv:1802.09284 [astro-ph.CO]].
  






    
  \bibitem{weinberg}
  S. Weinberg, Cosmology (New York: Oxford University Press, 2008).
  


\bibitem{Lyth:2007jh} 
  D.~H.~Lyth,
``The curvature perturbation in a box,''
  JCAP {\bf 0712}, 016 (2007)
  [arXiv:0707.0361 [astro-ph]].




  
  \bibitem{Felder:1999wt} 
  G.~N.~Felder, L.~Kofman and A.~D.~Linde,
``Gravitational particle production and the moduli problem,''
  JHEP {\bf 0002}, 027 (2000)
  [hep-ph/9909508].
  
  \bibitem{Traschen:1990sw} 
  J.~H.~Traschen and R.~H.~Brandenberger,
``Particle Production During Out-of-equilibrium Phase Transitions,''
  Phys.\ Rev.\ D {\bf 42}, 2491 (1990).
  \bibitem{Parry:1998pn} 
  M.~Parry and R.~Easther,
  ``Preheating and the Einstein field equations,''
  Phys.\ Rev.\ D {\bf 59}, 061301 (1999)
  [hep-ph/9809574].
  \bibitem{Bassett:1998wg} 
  B.~A.~Bassett, D.~I.~Kaiser and R.~Maartens,
``General relativistic preheating after inflation,''
  Phys.\ Lett.\ B {\bf 455}, 84 (1999)
  [hep-ph/9808404].
  
    \bibitem{Easther:1999ws} 
  R.~Easther and M.~Parry,
  ``Gravity, parametric resonance and chaotic inflation,''
  Phys.\ Rev.\ D {\bf 62}, 103503 (2000)
  [hep-ph/9910441].
  
  
  
\bibitem{Bassett:1999mt} 
  B.~A.~Bassett, F.~Tamburini, D.~I.~Kaiser and R.~Maartens,
  ``Metric preheating and limitations of linearized gravity. 2.,''
  Nucl.\ Phys.\ B {\bf 561}, 188 (1999)
  doi:10.1016/S0550-3213(99)00495-2
  [hep-ph/9901319].
  
\bibitem{Bassett:1999ta} 
  B.~A.~Bassett, C.~Gordon, R.~Maartens and D.~I.~Kaiser,
  ``Restoring the sting to metric preheating,''
  Phys.\ Rev.\ D {\bf 61}, 061302 (2000)
  doi:10.1103/PhysRevD.61.061302
  [hep-ph/9909482].

\bibitem{Tsujikawa:2002nf} 
  S.~Tsujikawa and B.~A.~Bassett,
  ``When can preheating affect the CMB?,''
  Phys.\ Lett.\ B {\bf 536}, 9 (2002)
  doi:10.1016/S0370-2693(02)01813-0
  [astro-ph/0204031].

  \bibitem{Afshordi:2000nr} 
  N.~Afshordi and R.~H.~Brandenberger,
  ``Super Hubble nonlinear perturbations during inflation,''
  Phys.\ Rev.\ D {\bf 63}, 123505 (2001)
  [gr-qc/0011075].
  
  \bibitem{Finelli:1998bu} 
  F.~Finelli and R.~H.~Brandenberger,
  ``Parametric amplification of gravitational fluctuations during reheating,''
  Phys.\ Rev.\ Lett.\  {\bf 82}, 1362 (1999)
  [hep-ph/9809490].
  
\bibitem{Renaux-Petel:2015mga} 
  S.~Renaux-Petel and K.~Turzy?ski,
  ``Geometrical Destabilization of Inflation,''
  Phys.\ Rev.\ Lett.\  {\bf 117}, no. 14, 141301 (2016)
  [arXiv:1510.01281 [astro-ph.CO]].
   
     \bibitem{Lozanov:2016pac} 
  K.~D.~Lozanov and M.~A.~Amin,
  ``The charged inflaton and its gauge fields: preheating and initial conditions for reheating,''
  JCAP {\bf 1606}, no. 06, 032 (2016)
  [arXiv:1603.05663 [hep-ph]].


\bibitem{Chen:2014cwa} 
  X.~Chen, M.~H.~Namjoo and Y.~Wang,
  ``Models of the Primordial Standard Clock,''
  JCAP {\bf 1502}, no. 02, 027 (2015)
  [arXiv:1411.2349 [astro-ph.CO]].

\bibitem{Chen:2015lza} 
  X.~Chen, M.~H.~Namjoo and Y.~Wang,
  ``Quantum Primordial Standard Clocks,''
  JCAP {\bf 1602}, no. 02, 013 (2016)
  [arXiv:1509.03930 [astro-ph.CO]].


  
 
  


\bibitem{Freese:2017ace} 
  K.~Freese, E.~I.~Sfakianakis, P.~Stengel and L.~Visinelli,
``The Higgs Boson can delay Reheating after Inflation,''
  JCAP {\bf 1805}, no. 05, 067 (2018)
  [arXiv:1712.03791 [hep-ph]].
  
  
  \bibitem{Adshead:2017xll} 
  P.~Adshead, J.~T.~Giblin and Z.~J.~Weiner,
``Non-Abelian gauge preheating,''
  Phys.\ Rev.\ D {\bf 96}, no. 12, 123512 (2017)
  [arXiv:1708.02944 [hep-ph]].
  
  
  \bibitem{Enqvist:2015sua} 
  K.~Enqvist, S.~Nurmi, S.~Rusak and D.~Weir,
``Lattice Calculation of the Decay of Primordial Higgs Condensate,''
  JCAP {\bf 1602}, no. 02, 057 (2016)
  [arXiv:1506.06895 [astro-ph.CO]].
    
  \bibitem{Lindealpha1} R. Kallosh and A. Linde, ``Non-minimal inflationary attractors," JCAP 1310 (2013): 033, arXiv:1307.7938 [hep-th].






\bibitem{Bezrukov:2014bra} 
  F.~Bezrukov and M.~Shaposhnikov,
``Higgs inflation at the critical point,''
  Phys.\ Lett.\ B {\bf 734}, 249 (2014)
  doi:10.1016/j.physletb.2014.05.074
  [arXiv:1403.6078 [hep-ph]].


\bibitem{Ezquiaga:2017fvi} 
  J.~M.~Ezquiaga, J.~Garcia-Bellido and E.~Ruiz Morales,
``Primordial Black Hole production in Critical Higgs Inflation,''
  Phys.\ Lett.\ B {\bf 776}, 345 (2018)
  doi:10.1016/j.physletb.2017.11.039
  [arXiv:1705.04861 [astro-ph.CO]].


\bibitem{Masina:2018ejw} 
  I.~Masina,
``Ruling out Critical Higgs Inflation?,''
  arXiv:1805.02160 [hep-ph].

\bibitem{Enckell:2018kkc} 
  V.~M.~Enckell, K.~Enqvist, S.~Rasanen and E.~Tomberg,
  ``Higgs inflation at the hilltop,''
  JCAP {\bf 1806}, no. 06, 005 (2018)
  doi:10.1088/1475-7516/2018/06/005
  [arXiv:1802.09299 [astro-ph.CO]].

\bibitem{Rasanen:2017ivk} 
  S.~Rasanen and P.~Wahlman,
  ``Higgs inflation with loop corrections in the Palatini formulation,''
  JCAP {\bf 1711}, no. 11, 047 (2017)
  doi:10.1088/1475-7516/2017/11/047
  [arXiv:1709.07853 [astro-ph.CO]].


\end{thebibliography}
\end{document}